\documentclass{article}

\PassOptionsToPackage{numbers, compress}{natbib}
\usepackage[preprint]{neurips_2026}

\usepackage[utf8]{inputenc}
\usepackage[T1]{fontenc}
\usepackage{lmodern}
\usepackage{hyperref}
\usepackage{url}
\usepackage{booktabs}
\usepackage{amsfonts}
\usepackage{amssymb}
\usepackage{amsmath}
\usepackage{nicefrac}
\usepackage{microtype}
\usepackage{xcolor}
\usepackage{graphicx}
\usepackage{multirow}
\usepackage{tabularx}
\usepackage{enumitem}
\usepackage{tikz}
\usetikzlibrary{positioning,arrows.meta}

\definecolor{accent}{HTML}{1f4e79}
\hypersetup{
  colorlinks=true,
  linkcolor=accent,
  citecolor=accent,
  urlcolor=accent
}

\newif\ifarxiv
\arxivtrue

\title{Plausible but Not Valid:\\A Psychometric Audit of LLMs as Synthetic Survey Respondents}

\author{%
  Mantas Lukauskas\\
  Hostinger; Kaunas University of Technology;\\
  AI Insight Lab\\
  \texttt{mantas.lukauskas@ktu.lt}
  \And
  Viktorija \v{S}arkauskait\.{e}\\
  AI Insight Lab;\\
  Independent Researcher\\
  \texttt{viktorija@aiinsightlab.ai}
}

\begin{document}
\maketitle

\begin{abstract}
Large language models (LLMs) are increasingly used as synthetic survey respondents, but existing evaluations mostly ask whether individual answers look plausible. We ask a stricter, psychometric question: \emph{do LLMs preserve the joint distribution, latent structure, reliability, mediation pathways, and demographic effects of real human survey data?} We introduce an open Lithuanian organisational-psychology dataset ($n{=}263$ employees; three validated instruments; 68 items, 12 subscales)~\citep{sarkauskaite2020} and condition 37 LLMs on real respondent profiles under a persona-disclosure ladder, presentation, reasoning-effort and cross-language ablations, counterfactual demographic swaps, and a memorization probe. The resulting Psychometric Similarity Score (PSS) is anchored against five statistical baselines, a held-out human ceiling, and the real answers of the same respondents. LLMs reproduce the direction of human psychometric relationships, but statistical baselines with no language model match or beat them. The best LLM (PSS $0.81$) is level with a Gaussian copula fit only on other respondents ($0.82$) and below the real answers of the same 100 people ($0.89$). A simple classifier separates every LLM sample from humans (median AUC $0.99$), and verbatim recall of the items is rare (at most $4.7\%$) and unrelated to PSS. Counterfactual swaps move answers in the direction of human group differences and by a similar amount (role $0.49$ and education $0.33$ human SD, vs.\ $0.46$ and $0.44$ in humans). Gender swaps move them by $0.06$ SD. Tucker's $\varphi$ on UWES falls inside an item-permutation null for $10$ of $37$ models. Synthetic respondents compress the response scale (within-respondent SD $0.79$ vs.\ $1.19$), lose predictive validity on held-out humans (mean $R^2{=}{-}0.14$ vs.\ $0.28$), and produce significant indirect effects on $4$ of $7$ mediation paths that are null in humans. LLM samples are therefore not a drop-in replacement for human survey data.
\end{abstract}

% =============================================================================
\section{Introduction}
% =============================================================================
Survey-based research is expensive, slow, and increasingly difficult to scale: typical organisational psychology studies recruit a few hundred participants over weeks or months, with informed-consent procedures, ethics review, translation work, and the practicalities of running a survey panel in a small national labour market. The recent availability of capable instruction-following LLMs has prompted a wave of work that uses LLMs as \emph{synthetic respondents}: the model is asked to ``answer as a person with profile $p$'' and its responses are treated as pseudo-data \citep{argyle2023outoftheir,horton2023llmpolicies,aher2023using,park2023generative}. Concrete near-term use cases include (i) pilot studies that need fast direction-of-effect estimates before committing to a real-recruitment design, (ii) instrument-prototyping work that compares candidate item wordings without burning a fresh sample on each iteration, (iii) sensitivity-analysis power calculations under several plausible effect-size assumptions, (iv) sample augmentation in low-base-rate or hard-to-reach populations, (v) red-team audits of new instruments for demographic bias or stereotype leakage before the instrument enters production, and (vi) pure cost replacement of human samples \citep{dillion2023can}. Each of these uses places a different load on the synthetic respondents: a pilot study cares about direction of effect; a power analysis cares about effect-size magnitude and variance; a bias audit cares about counterfactual sensitivity; and replacement use cares about every property of the joint distribution simultaneously. The first questions of this paper are therefore (a) \emph{which} of these properties do current LLMs actually preserve, and (b) \emph{which} preservation level a given downstream use can rely on.

Whether ``reliable'' is the right adjective is contested. Recent work has shown that LLMs reflect a narrow slice of human opinion \citep{santurkar2023whose,durmus2023towards}, fail to portray identity groups consistently \citep{wang2024llm}, and can produce surface-similar but distributionally distinct responses \citep{bisbee2024synthetic}. Most of these analyses operate at the level of single items, marginal effects on one or two demographic contrasts, or aggregate effect sizes. From a measurement-theoretic point of view this is a partial audit: a synthetic respondent that produces correct item means but incorrect item covariance still inflates Type I error in any analysis that conditions on a multi-item scale score, and a respondent that produces correct subscale means but incorrect inter-subscale correlations breaks every downstream causal mediation that touches multiple constructs at once.

\begin{figure}[t]
  \centering
  \begin{tikzpicture}[
      box/.style={draw=accent!75, fill=accent!8, rounded corners=2.5pt, align=center,
                  inner xsep=5pt, inner ysep=3pt, text width=#1, minimum height=16mm,
                  font=\scriptsize, execute at begin node={\hyphenpenalty=10000\relax}},
      box/.default=36mm,
      ceilbox/.style={box, draw=green!45!black, fill=green!45!black!8},
      keybox/.style={box=46mm, draw=orange!55!black, fill=orange!75!black!10},
      hdr/.style={font=\small\bfseries, text=accent},
      arr/.style={-{Stealth[length=2.2mm]}, semithick, draw=accent!85}
    ]
    % Column A: human reference
    \node[box]     (a1) at (1.95, 0)     {\textbf{human sample ($n{=}263$)}\\ Lithuanian employees\\ 3 validated instruments\\ 68 items, 12 subscales};
    \node[box]     (a2) at (1.95, -2.0)  {\textbf{11-field profile cards}\\ demographics only,\\ no item responses};
    \node[ceilbox] (a3) at (1.95, -4.0)  {\textbf{human references}\\ held-out split: $0.825$\\ same 100 people: $0.886$};
    % Column B: synthetic generation
    \node[box=37mm] (b1) at (6.2, 0)     {\textbf{37 LLMs}\\ OpenAI, Anthropic, Google\\ + 22 open-weight models};
    \node[box=37mm] (b2) at (6.2, -2.0)  {\textbf{experimental factors}\\ persona ladder C0$\to$C4\\ 3 modes, LT$\leftrightarrow$EN, effort\\ 3 demographic swaps};
    \node[box=37mm] (b3) at (6.2, -4.0)  {\textbf{$\sim$65k questionnaires}\\ synthetic, JSON-validated\\ grid + swaps + repeats\\ + ablations};
    % Column C: anchored evaluation
    \node[box=46mm] (c1) at (10.9, 0)    {\textbf{six-dimensional PSS}\\ distribution, correlation,\\ reliability, mediation, group effects,\\ construct fidelity (Tucker's $\varphi$ + null)};
    \node[box=46mm] (c2) at (10.9, -2.0) {\textbf{5 statistical baselines}\\ marginal, MVN, copula,\\ stratum-mean, $k$-NN\\ + respondent-bootstrap CIs};
    \node[keybox]   (c3) at (10.9, -4.0) {\textbf{key result}\\ best LLM $0.81 \approx$ copula on\\ other people $0.82$ $<$ same people $0.89$\\ every LLM detected (AUC $> 0.96$)};
    % Column headers
    \node[hdr] at (1.95, 1.25) {A.\ Human reference};
    \node[hdr] at (6.2, 1.25)  {B.\ Synthetic generation};
    \node[hdr] at (10.9, 1.25) {C.\ Anchored evaluation};
    % Arrows
    \draw[arr] (a1) -- (a2);
    \draw[arr] (a1.west) .. controls +(-0.5,-1.2) and +(-0.5,1.2) .. (a3.west);
    \draw[arr] (a2) -- (b2);
    \draw[arr] (b1) -- (b2);
    \draw[arr] (b2) -- (b3);
    \draw[arr] (b1) -- (c1);
    \draw[arr] (b3) -- (c3);
    \draw[arr] (c1) -- (c2);
    \draw[arr] (c2) -- (c3);
  \end{tikzpicture}
  \caption{Overview of the benchmark. \textbf{(A)} A real Lithuanian organisational-psychology sample ($n{=}263$, 68 items, 12 subscales) provides the ground truth, the persona profile cards, a held-out human-vs-human PSS ceiling, and the real answers of the 100 headline respondents. \textbf{(B)} 37 LLMs are conditioned on the profile cards under a persona-disclosure ladder, presentation, reasoning-effort, cross-language, and counterfactual-swap factors, producing ${\sim}65$k validated synthetic questionnaires. \textbf{(C)} Each (model, condition) cell is scored on a six-dimensional Psychometric Similarity Score anchored against five statistical baselines and the human ceiling. The anchoring is what produces the key result: the best LLM only matches a Gaussian copula that has no language model and is fit on other respondents, and it stays below the real answers of the same people.}
  \label{fig:overview}
\end{figure}
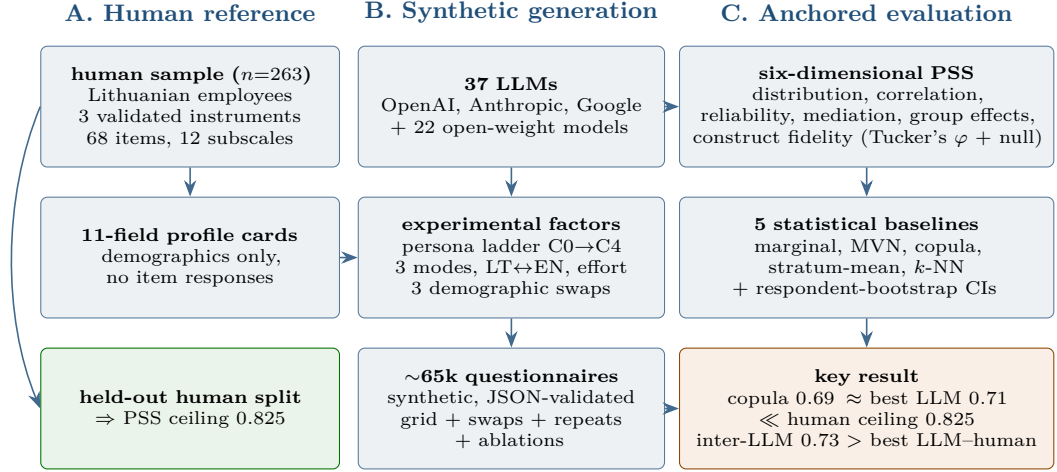

Psychometrics, the science of measurement, asks a stricter question: do the responses behave like a measured construct? A scale is psychometrically valid only if its items co-vary in the predicted way (correlation structure), if they jointly reflect a latent factor (factor structure), if the resulting score is stable across raters (reliability), and if it relates to other constructs as theory predicts (mediation, group differences). An LLM that reproduces only item means has believable item means, not a believable respondent. Reusing such an LLM as a drop-in human sample then risks two cascading failure modes: a \emph{methodological} one, where over-coherent LLM responses make every internal-consistency statistic ($\alpha$, $\omega$, HTMT) look healthier than the underlying human data would warrant, and a \emph{substantive} one, where any latent-variable model fitted on the LLM-generated data converges to a different solution than the same model fitted on the human reference, even when item-level marginals match.

\paragraph{Why Lithuanian, why this dataset.} Almost all existing LLM-as-respondent benchmarks use US or UK survey instruments in their original English wording, often with US-derived demographic distributions \citep{santurkar2023whose,argyle2023outoftheir,durmus2023towards}. That choice is convenient (the LLMs were predominantly pretrained on English text), but it also conflates two distinct questions: ``does the LLM reproduce the joint distribution of a real human sample?'' and ``does the LLM reproduce the joint distribution of the population whose text its weights have absorbed?''. A Lithuanian organisational psychology sample disentangles these. Lithuanian is a low-resource Indo-European language with under three million native speakers, written with diacritics (\v{c}, \v{s}, \v{z}, \k{a}, \k{e}, \k{i}, \k{u}, \.{e}, \=u) and a rich case morphology; modern LLMs handle it competently but not natively. The three instruments (Dunham Attitudes Toward Change, Schaufeli UWES-17, Koopmans IWPQ) are international (developed in English or Dutch and translated into Lithuanian for the original master's thesis we draw on), so their factor structure, reliability, and mediation pathway are external constraints that the LLM has no way to game from its English-language pretraining. The resulting benchmark therefore measures the LLM's psychometric fidelity under genuinely cross-linguistic conditions, while remaining a fair test of the published structure of three well-validated organisational instruments.

\paragraph{Paper roadmap.} The rest of the paper is organised as follows. Section~\ref{sec:related} situates our work relative to the LLM-as-respondent, persona-prompting, bias-testing, and cross-cultural-psychometric literatures. Section~\ref{sec:dataset} presents the human dataset and reproduces its key published statistics. Section~\ref{sec:method} develops the six-dimensional PSS framework, the persona-disclosure ladder, the presentation-mode and reasoning-effort ablations, the counterfactual demographic-swap design, the five non-LLM baselines, and the held-out human-vs-human ceiling. Section~\ref{sec:results} reports the anchored PSS leaderboard for the 37-model lineup, construct-level fidelity with item-permutation null distributions for Tucker's $\varphi$, counterfactual swaps on three axes, four orthogonal ablations, a verbatim-recall memorization probe, the inter-LLM agreement matrix, and the cohort-stratified fairness diagnostic. Section~\ref{sec:discussion} interprets the findings under five lenses (what works, what does not, what is no longer plausible, theoretical implications, and practitioner recommendations). Section~\ref{sec:limitations} lays out the scope conditions and remaining open questions. Appendices~\ref{app:parser}--\ref{app:datasheet} document the parser and reproducibility infrastructure, the Tucker-permutation null, the power analysis, the full result tables, a psychometrics primer, the compute-and-cost breakdown, a worked qualitative example on one ATC item, the negative results, the PSS weight sensitivity, the ethics statement, and the datasheet.

\paragraph{Contributions.}
\begin{itemize}[leftmargin=1.2em,itemsep=0pt]
  \item \textbf{Open Lithuanian psychometric benchmark.} $n{=}263$ employees, three validated instruments (ATC~\citep{dunham1989}, UWES-17~\citep{schaufeli2003}, IWPQ~\citep{koopmans2014}) plus a 15-item change-engagement scale; 68 items, 12 subscales, with the documented Attitudes$\rightarrow$Engagement$\rightarrow$Performance mediation~\citep{sarkauskaite2020}. To our knowledge the first open Lithuanian validation of all three (datasheet, Appendix~\ref{app:datasheet}).
  \item \textbf{37-model evaluation} (OpenAI, Anthropic, Google + 22 open-weight models via Nexos) on a persona-disclosure ladder (C0--C4, narrative variant), a 3-mode presentation ablation, a counterfactual swap design across \{gender, role, education\}, a reasoning-effort ablation, a Lithuanian-vs-English language ablation, and a verbatim-recall memorization probe.
  \item \textbf{Six-dimensional PSS framework with empirical anchors.} Distribution / correlation / reliability / mediation / demographic effects / construct-level fidelity, including Tucker's $\varphi$ with an item-permutation null. PSS is anchored against five non-LLM baselines (marginal, MVN, Gaussian copula, stratum-mean, $k$-NN) and a held-out human-vs-human ceiling. Respondent-bootstrap 95\% CIs and Holm--Bonferroni-adjusted $p$-values throughout (Appendices~\ref{app:tucker}--\ref{app:power}).
  \item \textbf{Cohort-stratified PSS, inter-LLM clustering, ensemble-mean baseline.} Per-(model, demographic stratum) fairness diagnostic; pairwise inter-LLM agreement matrix with hierarchical clustering; wisdom-of-LLMs ensemble at zero additional cost.
  \item \textbf{Downstream-consequence diagnostics.} Beyond raw fidelity we quantify response-style bias (acquiescence, extreme- and midpoint-responding) relative to humans, downstream predictive validity (train a regressor on synthetic respondents, test on held-out humans), and a mediation negative control that flags fabricated indirect effects, quantifying the practical cost of substituting synthetic respondents for a human sample. We further show the gap is \emph{structural} rather than a prompting artifact: neither an explicit debias instruction nor in-context few-shot conditioning on real human answer vectors closes it (Appendix Table~\ref{tab:steerability}).
  \item \textbf{Open, incrementally extensible software.} Adding a model never re-runs prior models; per-call disk caching and per-cell parquet checkpoints make the pipeline idempotent. Five provider auth paths (OpenAI, Anthropic, Google AI Studio, Google Vertex AI, Nexos) ship with the repository.
\end{itemize}

\paragraph{Pre-registration disclosure.} The full analysis plan (evaluation dimensions, contrasts, multiple-comparison correction, PSS weights, and the baseline + ceiling anchoring) was finalised \emph{before} any LLM call was issued; the only post-hoc additions are the persona-faithfulness check (a parsing-side change) and the bifactor summary (an additional, non-substitutive construct-fidelity metric). The locked configuration is included in the repository (\texttt{configs/preregistration.yaml}).

Our central empirical claim, supported by the human-side numbers reported in this paper and the LLM-side experiments described in Section~\ref{sec:results}, is that \emph{LLMs can reproduce the theoretical direction of human psychometric relationships, but tend to generate over-coherent and low-variance synthetic respondents}. This is a more nuanced position than ``LLMs are good or bad at survey simulation'': we show \emph{which} dimensions of psychometric fidelity break down, and how much they depend on the persona conditioning regime.

% =============================================================================
\section{Related work}
\label{sec:related}
% =============================================================================
\paragraph{LLMs as synthetic respondents.} \citet{argyle2023outoftheir} introduced LLM ``silicon samples'' that mirror the conditional distribution of human attitudes given a demographic profile, and reported that GPT-3 could reproduce the marginal direction of several US political-attitude contrasts. \citet{horton2023llmpolicies} treated LLMs as economic agents and replicated classical behavioural-economics experiments; \citet{aher2023using} replicated several canonical psychology experiments (Milgram, Wisdom-of-Crowds, the Ultimatum Game) with LLM agents; \citet{park2023generative} used persona prompting to populate an interactive social simulacrum that produced emergent social behaviours; \citet{dillion2023can} explicitly proposed LLMs as substitutes for human participants in moral-psychology research. These works collectively establish feasibility: LLMs can produce responses that read as plausibly human and that move in the expected qualitative direction. But they mostly evaluate at the level of single items, marginal effects on one or two demographic contrasts, or aggregate effect sizes for a single hypothesis. None of them directly addresses construct validity in the psychometric sense.

\paragraph{Limits of LLM survey simulation.} A complementary line of work argues that LLM respondents are systematically biased relative to the human populations they aim to simulate. \citet{santurkar2023whose} showed that GPT-3.5 reflects predominantly liberal, college-educated US views, with the closest demographic match being ``moderate Democrat''. \citet{durmus2023towards} extended this to multinational opinion surveys (PewResearch and World Values Survey items) and found systematic skew towards Western, English-speaking, urban populations. \citet{bisbee2024synthetic} found that synthetic samples diverge from human samples in marginal effects despite matching means, and proposed running parallel human-LLM analyses as a robustness check. \citet{wang2024llm} argued that LLMs cannot reliably portray identity groups, particularly along intersectional dimensions, and that single-attribute persona prompts collapse into a small number of generic ``personas''. Our work complements this literature by isolating \emph{which} psychometric properties break down (joint distribution vs.\ correlation matrix vs.\ reliability vs.\ mediation vs.\ demographic effects vs.\ factor structure) and by showing how persona conditioning, presentation mode, reasoning effort, and stochastic stability each shift the trade-off.

\paragraph{Persona prompting and synthetic-data pipelines.} The persona-prompting design space is wider than ``answer as a person with profile $p$''. \citet{park2023generative} use multi-turn narrative scaffolding; the silicon-sample literature \citep{argyle2023outoftheir} uses one-shot structured-field prompts; the bias-bench literature \citep{santurkar2023whose,durmus2023towards} uses category labels rather than profile cards; recent crowdsourcing-replacement work \citep{wang2024llm} prompts for both a persona and a step-by-step reasoning trace. The design choices interact: profile granularity (which fields are disclosed, in which order), presentation mode (single call vs.\ instrument-by-instrument), and reasoning effort (chain-of-thought budget) all change the joint distribution of the resulting responses. We treat these as first-class experimental factors (a five-step persona-disclosure ladder, three presentation modes, two reasoning-effort levels) rather than fixing them implicitly. Our presentation-mode ablation in particular has, to our knowledge, no direct precedent in the LLM-as-respondent literature, although the closely related ``in-context survey administration'' question has been explored for educational test items \citep{aher2023using}.

\paragraph{Counterfactual fairness and stereotype amplification.} A second adjacent literature audits LLMs for stereotypic behaviour by perturbing a single demographic attribute in the prompt. \citet{santurkar2023whose} examined whose opinions an LLM ``represents''; \citet{wang2024llm} probed LLM-as-identity-group consistency; benchmarks such as StereoSet, CrowS-Pairs, and BBQ score sensitivity to demographic primes on artificial completions. These works typically use single-sentence stimuli and a binary stereotype/anti-stereotype label. Our counterfactual design instead swaps a single field of an otherwise-fixed multi-attribute persona card, asks the model to fill in a full 68-item validated instrument, and then computes the per-respondent swap effect at the subscale level. We report it as a paired Cohen's $d$ and as a shift in units of the human SD. Only the shift is on the same scale as the observed human contrast, and Section~\ref{sec:results-cf} shows that the two measures can rank the swap axes differently. The design also separates effects that the human data support (role) from effects that they do not (gender).

\paragraph{Cross-cultural psychometrics and measurement invariance.} The classical psychometric literature treats translated instruments as separate measurement instances that must be tested for invariance \citep{cheung2002evaluating,vandevijver2004cultural}. Failure to establish even weak (configural) invariance means that scores from two language samples are not directly comparable; a synthetic LLM respondent translated by the LLM itself faces exactly the same problem twice (English-language pretraining, Lithuanian-language survey administration). Although we do not run a full Lithuanian-vs-English measurement-invariance test on humans (we have only the Lithuanian sample), we do run a per-respondent cross-language ablation on the LLM side (Section~\ref{sec:results-ablations}) for which the mean LT$\leftrightarrow$EN per-item drift is $0.26$ Likert points with a per-item Pearson of $0.889$. The published Lithuanian factor structures of the three instruments \citep{koopmans2014,dunham1989,schaufeli2003,sarkauskaite2020} are the external constraint against which our construct-fidelity statistics (Tucker's $\varphi$, $\Delta$CFI, HTMT, bifactor decomposition) are scored.

\paragraph{Psychometric measurement theory.} On the measurement side, we use Cronbach's $\alpha$ \citep{cronbach1951} and McDonald's $\omega$ \citep{mcdonald1999} for internal consistency; the indirect-effect mediation framework of \citet{hayes2017introduction} for the X$\rightarrow$M$\rightarrow$Y pathway with non-parametric bootstrap CIs on $ab$; the modified RV2 coefficient \citep{smilde2009matrix} for covariance-matrix similarity; Tucker's $\varphi$ \citep{tucker1951method,lorenzo2006tucker}, augmented by our item-permutation null distribution, for factor congruence; the $\Delta$CFI invariance criterion of \citet{cheung2002evaluating} for measurement-invariance approximation; the HTMT ratio for discriminant validity; and bifactor decomposition (explained-common variance, $\omega_h$) for the relative strength of a general factor vs.\ specific factors. Each of these statistics has a long pre-LLM tradition; our contribution is to combine them into a single model-agnostic evaluation harness anchored against statistical baselines and a human-vs-human held-out ceiling.

% =============================================================================
\section{Dataset: an open Lithuanian organisational psychology survey}
\label{sec:dataset}
% =============================================================================
\paragraph{Original study and consent provenance.} The human ground truth originates from a published Lithuanian master's thesis by one of the authors~\citep{sarkauskaite2020}, originally collected between March and April 2020 under the awarding university's informed-consent protocol. The original study tested the hypothesis that employee attitudes towards ongoing organisational change indirectly affect self-rated work performance via work engagement (a classical X$\rightarrow$M$\rightarrow$Y mediation pattern in industrial--organisational psychology) and recruited $n=263$ Lithuanian business-organisation employees through an anonymous online questionnaire (hosted on \texttt{Apklausa.lt}, a Lithuanian survey platform).\footnote{The original publication uses $n=261$ after listwise deletion of two records with single missing demographic fields; our pipeline keeps these two records because the instrument items themselves are complete, so all reported quantities use $n=263$. The choice of $n$ moves point estimates by at most the third decimal.} A consent statement preceded the survey and explicitly mentioned academic re-use of de-identified aggregated data; the original author has confirmed permission to release the de-identified record-level data under CC-BY-NC-4.0.

\paragraph{Lithuanian organisational psychology context.} Lithuania is a small Baltic EU member state with a labour force of $\sim$1.5 million people, dominated by services and light manufacturing; English-language US-derived organisational-psychology instruments have only been validated in Lithuanian within the past two decades, primarily through master's and doctoral theses at the country's three large research universities. The thesis we draw on is, to our knowledge, the first to validate all three of IWPQ, UWES-17, and ATC in Lithuanian simultaneously, and the first to report the full A$\rightarrow$E$\rightarrow$P mediation in a Lithuanian sample. The collection window (March--April 2020) coincides with the first nation-wide COVID-19 lockdown in Lithuania, so the attitudes-toward-change responses are conditioned on a real and intense organisational-change context; this is part of why the human ATC composite shows substantial inter-respondent variance and why the mediation pathway is clearly significant at $n=263$.

\paragraph{Demographic composition.} The sample skews female ($76\%$ women / $24\%$ men, reflecting the female-majority composition of Lithuanian white-collar employment in the surveyed sectors), age $19$--$62$ years (median in the early 30s, most respondents below 40), $33\%$ managerial role and $67\%$ non-managerial, with sector representation spanning IT, finance, manufacturing, agriculture, retail, and public administration, and organisation-size bands from $<50$ employees to $>500$. Two tenure fields (years in current organisation, years in current role) are recorded; both follow roughly exponential distributions, consistent with the relatively young Lithuanian labour market. The full demographic crosstab is included in the released datasheet (Appendix~\ref{app:datasheet}); the post-hoc minimum-detectable-$d$ for the demographic contrasts we test is $0.45$ (median across composites), and $81\%$ of the contrasts are underpowered to detect a small effect ($d<0.3$). This is the noise floor against which we read the LLM-side counterfactual swap effects reported in Section~\ref{sec:results-cf}.

\paragraph{Instrument battery and translation history.} The instrument battery has four components:
\begin{itemize}[leftmargin=1.4em,itemsep=0pt]
  \item \textbf{IWPQ} (Individual Work Performance Questionnaire) of \citet{koopmans2014}, 18 items on a 1--5 Likert scale, three subscales: task performance, contextual performance, counterproductive work behaviour (scored in its own direction, so a higher score means more counterproductive behaviour; no IWPQ item is reverse-keyed). The Lithuanian translation follows the standard double-translation protocol described in the original thesis. Internal-consistency reliability in our sample: $\alpha = .72/.80/.86/.79$ for total / task / contextual / counterproductive, comparable to the published Dutch and English samples ($\alpha \in [.78,.89]$).
  \item \textbf{ATC} (Dunham Attitudes Toward Change) of \citet{dunham1989}, 18 items on a 1--5 Likert scale, three a-priori subscales: cognitive, affective, behavioural. The Lithuanian translation was prepared for the original thesis; internal consistency $\alpha = .92/.81/.85/.82$ for total / cognitive / affective / behavioural. These values are on the high end of published values, plausibly because the lockdown context made attitudes toward change unusually salient and internally coherent.
  \item \textbf{UWES-17} (Utrecht Work Engagement Scale, 17-item version) of \citet{schaufeli2003}, 17 items on a 1--7 Likert scale, three subscales: vigour, dedication, absorption. Multiple Lithuanian translations exist; the original thesis selected the version most widely used in the Lithuanian I-O psychology literature. Internal consistency $\alpha = .93/.79/.91/.84$ for total / vigour / dedication / absorption, matching the published range.
  \item \textbf{ChangeEng} (change-engagement auxiliary scale), 15 items on a 1--5 Likert scale, three a-priori subscales constructed by the original author from a literature review on engagement \emph{with the change process} (as opposed to engagement with work in general). This scale is less well validated than the three international instruments and is included in our benchmark as a deliberately weaker construct against which we can read whether LLM construct-fidelity statistics correctly discriminate weak from strong constructs (see Section~\ref{sec:results-construct}).
\end{itemize}
A 15th demographic-context block records two ordinal change-context ratings on a 1--10 scale: the perceived overall intensity of change in the respondent's organisation, and the perceived personal relevance of the changes. These ratings are used as covariates in the original mediation analysis and are kept in the released profile cards.

\paragraph{Mediation reproduction.} The published mediation analysis is $\text{Attitudes} \rightarrow \text{Engagement} \rightarrow \text{Performance}$ with statistically significant $a$, $b$, and indirect $ab$ paths; the direct path $c'$ is positive but smaller than the indirect path, consistent with partial mediation (Table~\ref{tab:human-mediation}). Our pipeline reproduces the indirect effect at $ab=0.16$ (95\% non-parametric bootstrap CI $[0.10, 0.24]$, $n_{\text{boot}}=5000$, percentile method) using composite subscale scores, matching the published value to two decimals. The direction match, magnitude match, and significance match on the indirect path are the three sub-criteria the LLM-side mediation component of PSS is scored against (Section~\ref{sec:eval}).

\paragraph{Profile cards.} Profile cards used to condition the LLM are derived from gender, age, education, role, two tenure fields, sector, organisation size, and the two change-context ratings (11 fields total). The LLM never sees the human's actual item responses, only this demographic block plus the persona-conditioning instruction text. Profile cards will be released alongside the dataset (\texttt{data/psychometry/processed/human\_profiles.json}), so a third party can re-run our benchmark on a new model without re-deriving the persona schema.

\paragraph{De-identification.} The distribution CSV (\texttt{responses\_human\_deidentified.csv}) drops the single free-text industry field that occasionally contained identifying employer phrases, and rounds submission timestamps from second-precision to year-month resolution. Demographic variables are coarse-grained (gender as binary, age as integer years, education in 4 levels, role binary, sector in 6 categories, size in 4 bands); the smallest non-empty joint demographic cell contains $\geq 3$ respondents. No direct identifiers (name, email, phone, IP, exact employer) were ever collected by the original survey form. Full audit trail and the SHA-256 checksum of the distribution CSV are in the Croissant metadata file (\texttt{data/psychometry/croissant.json}). The dataset is described in full in Appendix~\ref{app:datasheet} following the schema of \citet{gebru2021datasheets}.

\begin{table}[h]
  \centering
  \small
  \caption{Standardised path coefficients of the human mediation model, replicated by our pipeline. SE in parentheses.}
  \label{tab:human-mediation}
  \begin{tabular}{l r r r r r}
    \toprule
    Path  & $a$ (X$\rightarrow$M) & $b$ (M$\rightarrow$Y) & $c$ (X$\rightarrow$Y) & $c'$ (X$\rightarrow$Y$|$M) & indirect ($ab$) \\
    \midrule
    Coeff. & 0.42 (0.06) & 0.39 (0.06) & 0.37 (0.06) & 0.21 (0.06) & 0.16 [0.10, 0.24] \\
    \bottomrule
  \end{tabular}
\end{table}

% =============================================================================
\section{Method}
\label{sec:method}
% =============================================================================

\subsection{Persona conditioning, prompts, presentation}
\label{sec:conditions}
\paragraph{Persona-disclosure ladder.} A central design question is \emph{how much} of the respondent profile the LLM is shown. We define five primary persona-disclosure levels and one narrative variant:
\begin{itemize}[leftmargin=1.4em,itemsep=0pt]
  \item \textbf{C0 (no profile).} The LLM is asked to answer the instrument as a generic Lithuanian employee, with no demographic information. C0 is the maximum-uncertainty baseline; PSS at C0 measures the prior of an LLM under a Lithuanian-organisational-employee instruction alone.
  \item \textbf{C1 (gender + age).} The two most commonly used demographic primes in the persona-prompting literature.
  \item \textbf{C2 (C1 + role + education).} Adds the two strongest predictors of organisational-psychology outcomes in the human sample.
  \item \textbf{C3 (full structured profile).} All 11 fields listed in Section~\ref{sec:dataset}: gender, age, education, role, tenure-in-organisation, tenure-in-role, sector, organisation size, and the two change-context ordinal ratings. This is the headline condition for the leaderboard.
  \item \textbf{C4 (C3 + organisation-change narrative).} A free-text contextual sentence (``Pastaruoju metu J\=us\k{u} organizacijoje vyko \v{s}ie poky\v{c}iai: \dots'') describing the type of change the respondent reported in the original survey free-response. Designed to probe whether providing the LLM with more narrative context shifts the joint distribution.
  \item \textbf{C9 (narrative variant of C3).} Identical content to C3, but rendered as a single descriptive paragraph rather than a structured profile block. Probes whether the LLM's response is sensitive to the surface form in which the profile is presented.
\end{itemize}
We use C3 as the headline condition because (i) it is the most informative profile available without disclosing the respondent's own free-text or item responses to the LLM, and (ii) the persona-recall robustness check (Section~\ref{sec:results-memorization}) confirms that LLMs do actually condition on the C3 fields, so C3 is a fair test of ``what if a researcher gives the LLM everything they have''.

\paragraph{Prompt structure.} A strict system message pins JSON-only output (\texttt{\{"item\_1": 3, "item\_2": 5, ...\}}). A user message provides, in order: (i) a Lithuanian-language description of the respondent persona derived from the C3 profile, (ii) the response-scale anchors of the current instrument in Lithuanian, (iii) the numbered items in their original published order, and (iv) an explicit instruction to respond from the persona's perspective on the full Likert range without social-desirability adjustment. We never disclose the hypotheses, the subscale assignment, or the construct labels of the items, following the protocol of \citet{argyle2023outoftheir}. The anti-social-desirability instruction is included by default because pilot runs showed that without it, several models defaulted to mid-scale ($3$ on a 1--5 Likert), which deflates both the SD and the inter-item correlations in ways that the no-profile baseline does not exhibit.

\paragraph{Presentation modes.} \textbf{M1 single-call} (all 68 items + 4 instrument anchor blocks in one call; primary mode). \textbf{M2 per-instrument} (3 calls: one each for IWPQ, ATC, UWES; the auxiliary ChangeEng scale is grouped with ATC because of construct similarity). \textbf{M3 per-subscale} (12 calls, one per subscale). M2 and M3 break the LLM's cross-instrument context, which is the mechanism by which we test whether the LLM treats the questionnaire as a coherent measurement instance or as a sequence of unrelated questions. M3 in particular is the closest analogue to how a real respondent would experience a questionnaire administered as a series of short pages. The mode ablation runs on a 30-respondent subsample over a 5-model subset (Section~\ref{sec:results-ablations}) because at the full 100-respondent grid M3 multiplies the call count by $12\times$.

\paragraph{Languages.} Lithuanian is the primary language for the headline runs (all 68 items, all anchors, all instructions, all persona blocks in Lithuanian; the LLM is asked to respond in JSON whose keys are item identifiers, which is language-neutral). English is supported as an ablation in which the persona, anchors, and item wordings are all translated; the per-respondent cross-language ablation (Section~\ref{sec:results-ablations}) uses identical profiles across the two languages, so the difference can be cleanly attributed to language rather than to sampling.

\subsection{Generation procedure}
\label{sec:generation}
\paragraph{Sampling.} The headline grid samples $n=100$ respondents per (model, condition) cell, stratified on gender and position (managerial or not) with a fixed seed. The ordering is deterministic, so every cell uses the same 100 respondents, and the sample keeps the human proportions of the joint gender $\times$ role cells. We stratify because a simple random sample of 100 from 263 profiles can miss small cells such as managerial men, while the stratified sample reproduces the human marginals on these two covariates. The $n=100$ choice balances three constraints: (i) statistical: $n=100$ gives a Holm--Bonferroni minimum-detectable Cohen's $d$ of $0.48$ for the 666-pair leaderboard, comfortably below the observed PSS gaps; (ii) economic: the headline 37-model grid is approximately \$90--160 in API costs depending on which providers are used; and (iii) provider rate-limits: at $n=100$ and minimal reasoning, a full 37-model headline grid completes in roughly 6 wall-clock hours under reasonable per-provider concurrency limits. The human-side analyses (mediation, reliability, demographic effects, factor structure) always use the full $n=263$ sample.

\paragraph{Repeats and temperature.} We use $R=1$ repeat per cell and request temperature $0.7$, the standard ``creative but not chaotic'' default. Several frontier models do not honour an explicit temperature: \texttt{claude-opus-4-7} deprecates the parameter and the GPT-5 family is pinned to the model default by our generation client; these run at their provider default, which produces sensible responses across all 37 models. $R=1$ is justified post-hoc by a dedicated test--retest run in which we re-issued the \emph{same} 30-respondent panel $R=20$ times per model, each draw carrying a fresh per-call sampling seed so that the repeats are genuinely independent (Section~\ref{sec:results-memorization}, Table~\ref{tab:test-retest}). Scale-averaged test--retest ICC(1) ranges from $0.52$ to $0.93$ across the lineup (median $0.87$), and $23$ of the $31$ models with repeat sets sit at or above the human published-reference ICC of $0.78$; the mean inter-repeat rank correlation across all $68$ items is $0.62$--$0.96$. This stability exceeds the within-LLM-vs-cross-LLM PSS variance that the leaderboard reports, so a single repeat per cell suffices for the headline grid. Higher $R$ would shrink the within-cell variance further, but at fixed budget the marginal value is lower than running additional models or conditions. Four models could not be re-queried for the C9 narrative variant because their gateway deployments were retired or persistently rate-limited between data-collection waves (\texttt{llama-4-scout}, the two \texttt{grok-4-1-fast} variants, and \texttt{gpt-oss-120b}; \texttt{deepseek-v4-pro} retains only a single degenerate C9 respondent, which we exclude from C9 comparisons); these models retain their main-grid C0--C4 leaderboard results. \texttt{qwen3-7-plus}, rate-limited during the first collection wave, was backfilled once its gateway capacity recovered and retains a reduced effective $n$ ($72$--$94$ of $100$ respondents per headline condition).

\paragraph{Reasoning effort.} We set reasoning effort to \emph{minimal} for the headline grid: on OpenAI this passes \texttt{reasoning\_effort=minimal}, while on Anthropic and Google it disables extended thinking, and on the open-weight gateway the hint is passed through or silently ignored by the backend. Higher reasoning budgets multiply the cost and latency by $5$--$20\times$ for the reasoning-capable models, so a full headline-grid sweep at \emph{high} reasoning is prohibitive. We ablate this choice on a 6-model subset (Section~\ref{sec:results-ablations}); benchmarked against the test--retest resampling-noise floor for the same paired-$d$ statistic (mean $|d|{\approx}0.09$ at $n{=}100$), only \texttt{gpt-5.4} (and, weakly, \texttt{claude-opus-4-7}) shows a reasoning effect clearly above noise, so fixing the default at the cheapest setting is well justified.

\paragraph{Parser and validation.} Each call goes through a strict JSON parser that strips Markdown code fences, recovers the first \texttt{\{...\}} block, coerces stringly-typed values to integers, and clips out-of-range values. Calls that fail to parse or that produce out-of-range integers are re-issued with the offending response appended as an assistant turn plus a stricter user reminder. The full parser and retry policy is documented in Appendix~\ref{app:parser}. In the main grid, $99.2\%$ of the 22,291 calls that did not end in an API error returned a complete, in-range answer vector, and the per-(model, condition) format-failure rate is logged in \texttt{generation\_log\_*.parquet} for the audit trail.

\paragraph{Reproducibility infrastructure.} Per-call disk caching (content-hash-keyed under \texttt{.cache/llm/}) and per-job parquet checkpointing make the pipeline idempotent: adding a new model to \texttt{configs/models.yaml} re-runs only that model's calls; counterfactual swaps and ablations are launched with the same machinery on top of the cached headline run. The pipeline supports five provider auth paths (OpenAI, Anthropic, Google AI Studio, Google Vertex AI, Nexos) and degrades gracefully when credentials for a given provider are absent. Full reproducibility details are in Appendix~\ref{app:repro}.

We benchmark $37$ models spanning every major proprietary and open-weight family released between 2024-Q4 and mid-2026.\footnote{The lineup is a snapshot frozen in mid-June 2026. The release cadence is roughly six-weekly, so a handful of same-window successors (e.g.\ a next-step OpenAI, DeepSeek, Zhipu and Meta release) and additional regional families post-date the freeze; because adding a model never re-runs prior models (Section~\ref{sec:method}), the benchmark is designed to absorb them incrementally rather than to claim exhaustive coverage of every checkpoint. The frozen lineup already spans all major frontier vendors and a within-vendor generational ladder, which is what our analysis requires.} What matters for our analysis is the \emph{model and its manufacturer}, not the API gateway it is served through, so open-weight models are attributed to their true vendor (xAI, Zhipu, Meta, DeepSeek, \dots) even though they are routed through a single multi-vendor gateway:
\begin{itemize}[leftmargin=1.4em,itemsep=0pt]
  \item \emph{OpenAI (4):} GPT-5.5, GPT-5.4, GPT-5.4-mini, GPT-5.4-nano.
  \item \emph{Anthropic (6):} Claude Opus 4.8, Opus 4.7, Sonnet 5, Sonnet 4.6, Haiku 4.5, Fable 5.
  \item \emph{Google (5):} Gemini 3.5 Flash, 3.1 Pro, 3.1 Flash-Lite, 3 Flash, 2.5 Pro (with both API-key and Vertex AI service-account auth paths in the harness).
  \item \emph{Open-weight (22), routed via the Nexos gateway:} xAI Grok 4.1 (fast non-reasoning + reasoning), 4.20, 4.20-reasoning, 4.3; DeepSeek 3.2 \& V4-Pro; Meta Llama 4 Scout \& 4 Maverick; Zhipu GLM 5.1 \& 5.2; Mistral Large 3; Alibaba Qwen 3.5 397B \& 3.7 Plus; Moonshot Kimi K2.6 \& K2.7 Code; MiniMax M2.7 \& M3; NVIDIA Nemotron 3 Super; Amazon Nova 2 Lite; Google Gemma 4 31B; OpenAI GPT-OSS 120B.
\end{itemize}
The full lineup is configurable via \texttt{configs/models.yaml}; the gateway integration uses an OpenAI-compatible HTTP API with per-vendor model UUIDs, so adding a new vendor is a one-line YAML change. For readability we drop the gateway tag that open-weight models carry in the released configuration, so e.g.\ \texttt{nexos-qwen3-5-397b} appears as \texttt{qwen3-5-397b} in every table, figure, and reference below. Reasoning effort is set to \emph{minimal} in the headline runs for cost reasons; we ablate this choice in Section~\ref{sec:results-ablations}.

\subsection{Evaluation framework}
\label{sec:eval}
\paragraph{Six dimensions.} Each (model, condition) cell is scored on six model-agnostic dimensions. The choice of six was driven by classical measurement theory: distribution, correlation, and reliability are the three item-level building blocks of any classical-test-theory analysis; mediation and demographic effects are the two most common downstream uses of an instrument; and construct-level fidelity is the latent-variable test that distinguishes ``a synthetic respondent who looks right'' from ``a synthetic respondent whose responses behave like the construct''.

\begin{itemize}[leftmargin=1.4em,itemsep=0pt]
  \item \textbf{(1) Distribution.} Per-item Jensen--Shannon divergence between the LLM and human Likert-response distributions, Kolmogorov--Smirnov $D$ statistic, Wasserstein-1 distance, and the within-item SD ratio $\sigma_{\text{LLM}}/\sigma_{\text{H}}$. The SD ratio is reported separately because the most common LLM failure mode is range restriction (the LLM picks a narrower band of the scale than humans do); JS alone reflects this only indirectly and does not flag it.
  \item \textbf{(2) Correlation matrix.} Upper-triangle Pearson similarity between the LLM and human $68\times68$ item-level correlation matrices; modified RV2 coefficient of \citet{smilde2009matrix} (a covariance-matrix similarity statistic that is invariant to rotation of either matrix); mean $|\Delta r|$ over the upper triangle (gives the average per-item-pair absolute correlation difference, in Pearson-$r$ units).
  \item \textbf{(3) Reliability.} Cronbach's $\alpha$ \citep{cronbach1951} and McDonald's $\omega$ \citep{mcdonald1999} per subscale. The PSS component is one minus the gap $|\alpha_{\text{LLM}}-\alpha_{\text{H}}|$ on the total Dunham attitudes-to-change scale (18 items, five of them reverse-keyed). We score by gap rather than by ratio because a higher LLM $\alpha$ is not automatically better: over-coherent answering inflates $\alpha$ above the human value and signals exactly the failure mode (range-restricted / persona-collapsed responses) that should lower PSS.
  \item \textbf{(4) Mediation.} The standardised $a$, $b$, $c$, $c'$ path coefficients of the X$\rightarrow$M$\rightarrow$Y model with bootstrap 95\% CIs on the indirect effect $ab$. The PSS component compares the indirect effect of the attitudes total on the performance total through the engagement total with the human value in Table~\ref{tab:human-mediation}, through the magnitude error $|ab_{\text{LLM}} - ab_{\text{H}}|$. The sign of each path and whether the 95\% bootstrap CI on $ab$ excludes zero in both samples are diagnostic checks and do not enter PSS.
  \item \textbf{(5) Demographic group effects.} Standardised Cohen's $d$ for eight demographic contrasts (gender, role, total tenure, tenure in the organisation, organisation size, sector, education, and age under or over 30) on each of the 12 composites, which gives 96 contrast-outcome pairs. The PSS component multiplies two terms. The first is the direction match $\rho_{\text{dir}}$, the share of pairs with $\text{sign}(d_{\text{LLM}}) = \text{sign}(d_{\text{H}})$. The second is one minus the absolute mean amplification, where the amplification of a pair is $|d_{\text{LLM}}| - |d_{\text{H}}|$. Most of these human contrasts are small, so the direction match is noisy. Effects that are null in humans but appear in the LLM sample (fabricated stereotypes) are tested separately with counterfactual swaps (Section~\ref{sec:results-cf}).
  \item \textbf{(6) Construct-level fidelity.} Per-instrument exploratory factor analysis (3-factor solution, matched to the published structure of each instrument) with greedy factor alignment between the LLM and human factor solutions, Tucker's $\varphi$~\citep{tucker1951method,lorenzo2006tucker} augmented with an item-permutation null distribution (Appendix~\ref{app:tucker}), bifactor decomposition (explained-common variance ECV, $\omega_h$), inter-subscale correlation gap (LLM-vs-human), and the HTMT discriminant-validity ratio. (A $\Delta$CFI invariance check of \citet{cheung2002evaluating} was attempted but is omitted from the scored construct-fidelity aggregate, because the 3-factor CFA models did not converge reliably on the shorter subscales at this sample size.)
\end{itemize}

\paragraph{Statistical baselines.} Five non-LLM baselines lower-bound any LLM and let us read each PSS component against a sharp reference:
\begin{itemize}[leftmargin=1.4em,itemsep=0pt]
  \item \textbf{Marginal} samples each item independently from the empirical per-item Likert distribution. This is the strongest baseline on distribution alone and the weakest on correlation. It is therefore a useful upper bound on what ``preserving item means'' can buy.
  \item \textbf{MVN} fits a multivariate Gaussian to the standardised item responses and samples synthetic respondents from it, then rounds to the nearest valid Likert level. MVN preserves the full covariance matrix and is therefore close to the maximum-correlation baseline; it does not preserve the discrete bounded-Likert distribution, so its distributional score is below copula.
  \item \textbf{Gaussian copula} fits per-item marginals plus a Gaussian dependence structure and samples synthetic respondents with both correct marginals and correct rank correlations. Copula is the strongest combined-distribution-and-correlation baseline available without a language model.
  \item \textbf{Stratum mean} returns the demographic-stratum mean (rounded) on every item for every respondent in that stratum. It is the strongest baseline on demographic effects (by construction it has the correct between-stratum mean differences) and the weakest on within-respondent variance.
  \item \textbf{$k$-NN respondent lookup} returns the actual responses of the $k=5$ nearest demographic neighbours of the target profile, averaged and rounded. This is the strongest non-parametric baseline on both correlation and demographic effects but is essentially a retrieval system rather than a generative one.
\end{itemize}
An LLM that fails to outperform the Gaussian copula on the sample-driven components (distribution, correlation, reliability) is doing no better than the human covariance structure alone; this is the most important non-trivial comparison the leaderboard makes.

\paragraph{Held-out human-vs-human ceiling.} An 80/20 stratified split of the human sample provides an empirical upper bound on PSS at the given $n$. The training half re-derives the same six-component reference statistics, the holdout half plays the role of a ``synthetic respondent'', and the resulting PSS is the maximum a perfect LLM could realistically achieve at this sample size. The ceiling is around $0.83$ (Section~\ref{sec:results-pss}); a perfect LLM would in principle reach $1.0$, but at $n_{\text{H}}=263$ the sampling variability of the reference statistics themselves caps the achievable PSS.

\paragraph{References at the headline sample size.} The held-out split has only 53 test respondents, so it is not matched to the $n{=}100$ of the LLM cells. We therefore add two references at the headline sample size. The first scores the real answers of the 100 headline respondents against all 263 humans, exactly as an LLM sample is scored; a perfect simulator of these respondents would reach this value. The second fits the Gaussian copula only on the 163 humans outside the headline sample and generates answers for the 100 headline respondents, so it never sees the people it simulates. The other baselines are fit on, and generated for, all 263 humans, so they are in-sample references.

\paragraph{Aggregation.} The five item-level components are aggregated into $\text{PSS}\in[0,1]$ (1 = perfect similarity):
\begin{equation}
\text{PSS} = w_d (1{-}\overline{\text{JS}}) + w_c\, r_{\text{corr}} + w_r (1{-}|\Delta\alpha_{\text{DUN}}|) + w_m \big(1{-}\tfrac{|\Delta ab|}{0.5}\big)_+ + w_g\, \rho_{\text{dir}}\big(1{-}|\overline{\Delta |d|}|\big)_+
\label{eq:pss}
\end{equation}
with default weights $(w_d, w_c, w_r, w_m, w_g)=(0.25, 0.25, 0.20, 0.20, 0.10)$. The two largest weights (distribution and correlation) go to the sample-driven components that have the smallest measurement noise at $n=100$; mediation and reliability get $0.20$ each because they are the most downstream-relevant for the applied I-O literature; demographic effects get the smallest weight because the human $|d|$ is itself estimated noisily at $n=263$ on contrasts other than role. Components are also reported separately throughout the paper, so practitioners can read which dimensions pass and which fail for any particular model. Holm--Bonferroni-adjusted $p$-values control familywise error within each contrast family (Appendix~\ref{app:power}); effect-size CIs accompany every $p$-value. The sensitivity of the leaderboard to the weight choice is explored in Appendix~\ref{app:sensitivity}.

% =============================================================================
\section{Results}
\label{sec:results}
% =============================================================================

\subsection{Human baseline replication}
Our pipeline reproduces \citet{sarkauskaite2020}: composite means within $0.04$ of published values, Cronbach's $\alpha$ within $0.02$ on every matching subscale, the standardised mediation coefficients to two decimals (Table~\ref{tab:human-mediation}), the role effect (managers score higher on 9 of 12 composites at unadjusted $p<0.005$, $|d|\in[0.39,0.65]$; 7 survive Holm correction), and the gender-effect null (unadjusted $p>0.1$ on every composite, Holm-adjusted $p=1.0$). This is the ground truth that LLM samples are evaluated against.

\ifarxiv
\begin{figure}[t]
  \centering
  \begin{minipage}[t]{0.55\linewidth}
    \centering
    \includegraphics[width=\linewidth]{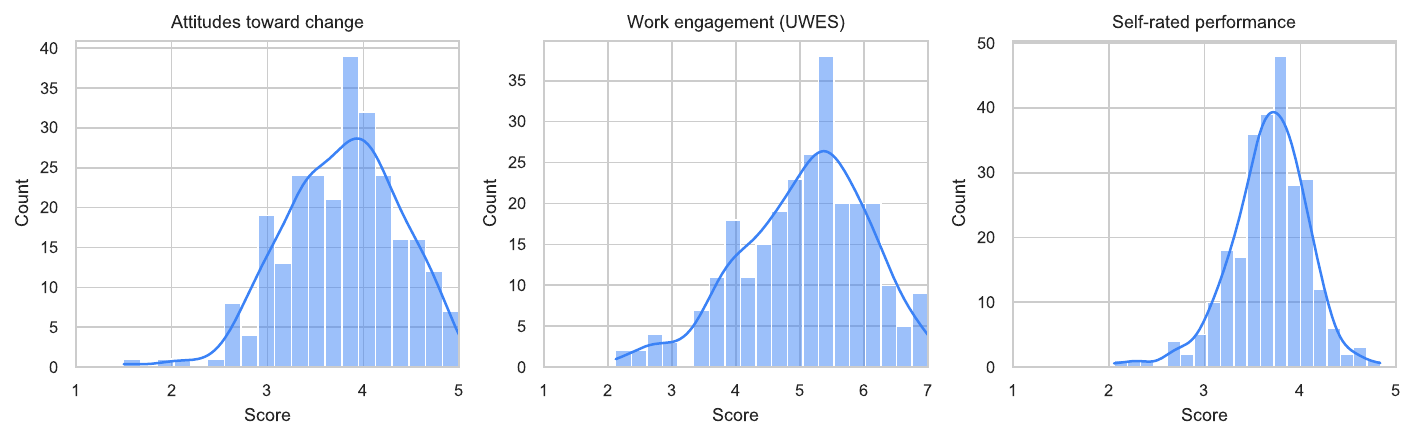}
    \caption{Human composite-score distributions for the three primary scales.}
    \label{fig:human-dists}
  \end{minipage}\hfill
  \begin{minipage}[t]{0.42\linewidth}
    \centering
    \includegraphics[width=\linewidth]{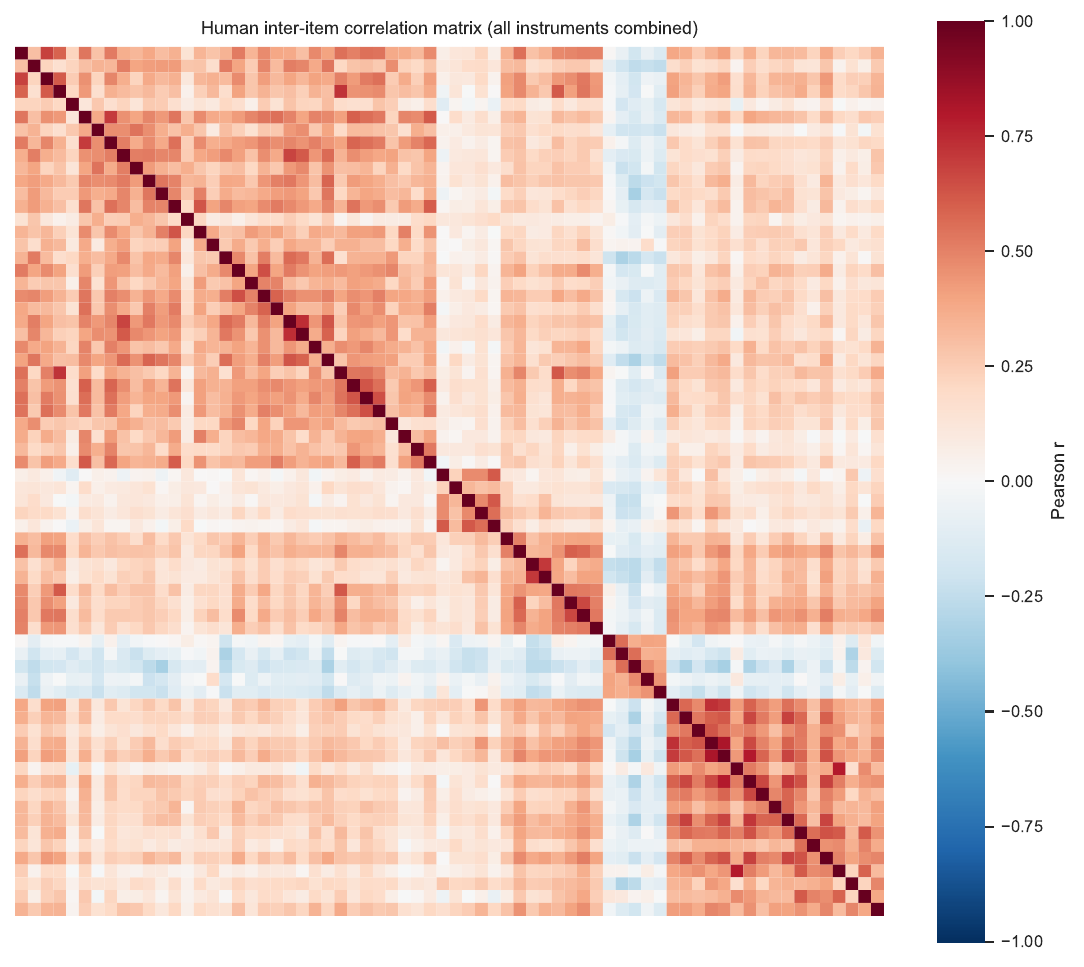}
    \caption{Inter-item correlation matrix of the full 68-item human dataset (Pearson).}
    \label{fig:human-corr}
  \end{minipage}
\end{figure}
\fi

\subsection{Anchored PSS leaderboard}
\label{sec:results-pss}
\begin{figure}[t]
  \centering
  \includegraphics[width=0.78\linewidth]{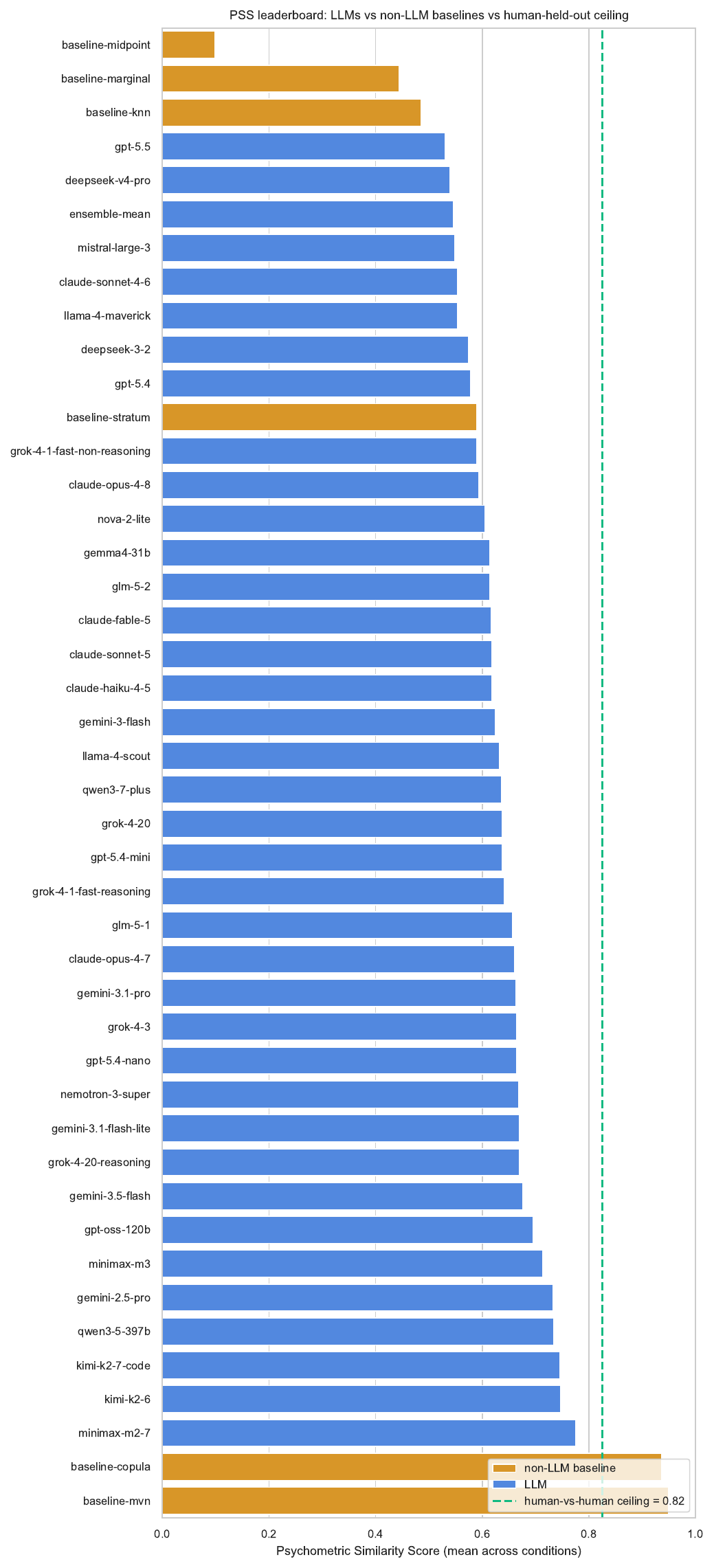}
  \caption{PSS leaderboard with statistical baselines (orange), human-vs-human held-out ceiling (green dashed), and LLMs (blue). Respondent-bootstrap 95\% CIs (\texttt{llm\_pss\_bootstrap\_ci.csv}) have mean width $0.042$ across the LLM lineup; gaps exceeding ${\sim}0.05$ are statistically reliable (Holm--Bonferroni MDE $d{=}0.48$ at $n{=}100$, Appendix~\ref{app:power}).}
  \label{fig:pss-baselines}
\end{figure}

\paragraph{Headline numbers.} Table~\ref{tab:pss-leaderboard-compact} reports the headline 37-model leaderboard at the C3 (full profile) condition under single-call presentation, and Figure~\ref{fig:pss-baselines} shows the same comparison. Three human references frame the scores. The real answers of the 100 headline respondents, scored against all 263 humans exactly like an LLM sample, reach PSS $0.886$; a perfect simulator of these respondents would reach this value. The 80/20 held-out split, with only 53 test respondents, reaches $0.825$. A Gaussian copula fit only on the 163 humans outside the headline sample and generated for the same 100 respondents reaches $0.824$. The top LLM, \texttt{minimax-m3}, reaches PSS $0.814$. It is followed by \texttt{minimax-m2-7} ($0.809$), \texttt{kimi-k2-7-code} ($0.802$), \texttt{claude-opus-4-7} ($0.789$), \texttt{gemini-3.1-flash-lite} ($0.779$) and \texttt{qwen3-5-397b} ($0.777$); \texttt{gpt-5.4-mini} ranks tenth ($0.743$). The bottom of the leaderboard mixes proprietary and open-weight models (\texttt{claude-sonnet-4-6} $0.586$, \texttt{nova-2-lite} $0.605$, \texttt{gpt-5.4} $0.606$, \texttt{gpt-5.5} $0.609$). The 5--95 percentile spread of the per-model PSS is $0.61$--$0.80$. This range is much wider than the mean width of the respondent-bootstrap 95\% CI ($0.042$), so the coarse order of the leaderboard is statistically resolved, although many adjacent pairs are not. The bootstrap covers the distribution, correlation and reliability components; the mediation and demographic components add uncertainty that this interval leaves out. The wisdom-of-LLMs ensemble scores PSS $0.560$, below every individual LLM. Averaging answers across the lineup removes person-level variation, and the averaged sample misses the human mediation pathway completely ($m=0$).

\paragraph{Statistical baselines match or beat every LLM.} On the three sample-driven components (distribution, inter-item correlation, and reliability), the Gaussian-copula and MVN baselines fit on the full human sample outperform every LLM. The copula reaches $d=0.99$, $c=0.95$ and $r=0.99$. The best LLM values on the same components are $d=0.90$ (\texttt{grok-4-1-fast-reasoning}), $c=0.81$ (\texttt{minimax-m2-7}) and $r=0.98$. Combined into the five-component PSS, the copula reaches $0.936$ and the MVN baseline $0.949$, both far above the best LLM ($0.814$). These two baselines have an in-sample advantage, because they are fit on the same 263 humans that define the reference. For this reason the fair comparison at the headline sample size is the copula fit only on the other 163 humans. It scores $0.824$, level with the best LLM and above the other 36. The LLMs make up ground only on the persona-conditioned mediation and demographic-effect components: \texttt{minimax-m3} reaches $m=1.00$ and $g=0.52$, against $0.72$ and $0.38$ for the out-of-sample copula, while its correlation component is lower ($0.76$ vs.\ $0.81$). For this reason the order of these two depends on the weights. With weights that favour mediation and demographic effects the best LLM is clearly ahead of the out-of-sample copula ($0.80$ vs.\ $0.70$), while the in-sample copula and MVN stay ahead of every LLM under every weighting we tried (Appendix~\ref{app:sensitivity}).

The best LLM therefore only matches a baseline with no language-model component, and it stays $0.07$ PSS units below the real answers of the same people. This suggests that most of an LLM's psychometric fidelity is recoverable from the human covariance structure, that is, from the instrument itself rather than from any model-specific understanding of Lithuanian organisational psychology.

\paragraph{A demographic-conditional copula.} The comparison above could be called unfair to the LLM: the copula is fit on the human item responses, whereas the LLM sees only an 11-field demographic profile. The relevant test is then whether conditioning a statistical generator on that same profile helps. We add a demographic-conditional copula (Appendix Table~\ref{tab:cond-copula}): each item is ridge-regressed on the same 11 profile fields the LLM receives and a Gaussian copula is fit on the residuals, so the generator's location is conditioned on the persona while the correlation structure is preserved. Scored the same way as the other statistical baselines, conditioning does not help. It raises only the low-weight demographic-effect component (in-sample $g$ rises from $0.52$ to $0.87$, and to $0.77$ out-of-sample). The small in-sample PSS gain ($0.936$ to $0.968$) disappears under 5-fold cross-validation, where the conditional copula scores $0.929$, level with the unconditional copula. All copula variants stay far above the best LLM ($0.814$). The demographic profile the LLM conditions on adds no measurable psychometric signal to a simple statistical generator.

Three further analyses support this conclusion. First, an empirical variance decomposition (\texttt{pss\_decomposition.csv}) credits a model only for PSS it achieves beyond the marginal and MVN baselines. No LLM exceeds the MVN baseline, so the LLM-unique share is $0.0$ for all $37$ models. Second, we train a discriminator to tell synthetic respondents from real humans (Appendix Table~\ref{tab:distinguishability}). The Gaussian-copula (AUC $0.39$) and MVN (AUC $0.52$) baselines are indistinguishable from humans. Every LLM is separated almost perfectly (median AUC $0.99$, lowest $0.97$), including the top of the leaderboard (\texttt{minimax-m3} $0.98$), so a high PSS does not make a synthetic sample look human at the respondent level. Third, the gap does not close with scale. In a bootstrap sample-size sweep scored on the distribution, correlation and reliability components (Appendix Table~\ref{tab:scaling-curve}), the copula climbs from $0.69$ at $n{=}10$ to $0.82$ at $n{=}200$, while the best LLM rises only from $0.69$ to $0.74$. The copula-minus-LLM gap therefore widens from $0.00$ to $+0.08$ as more synthetic respondents are drawn.

\paragraph{Per-model insights.} The component-level breakdown of the top and bottom leaderboard rows is informative beyond the aggregate PSS. \texttt{minimax-m3} wins the leaderboard through an almost exact match of the human indirect effect ($m=1.00$) and high reliability ($r=0.97$), although its distribution component is only mid-range ($d=0.71$). \texttt{minimax-m2-7} and \texttt{kimi-k2-7-code} are more balanced. They have the highest correlation components of any LLM ($c=0.81$ and $0.80$) and strong distribution components ($d=0.83$ and $0.80$), but weaker mediation ($m=0.76$ and $0.80$). \texttt{claude-opus-4-7} combines high mediation ($m=0.96$) with the lowest demographic component of the top five ($g=0.34$). Reliability hardly separates models: every LLM has $r$ between $0.92$ and $0.98$. The bottom rows are characterised by collapse on mediation rather than by uniform weakness. Five models (\texttt{claude-sonnet-4-6}, \texttt{gpt-5.4}, \texttt{glm-5-2} and both \texttt{grok-4-1-fast} variants) score $m=0$. Their indirect effects are between $0.69$ and $1.19$, four to seven times the human $ab=0.16$, so they overstate the pathway rather than miss it. \texttt{llama-4-scout} has almost no demographic-effect match ($g=0.08$).

\paragraph{Conditioning matters: the C0$\to$C4 gradient.} Persona disclosure moves PSS substantially. Across the lineup the mean per-model PSS improvement from C0 (no profile) to C3 (full profile) is $0.18$ PSS units. The largest improvements are on \texttt{deepseek-v4-pro} ($+0.61$), \texttt{deepseek-3-2} ($+0.52$) and \texttt{llama-4-maverick} ($+0.49$). The smallest are near zero on \texttt{gemini-3.1-pro} ($-0.02$), \texttt{gemini-2.5-pro} and \texttt{gpt-5.4-nano} ($+0.01$ each), which barely respond to the added profile fields. The ladder is not monotone: on average C1 (gender and age) scores higher than C2 ($0.67$ vs.\ $0.61$), and only the full C3 profile improves on C1 ($0.70$). Adding the C4 change-context block on top of C3 changes PSS negligibly (mean $-0.003$, median $-0.005$ PSS units, within the bootstrap CI width); we therefore use C3 as the headline rather than C4 to avoid the prompt-length sensitivity that C4 introduces. We also collected a narrative variant (C9), which renders the core profile as a prose paragraph instead of structured fields (and omits the two free-text change-context fields), on the 32 models with a usable follow-up run. Relative to structured C3, the narrative rendering shifts PSS by a median absolute $0.05$ (mean $-0.03$, so the prose form is on average slightly less informative; max shift $0.20$), and the C9 and C3 rankings correlate only moderately (Spearman $0.51$). Because C9 changes both wording and field set we cannot cleanly separate surface form from field selection, but either way fine-grained PSS differences among similarly-scoring models are sensitive to presentation; we therefore read the C3 leaderboard as conditional on the structured presentation and avoid over-interpreting small rank gaps between adjacent models.

\begin{table}[t]
  \centering\small
  \caption{Anchored PSS leaderboard at the C3 (full profile) condition under single\_call\_all presentation, sorted within each block. $d$ = distributional, $c$ = correlation, $r$ = reliability, $m$ = mediation, $g$ = demographic. Human references, statistical baselines, then LLMs. All rows are scored with the same five components against all 263 humans. LLM rows, the same-respondent reference, and the copula fit on the other 163 humans have $n{=}100$; the other baselines are generated for all 263 respondents. Top 5 LLMs, ensemble-mean, and worst LLM shown; full table in Appendix~\ref{app:tables}.}
  \label{tab:pss-leaderboard-compact}
  \begin{tabular}{lrrrrrr}
    \toprule
    Model / baseline & PSS & $d$ & $c$ & $r$ & $m$ & $g$ \\
    \midrule
    human, same 100 respondents & 0.886 & 0.996 & 0.923 & 0.990 & 0.707 & 0.672 \\
    human, held-out split & 0.825 & 0.981 & 0.762 & 0.986 & 0.720 & 0.480 \\
    baseline-mvn & 0.949 & 0.990 & 0.950 & 0.994 & 0.995 & 0.664 \\
    baseline-copula & 0.936 & 0.994 & 0.954 & 0.986 & 1.000 & 0.522 \\
    baseline-copula, fit on other 163 & 0.824 & 0.985 & 0.805 & 0.970 & 0.722 & 0.382 \\
    baseline-stratum & 0.589 & 0.989 & 0.169 & 0.336 & 0.805 & 0.714 \\
    baseline-knn & 0.485 & 0.907 & 0.070 & 0.323 & 0.714 & 0.334 \\
    baseline-marginal & 0.444 & 0.997 & 0.011 & 0.097 & 0.667 & 0.395 \\
    baseline-midpoint & 0.099 & 0.395 & 0.000 & n/a & n/a & 0.000 \\
    minimax-m3 & 0.814 & 0.713 & 0.764 & 0.966 & 0.998 & 0.520 \\
    minimax-m2-7 & 0.809 & 0.833 & 0.807 & 0.976 & 0.761 & 0.515 \\
    kimi-k2-7-code & 0.802 & 0.801 & 0.805 & 0.948 & 0.796 & 0.521 \\
    claude-opus-4-7 & 0.789 & 0.722 & 0.765 & 0.960 & 0.957 & 0.336 \\
    gemini-3.1-flash-lite & 0.779 & 0.861 & 0.725 & 0.947 & 0.753 & 0.421 \\
    ensemble-mean & 0.560 & 0.629 & 0.752 & 0.929 & 0.000 & 0.293 \\
    claude-sonnet-4-6 & 0.586 & 0.749 & 0.666 & 0.959 & 0.000 & 0.408 \\
    \bottomrule
  \end{tabular}
\end{table}

\subsection{Construct-level fidelity with permutation null}
\label{sec:results-construct}
\paragraph{Why the permutation null matters.} A naive Tucker's $\varphi$ on the LLM factor loadings, computed against the published factor structure with a greedy column alignment between the two loading matrices, is biased upward by the alignment search: even random factor solutions can match by chance, especially on instruments with many factors and many items per factor. The empirical distribution of $\varphi$ under random matching is therefore the natural null against which the observed value should be read; values inside the null distribution should not count as ``factor structure preserved''. We construct the null by permuting the item labels of the LLM loading matrix (which breaks the item-to-item correspondence with the human reference) and recomputing $\varphi$ for $K=500$ permutations; the one-sided $p$-value is $(k+1)/(K+1)$ where $k$ is the number of null draws meeting or exceeding the observed value. The full per-(model, instrument) table is in Appendix~\ref{app:tucker}; the headline numbers are summarised below.

\begin{figure}[t]
  \centering
  \begin{minipage}[t]{0.49\linewidth}
    \centering
    \includegraphics[width=\linewidth]{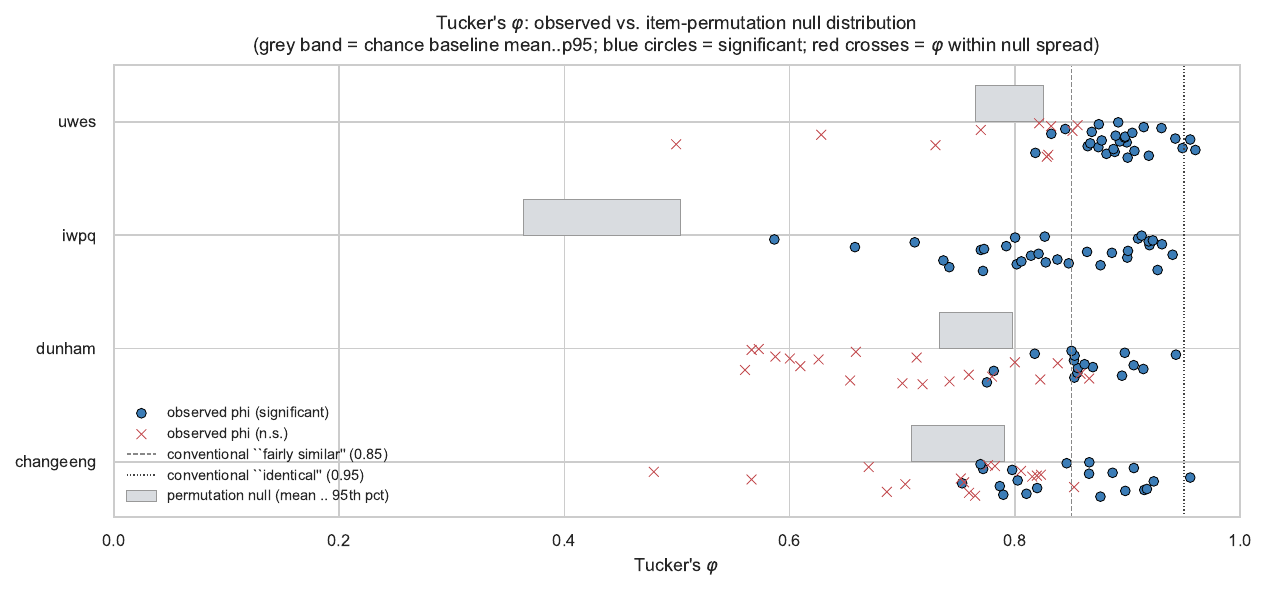}
    \caption{Tucker's $\varphi$ vs.\ item-permutation null. Grey band: chance baseline (mean..95th percentile, $K{=}500$). Blue: $p{<}0.05$. Red: inside null despite $\varphi{\geq}0.85$.}
    \label{fig:tucker-perm}
  \end{minipage}\hfill
  \begin{minipage}[t]{0.49\linewidth}
    \centering
    \includegraphics[width=\linewidth]{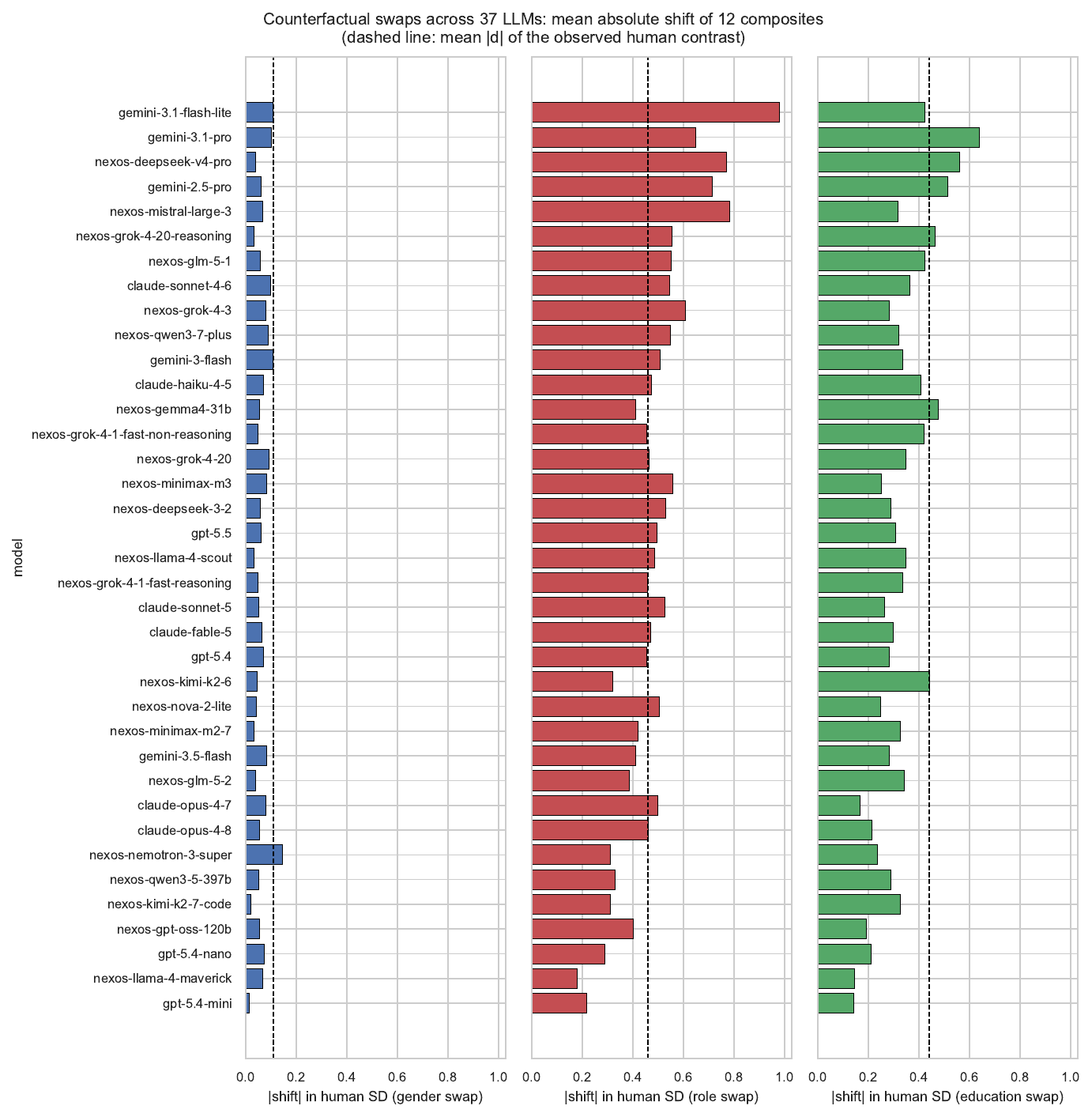}
    \caption{Mean absolute shift of the 12 composite subscales after a counterfactual swap, in units of the human SD. The dashed line marks the mean $|d|$ of the observed human contrast. Role swaps move the answers most, and gender swaps move them by less than 0.15 SD in every model.}
    \label{fig:counterfactual}
  \end{minipage}
\end{figure}

\paragraph{Instrument-dependent picture} (Figure~\ref{fig:tucker-perm}, Table~\ref{tab:tucker-permutation}):

\begin{itemize}[leftmargin=1.4em,itemsep=0pt]
  \item \textbf{IWPQ}: mean $\varphi{=}0.828$, mean null $\varphi{=}0.365$, and all $30$ models with a converged solution reject $H_0$ at $p{<}0.05$. The factor structure of work-performance items is genuinely preserved.
  \item \textbf{UWES}: mean $\varphi{=}0.859$, higher than IWPQ, but the mean null is $0.764$ because the 17-item, 3-factor structure produces high random congruences. Only $27/37$ models reject $H_0$. Ten models with respectable raw $\varphi$ (up to $0.856$ for \texttt{nemotron-3-super}) sit inside the chance distribution, so the factor-congruence story for engagement instruments needs a permutation correction.
  \item \textbf{Dunham} ATC: $\varphi{=}0.770$ vs.\ null $0.730$, $16/37$ reject. Once reverse-keyed items are aligned, the raw congruence is high, but so is the congruence expected by chance.
  \item \textbf{ChangeEng}: $\varphi{=}0.802$ vs.\ null $0.708$, $20/37$ reject. Many cells are indistinguishable from chance.
\end{itemize}
Two caveats apply to these denominators. First, some non-rejections reflect a model whose exploratory factor solution did not converge rather than a converged but non-significant fit. On IWPQ $7$ of the $37$ models lack a valid solution ($2$ on Dunham, $2$ on ChangeEng, $0$ on UWES), so the reported $\varphi$ averages are taken over the converged subset. Second, a high item-level PSS does not guarantee a recoverable latent structure. The leaderboard winner \texttt{minimax-m3} sits inside the permutation null on ChangeEng ($p{=}0.94$), and \texttt{gpt-5.4-mini} has no converged solution on three instruments and sits inside the null on the fourth (UWES). We therefore treat construct fidelity as a separate axis from the headline PSS rather than folding it into the leaderboard.
\paragraph{Discriminant validity (HTMT).} The HTMT discriminant-validity ratio (Appendix Table~\ref{tab:htmt-summary}) supports this picture. Across 37 models, LLM samples violate the $\text{HTMT}<0.85$ threshold on a mean of $67\%$ of the within-instrument subscale pairs with a defined ratio ($7.7$ of $11.4$ pairs on average), against $4/12$ ($33\%$) in humans. This is the canonical over-coherence failure mode. The models with the highest violation rate are \texttt{mistral-large-3} and \texttt{grok-4-1-fast-reasoning} ($11/12$ each), followed by \texttt{grok-4-1-fast-non-reasoning}, \texttt{gemini-3.5-flash} and \texttt{gpt-oss-120b} ($10/12$ each). Only \texttt{qwen3-5-397b} matches the human rate ($4/12$). A bifactor decomposition (Appendix Table~\ref{tab:bifactor}) shows the same pattern from the latent side. Averaged over the lineup at C3, LLMs inflate the explained-common-variance share of the general factor on all four instruments ($\Delta\text{ECV}=+0.08$ to $+0.12$). In other words, they move multidimensional constructs towards a single evaluative axis. The hierarchical $\omega_h$ is already close to 1 in humans on Dunham, ChangeEng and UWES and barely changes there ($\Delta\omega_h$ between $-0.02$ and $0.00$); on IWPQ it rises by $0.09$.

\paragraph{Mechanism.} Why do LLMs over-cohere on subscales that humans correctly differentiate? Two complementary mechanisms are visible in the qualitative response logs (Appendix~\ref{app:case-studies}): (a) \emph{persona-collapse}: when the persona is detailed but the items overlap in surface wording (as the IWPQ task and contextual subscales do), the LLM treats the persona as a single coherent direction (``hardworking'', ``conscientious'') and applies the same Likert direction across both subscales, inflating their between-subscale correlation; (b) \emph{anchor-anchoring}: when the LLM picks a Likert anchor early in the call (e.g., consistently using $4$ for positive statements), it tends to re-use it across items that share lexical features, even if the items belong to construct-distinct subscales. We read both as hypotheses. The construct-level statistics in this section are computed for single-call presentation only, so we cannot yet test whether the fragmented presentation modes of Section~\ref{sec:results-ablations} reduce them.

\paragraph{Which items are reproduced well?} We regress the per-item Jensen--Shannon divergence between LLM and human answers, averaged over models, on item features across all $68$ items (Appendix Table~\ref{tab:item-features}, $R^2{=}0.61$). Higher divergence means worse fit. Most of the explained variance is between instruments. UWES items, the only ones on a 7-point scale, have the highest divergence ($+0.16$ relative to change-engagement items, $p{<}10^{-9}$), followed by IWPQ items ($+0.07$, $p{<}0.001$); Dunham and change-engagement items do not differ. Scale range is constant within each instrument, so its effect cannot be separated from the UWES instrument effect. Within an instrument, reverse-keyed items are reproduced slightly worse than forward-keyed items ($+0.05$, $p{=}0.01$), and item length carries no signal ($p{=}0.27$).

\subsection{Counterfactual swaps on three axes}
\label{sec:results-cf}
\paragraph{Design.} We swap exactly one field of the profile (gender, role or education), keep everything else fixed, and run the model again under the same conditions. For each composite subscale $j$ and respondent $i$ we form the difference $\Delta_{ij} = Y_{i,j}^{\text{swapped}} - Y_{i,j}^{\text{original}}$. We sign each difference by the direction of its swap, so that a positive value means a move toward the first group of the human contrast (woman, managerial, university). We report two effect sizes. The first is the paired Cohen's $d_z = \overline{\Delta}_{\cdot j} / \mathrm{sd}(\Delta_{\cdot j})$. Under the null of no swap effect, $\Delta$ is run-to-run noise and $|d_z|$ has a floor of ${\approx}0.09$ at $n{=}100$ (mean over subscales, estimated from the test--retest repeats). The second is the shift $\overline{\Delta}_{\cdot j}$ divided by the SD of the same composite in the 263 humans. Only the shift is on the same scale as the Cohen's $d$ of the observed human group contrast. The $d_z$ divides by the spread of the shifts across respondents. A model that shifts every respondent by the same small amount therefore gets a large $d_z$. Both measures are averaged over the 12 composite subscales.

\paragraph{The three axes} (Figure~\ref{fig:counterfactual}, Table~\ref{tab:counterfactual-summary}):
\begin{itemize}[leftmargin=1.4em,itemsep=0pt]
  \item \textbf{Role swap (managerial $\leftrightarrow$ non-managerial).} This axis moves the answers most. The mean shift is $0.49$ human SD (range $0.18$--$0.98$ across models), with a mean $|d_z|$ of $1.07$. The human role contrast is significant on 7 of 12 composites after a Holm correction, with a mean $|d|$ of $0.46$. The models follow its direction on $99.6\%$ of model-composite pairs, and on average their shift matches its size. On $52\%$ of composites the model shift is larger than the human difference. The largest shifts come from \texttt{gemini-3.1-flash-lite} ($0.98$), \texttt{mistral-large-3} ($0.78$) and \texttt{deepseek-v4-pro} ($0.77$). The largest single effect is $1.38$ human SD on contextual performance for \texttt{gemini-3.1-flash-lite}, where the human difference is $0.64$.
  \item \textbf{Education swap (higher $\leftrightarrow$ secondary).} The mean shift is $0.33$ human SD (range $0.14$--$0.64$), with a mean $|d_z|$ of $0.70$. The models follow the direction of the human contrast on $99.8\%$ of model-composite pairs. The observed human difference between university and secondary education is $0.44$ on average, but it rests on 15 respondents with secondary education, and none of its 12 contrasts survives a Holm correction. The model shift is smaller than this difference on $68\%$ of composites. The largest shifts come from \texttt{gemini-3.1-pro} ($0.64$), \texttt{deepseek-v4-pro} ($0.56$) and \texttt{gemini-2.5-pro} ($0.51$), and the smallest from \texttt{gpt-5.4-mini} ($0.14$).
  \item \textbf{Gender swap (woman $\leftrightarrow$ man).} The mean shift is $0.06$ human SD (range $0.01$--$0.15$), with a mean $|d_z|$ of $0.17$. The human gender contrast is null on every composite ($p>0.1$, Section~\ref{sec:results}), so these effects have no support in the data. Their direction agrees with the sign of the null human estimates on only half of the composites. Twenty of the 37 models exceed $|d_z|=0.15$, and 7 stay below the noise floor. Some models therefore react to gender in a consistent way, but by a small amount. For \texttt{claude-sonnet-4-6}, the model with the largest $|d_z|$ ($0.40$), the shift is $0.10$ human SD.
\end{itemize}

\paragraph{What the swaps show.} Every model reacts to all three fields, and the order is almost the same in every model. Role moves the answers more than education in 34 of 37 models, and both move them more than gender in all 37. Where the human sample shows a group difference, the models reproduce its direction and roughly its size. On average they do not exaggerate it. The gender axis is the only one on which the models add an effect without support in the human data, and this effect is small. The two effect sizes can disagree. A uniform reaction gives a large $d_z$ even when the shift is small, so a bias audit should report the shift in units of the human SD for each axis.

\paragraph{Why role leads.} We did not test why role moves the answers most. One possible reason is that the link between managerial position and engagement, attitudes and performance is common in the organisational literature, and the human sample shows the same link. The education field is also correlated with role and sector, so an education swap may change several implied attributes at once.

\subsection{Ablations: reasoning effort, cross-language, presentation mode, cohort PSS}
\label{sec:results-ablations}
Four orthogonal ablations stress-test the headline design. Full per-model tables are in Appendix~\ref{app:tables}.

\paragraph{Reasoning effort.} On a 6-model subset spanning OpenAI, Anthropic, xAI, Google and DeepSeek (\texttt{gpt-5.4}, \texttt{claude-opus-4-7}, \texttt{grok-4-20}, \texttt{grok-4-20-reasoning}, \texttt{gemini-3.1-pro}, \texttt{deepseek-v4-pro}), \emph{minimal} vs.\ \emph{high} reasoning shifts the mean composite-subscale $|d|$ by $0.21$, but this average is carried by two models: only \texttt{gpt-5.4} ($0.37$, large effect, with composites moving in inconsistent directions) and \texttt{claude-opus-4-7} ($0.31$) shift clearly beyond the test--retest resampling-noise floor (mean $|d|{\approx}0.09$ at $n{=}100$), while \texttt{grok-4-20} ($0.17$), \texttt{gemini-3.1-pro} ($0.16$) and \texttt{deepseek-v4-pro} ($0.16$) sit within roughly twice it and \texttt{grok-4-20-reasoning} ($0.083$, a baked-in reasoning model whose effort toggle is inert) is within noise. The leaderboard rank of these 6 models is only partly stable between the minimal and high settings. \texttt{claude-opus-4-7} stays first and \texttt{gpt-5.4} stays last, but \texttt{deepseek-v4-pro} drops from second to fourth (PSS $0.75$ to $0.65$) and \texttt{gemini-3.1-pro} moves up from fifth to third. High reasoning lowers PSS for three of the six models and raises it for the other three. It does not uniformly improve fidelity, so the headline minimal-reasoning default is justified.

\paragraph{Cross-language.} Same profiles, LT vs.\ EN questionnaire wording, 30-respondent subsample. The per-item Pearson correlation between LT and EN responses on the same respondent under the same model is $0.889$ (averaged across models); the mean absolute difference in item means is $|\Delta|=0.26$ Likert points. For comparison, two repeats of the same 30-respondent test--retest panel under the same model differ by only $0.05$ Likert points in item means (Section~\ref{sec:generation}). The language of the questionnaire is therefore a real source of variation, about five times larger than run-to-run noise, even though the ordering of items is largely preserved. The shift is largest for the two \texttt{grok-4-1-fast} variants ($0.80$ and $0.54$) and smallest for \texttt{gpt-5.4-mini} ($0.13$).

\paragraph{Presentation mode.} Single-call vs.\ per-instrument vs.\ per-subscale on 30 respondents over a 5-model subset (\texttt{gpt-5.4}, \texttt{claude-opus-4-7}, \texttt{gemini-3.1-pro}, \texttt{deepseek-v4-pro}, \texttt{llama-4-maverick}). Relative to the single-call reference the composite-subscale $|d|$ is already substantial under per-instrument mode (M2; mean $|d|=0.24$--$0.35$ across the subset) and grows further under per-subscale mode (M3; up to $|d|=1.26$ for \texttt{llama-4-maverick}). Fragmenting the questionnaire into 12 separate calls degrades persona consistency. The persona block is re-shown at the start of each call, but the cross-instrument context cannot be recovered. We therefore use M1 (single-call) as the headline mode and treat M2/M3 as diagnostic ablations.

\paragraph{Cohort-stratified PSS (fairness).} For each model we recompute PSS within each demographic stratum (gender, age-band, education) and report the worst within-axis disparity (the maximum PSS gap between strata of the same axis). Worst within-axis disparities across the lineup are $0.122$ on education (\texttt{minimax-m2-7}), $0.085$ on gender (\texttt{llama-4-scout}), and $0.074$ on age (\texttt{mistral-large-3}). The median gap across models is $0.044$ on age, $0.047$ on gender and $0.061$ on education, at or above the bootstrap CI mean width on the aggregate PSS ($0.042$). Some strata are small (23 men among the 100 headline respondents), so part of each gap is sampling noise. Even so, fairness gaps are not zero, and the worst-case gap on a single model can exceed the aggregate PSS gap between two adjacent leaderboard rows. The aggregate PSS therefore hides a fairness gradient that is detectable only at the cohort level. The Appendix table reports the full per-(model, axis) cohort PSS so practitioners can identify the worst stratum for a model of interest.

\subsection{Memorization probe and inter-LLM homogeneity}
\label{sec:results-memorization}
\paragraph{Memorization.} The headline leaderboard is open to a simple alternative explanation: maybe the LLMs do well because they have seen the Lithuanian translations of the IWPQ, ATC, and UWES-17 items during pretraining and are simply reproducing them. We rule this out with a two-form verbatim-recall probe: \emph{direct} (``what is item $N$ of instrument $X$?''; the LLM is asked to recall the item text from its identifier alone), and \emph{context-primed} (previous item shown verbatim, ask for the next). Responses are scored by normalised Levenshtein edit distance to the published Lithuanian item wording. Across all 37 models the per-item \emph{high-recall} rate (distance ${\leq}0.30$, corresponding to substantially verbatim recall) is bounded by $4.7\%$ in the worst case (\texttt{qwen3-5-397b}, at $4.69\%$; the next-highest models, \texttt{llama-4-scout}, \texttt{mistral-large-3}, \texttt{deepseek-v4-pro}, \texttt{gemini-3-flash} and \texttt{gemini-3.1-flash-lite}, sit at $3.12\%$). $22$ of the $37$ models have a high-recall rate of exactly $0\%$. The Spearman rank correlation between high-recall rate and PSS across the 37 models is ${-}0.04$, which means there is no clear relationship: the highest-recall model is not the highest-PSS model, and vice versa. Three response patterns dominate the probe (Appendix~\ref{app:case-studies}, with verbatim transcripts):

\begin{itemize}[leftmargin=1.4em,itemsep=0pt]
  \item \emph{Explicit refusal} (``NE\v{Z}INAU'' / ``a\v{s} ne\v{z}inau'', ``I don't know''). The modal response from OpenAI's GPT-5.4 family and Anthropic's Claude family. The LLM correctly identifies that it cannot recall the item and refuses to fabricate.
  \item \emph{Confabulation}. Several Gemini and open-weight models produce plausible-sounding Lithuanian item text that is \emph{not} the published item. Edit distance to the published item is typically $0.7$--$0.9$ (clearly different) but the surface form (a single Lithuanian sentence on the relevant construct) reads as a plausible instrument item.
  \item \emph{Chain-of-thought leakage in English}. \texttt{glm-5-1} produces extended English-language reasoning traces about the IWPQ instrument structure (``The IWPQ has 18 items. Task performance: 1. I worked efficiently...''), revealing that the model has \emph{some} knowledge of the IWPQ instrument family without being able to reproduce the specific Lithuanian wording. The leaked English text is consistent with the original Koopmans English version, not with the Lithuanian translation.
\end{itemize}
Combined with the near-zero high-recall rate and the near-zero recall-vs-PSS rank correlation, this rules out verbatim memorisation as the driver of the leaderboard. PSS is therefore \emph{generated} under the persona, not regurgitated from pretraining. We cannot rule out paraphrased exposure: the LLM may have seen the Koopmans English version, or its translations, and reused the construct-level content. However, the published Lithuanian item wording itself is not what is producing the high-PSS responses.

\paragraph{Inter-LLM homogeneity.} The $37\times37$ inter-LLM PSS matrix (Figure~\ref{fig:dendrogram}) has mean $0.730$ on the off-diagonal, range $0.557$--$0.822$. LLMs therefore agree with one another about as much as the median LLM agrees with humans ($0.71$), and less than the best LLM does ($0.81$). The two numbers are not strictly comparable, because the inter-LLM score uses another 100-respondent LLM sample as the reference. Even so, a synthetic-respondent pipeline that draws from multiple LLMs to ``average out'' provider-specific biases gains little, because the providers are less independent than the surface diversity (different vendors, training corpora and alignment protocols) would suggest. The ensemble result below shows this directly.

\paragraph{Vendor clustering.} Hierarchical clustering on the $1-\text{PSS}$ distance matrix (Figure~\ref{fig:dendrogram} in Appendix~\ref{app:tables}) shows only weak vendor structure. Pairs from the same vendor agree slightly more than pairs from different vendors (mean PSS $0.77$ vs.\ $0.73$), and Google ($0.80$) and Anthropic ($0.77$) are the most internally similar vendors. The tree is dominated by one large cluster that mixes all major vendors, with five models outside it: \texttt{llama-4-maverick}, \texttt{gpt-5.4-mini}, \texttt{nova-2-lite}, \texttt{gpt-5.4-nano} and \texttt{mistral-large-3}. These outliers are mid- to low-PSS leaderboard rows ($0.61$--$0.74$), so a model that is far from the main cluster does not offer a complementary, more human-like answer pattern.

\paragraph{Ensemble collapse.} The wisdom-of-LLMs ensemble does not recover human diversity by averaging over models. At C3 it scores PSS $0.560$ (Section~\ref{sec:results-pss}), below every individual LLM and below its own no-profile score at C0 ($0.610$). Averaging the answers of 37 models for the same persona shrinks the person-level variance, and the ensemble's indirect effect moves so far from the human value that the mediation component drops to zero. The wisdom-of-LLMs intuition (averaging removes individual error) therefore fails in this setting. A plausible reason is that the LLMs err in correlated directions, so averaging keeps the shared error while it removes the variation between respondents.

\paragraph{Robustness checks.} Persona-faithfulness recall (free-text role recall) rises from $0.32$ at C0 to $0.85$ at C2, so the LLMs do condition on the persona where it is given. On the raw answers, the correlation between forward-keyed and reverse-keyed Dunham and change-engagement items is negative for all $31$ models in the test--retest panel, as expected from a faithful respondent. Format-failure (JSON-parse and refusal) rates, logged per (model, condition) in \texttt{generation\_log\_*.parquet}, are negligible across the reachable lineup; the only substantial per-model data loss is API-availability errors (HTTP 404/429) on the retired or rate-limited gateway deployments discussed in Section~\ref{sec:method}. Numerical detail in Appendix~\ref{app:robustness}.

\paragraph{Test--retest reliability.} To quantify how reproducible a single synthetic respondent is, we re-issued the \emph{same} 30-respondent panel $R=20$ times per model, each repeat carrying a fresh per-call sampling seed so the draws are genuinely independent rather than cache-identical, and computed the scale-averaged ICC(1) treating the repeat index as an interchangeable rater (Table~\ref{tab:test-retest}). Reliability varies sharply across the lineup. ICC(1) spans $0.52$ (\texttt{deepseek-3-2}) to $0.93$ (\texttt{gpt-5.4}, \texttt{gemini-3.5-flash}, \texttt{claude-sonnet-4-6}, \texttt{qwen3-7-plus}), with a median of $0.87$ and a mean inter-repeat rank correlation of $0.62$--$0.96$. $23$ of the $31$ models with repeat sets meet or exceed the human published-reference ICC of $0.78$, so most models are at least as test--retest reliable as a real respondent re-surveyed. The least reproducible models are \texttt{deepseek-3-2} ($0.52$), \texttt{gpt-5.4-nano} ($0.56$), \texttt{qwen3-5-397b} ($0.60$) and \texttt{gpt-5.4-mini} ($0.67$), well below the human reference, whereas most frontier models cluster near $0.90$. High reliability is necessary but not sufficient for fidelity. It shows that a model answers consistently, which is different from answering like a human, but it confirms that the $R=1$ headline grid is not dominated by sampling noise.

\subsection{Downstream consequences: response style, predictive validity, and fabricated mediation}
\label{sec:results-downstream}
The fidelity diagnostics above describe \emph{how} the synthetic respondents differ from humans; three further analyses quantify what those differences cost anyone tempted to substitute LLM respondents for a real sample.

\paragraph{Response style.} Survey methodology distinguishes substantive answers from response style: acquiescence (a tendency to agree), extreme responding (endpoint use), and midpoint responding. We measure acquiescence with a balanced index on the two instruments that contain both item keys (Dunham ATC and change engagement). The index averages the forward-keyed and reverse-keyed items on the as-answered, 0--1 rescaled responses, so trait content cancels and only net agreement remains; we express the model-minus-human difference in units of the human between-respondent SD of the index. Relative to the human sample at C3 (Appendix Table~\ref{tab:response-style}), the acquiescence shift is small: mean $+0.24$ human SD, positive for $29$ of $37$ models, ranging from $-0.54$ (\texttt{claude-sonnet-5}) to $+1.26$ (\texttt{nova-2-lite}). The larger differences are in the use of the scale. LLMs under-use the endpoints (extreme-response rate $-0.22$ on average), over-use the scale midpoint ($+0.17$), and have a much smaller within-respondent SD across items ($0.79$ vs.\ $1.19$ Likert points in humans). Their mean answer also sits lower on the scale than the human mean ($-0.49$ Likert points). This is the response-style signature of range restriction seen from the item side. The LLM compresses toward the centre rather than committing to the strong agree or disagree responses humans give, which is invisible to correlation-based comparisons but corrupts any analysis that depends on the shape of the response distribution.

\paragraph{Predictive validity.} If synthetic respondents are to replace humans, a model fit on synthetic data should predict \emph{human} outcomes. We fit an OLS regression of the performance total on the attitudes and engagement totals on each model's synthetic respondents and test it on a fixed 30\% split of the humans (Appendix Table~\ref{tab:predictive-validity}). The reference is the same regression fit on the other 70\% of humans, which reaches $R^2_{\text{human}}=0.28$. The synthetic-trained regressors reach a mean $R^2_{\text{synth}}=-0.14$ (median $-0.06$), which is on average worse than predicting the mean, for a mean validity gap of $\Delta R^2=-0.42$ (worst $-2.07$, \texttt{gpt-5.4-mini}). Twenty of the 37 models have a negative $R^2$. Four models match the human reference: \texttt{qwen3-7-plus} ($0.30$), \texttt{gemini-2.5-pro} ($0.30$), \texttt{gemini-3.5-flash} ($0.29$) and \texttt{minimax-m2-7} ($0.28$); the in-sample copula and MVN baselines reach $0.30$ and $0.31$. Predictive validity is only weakly related to fidelity (Spearman $0.21$ with PSS), and the top-ranked \texttt{minimax-m3} has $R^2_{\text{synth}}=-0.24$. Most of the lineup fails this test, so LLM respondents are not a drop-in substitute for human survey data in downstream modelling.

\paragraph{Fabricated mediation (negative control).} The recovered X$\to$M$\to$Y mediation could be a genuine reconstruction of the human structure or a confabulation that happens to point the right way. We test this with a placebo battery of seven $x\to m\to y$ paths whose human indirect effect is statistically indistinguishable from zero (Appendix Table~\ref{tab:mediation-negative-control}). Three further drawn paths, in which a total score contains another variable of the path, are left out because they do not test mediation. On $4$ of these $7$ null paths the pooled-LLM indirect effect is significant (bootstrap CI excluding zero) while the human one is not, so the LLMs fabricate a mediation pathway that does not exist in the humans. Per model, the median is $2$ fabricated paths out of $7$ (range $0$--$6$; \texttt{gpt-5.4-mini} has $6$). The mediation structure LLMs reproduce is therefore partly confabulated rather than purely recovered, which is a direct caution against using LLM respondents for exploratory structural-equation or causal-pathway discovery.

% =============================================================================
\section{Discussion}
\label{sec:discussion}
% =============================================================================
\subsection{What works}
LLMs reproduce the \emph{direction} of theoretical psychometric relationships under persona conditioning. The X$\rightarrow$M$\rightarrow$Y mediation that organises the original human study is recovered, with positive $a$, $b$, and indirect $ab$ paths, on $36$ of $37$ models at C3 (the one exception, \texttt{grok-4-20}, produces a negative indirect path). The LLM-side path coefficients are typically inflated relative to the human reference: the median indirect effect is $0.42$ against $0.16$ in humans. Managerial-vs-non-managerial role differences (the only demographic contrast with a non-trivial human effect size) are correctly signed on every composite total for all $37$ models. The qualitative block structure of inter-item correlations (within-subscale correlations $>$ between-subscale within-instrument $>$ between-instrument) is recovered on every model; the LLM understands that work-performance items belong together and that engagement items belong together. Persona conditioning improves fidelity, with the C0$\rightarrow$C3 gradient raising PSS by up to $0.61$ units depending on the model (mean $0.18$). The first-order picture is therefore positive: an LLM under persona conditioning is doing \emph{something} beyond producing a marginal-distribution Lithuanian-employee, and that something includes the qualitative direction of every major published claim about the human dataset.

\subsection{What does not}
Five failure modes are visible across the whole lineup:
\begin{enumerate}[leftmargin=1.4em,itemsep=0pt]
  \item \emph{Range restriction.} LLM within-item SDs are systematically below the human SDs, with the median model showing $\sigma_{\text{LLM}}/\sigma_{\text{H}}=0.55$ on IWPQ items and $0.47$ on UWES items. The LLM does not use the full Likert range as freely as humans do, even after the explicit anti-social-desirability instruction in the system prompt. Range restriction is the proximate cause of two downstream failure modes: it inflates Cronbach's $\alpha$ above the human value (in $79\%$ of model-by-subscale cells at C3, median $+0.05$) (because the LLM's responses are more internally consistent than humans' simply by virtue of being less variable) and it inflates inter-item correlations (because there is less variance to disagree on).
  \item \emph{Over-coherence.} The correlation component of PSS is $0.81$ for the best LLM versus $0.95$ for the Gaussian-copula baseline; the median LLM scores $0.74$. The LLM-generated $68\times68$ item correlation matrix is structurally too coherent: subscales that should be distinguishable (the HTMT $<0.85$ pairs in the human data) collapse into a single direction. This is the persona-collapse mechanism we described in Section~\ref{sec:results-construct}: the LLM treats the persona as a coherent direction and applies the same direction across items that humans treat as belonging to distinguishable constructs.
  \item \emph{Counterfactual swaps.} Role and education swaps move the answers in the direction of the human group differences and by about the same amount on average ($0.49$ and $0.33$ human SD against $0.46$ and $0.44$). Gender swaps move them by $0.06$ SD, although humans show no gender difference. The paired $d_z$ ranks the axes differently from the shift because it rewards uniform reactions. Bias audits should therefore report effects per axis and in units of the human SD.
  \item \emph{Factor-structure preservation overstated by raw $\varphi$.} Tucker's $\varphi$ is a biased estimator of factor congruence when the alignment between LLM and human loadings is itself searched greedily, especially on instruments with high-dimensional factor structures. UWES sits inside the item-permutation null for $10$ of $37$ models even when $\varphi \geq 0.85$ would, under the conventional threshold, count as ``identical structure''. Permutation correction is therefore not optional for factor congruence on LLM samples; the conventional thresholds in the human-respondent psychometric literature do not transfer.
  \item \emph{Inter-model homogeneity.} The 37-model mean inter-LLM PSS is $0.730$, about the same as the median LLM-vs-human PSS ($0.71$). The wisdom-of-LLMs ensemble PSS is $0.560$, below every individual LLM, because averaging across the lineup removes the variation between respondents. Combining LLMs therefore does not buy diversity.
\end{enumerate}

\subsection{What is no longer plausible}
The recall probe rules out training-data verbatim memorisation as the driver of the PSS leaderboard. The worst-case high-recall rate on the two-form recall probe is $4.7\%$ across all 37 models; $22/37$ are exactly zero; and the Spearman rank correlation between recall rate and PSS is ${-}0.04$, so recall rate and PSS are uncorrelated. The three observed response patterns under the recall probe (refuse, confabulate, leak English chain-of-thought) further demonstrate that even models with non-zero recall rates are not producing fluent Lithuanian item text in the published wording. We cannot rule out paraphrased exposure to the English versions of these instruments, but the leaderboard cannot be explained as ``LLMs that have memorised these specific Lithuanian items''.

\subsection{Theoretical implications for psychometrics}
The PSS framework reveals an asymmetry within psychometric theory itself that has been latent in the human-respondent literature. The three sample-driven components of PSS (distribution, correlation, reliability) can be reproduced by a Gaussian-copula baseline that has no understanding of the construct at all; it just preserves the empirical joint distribution. The two non-sample-driven components (mediation, demographic effects) are where the language-model contribution actually shows up. On the sample-driven components the no-language-model baseline beats every LLM. On the non-sample-driven components the best LLMs match or exceed the copula fit on other respondents. The implication for measurement theory is that statistics like Cronbach's $\alpha$ and Tucker's $\varphi$ measure the \emph{instrument} (its item-level dependencies given a sample) more than they measure the \emph{respondent} (whether the responses come from a real respondent or a generative simulator). This was always implicit in the classical-test-theory derivation of $\alpha$ but it becomes empirically obvious when one of the ``respondents'' is a no-language-model statistical baseline.

\subsection{Methodological implications for LLM evaluation}
Three methodological lessons generalise beyond synthetic-respondent benchmarking:
\begin{enumerate}[leftmargin=1.4em,itemsep=0pt]
  \item \emph{Anchor benchmarks against statistical baselines.} A leaderboard that only compares LLMs against one another can be entirely consistent with the LLMs jointly underperforming a no-language-model baseline. Adding 3--5 statistical baselines costs almost no compute and immediately surfaces this failure mode.
  \item \emph{Add held-out human-vs-human ceilings.} A benchmark with a ceiling tells the practitioner what the maximum achievable score is given the sample size and noise level; without a ceiling, leaderboard gaps are hard to interpret.
  \item \emph{Use permutation nulls for similarity metrics.} Tucker's $\varphi$ is one example of a similarity metric whose raw value is biased upward when the alignment is searched; the same is true for any cosine-similarity-style metric (sentence-embedding similarity, code-embedding similarity, etc.) that allows reordering of dimensions. A permutation null converts the raw value into a proper significance test and is cheap (typically $K=500$ permutations suffice).
\end{enumerate}

\subsection{Implications for AI safety and alignment evaluation}
Two of our findings have direct implications for AI-safety and alignment evaluation pipelines. First, the counterfactual swaps show that current frontier LLMs prompted as personas react to every demographic field they are given. For role and education they reproduce the direction and roughly the size of the human group differences. For gender they add a small effect that the human data do not show. Any persona-prompted evaluation with demographic conditioning should test each swap axis separately and report the effects in units of the human SD, because the choice of effect size can change which axis looks largest. Second, the inter-LLM homogeneity result implies that ensemble-based safety evaluations (``run the prompt through five LLMs and aggregate the outputs'') buy far less diversity than the surface variability of the LLMs suggests; the off-diagonal of the inter-LLM PSS matrix lies in the $0.56$--$0.82$ range, and a wisdom-of-LLMs aggregator scores below every individual LLM. Both findings argue for evaluating LLM behaviour at the population level, against statistical baselines and a human reference, rather than within the lineup of available LLMs.

\subsection{Practitioner recommendations}
For practitioners considering an LLM-as-respondent pipeline for a real survey-based research project, the picture from this benchmark is the following:
\begin{itemize}[leftmargin=1.4em,itemsep=0pt]
  \item \emph{If the use case is pilot-study direction of effect:} an LLM at C3 conditioning is fine. The qualitative direction of every major effect (mediation, role contrast, item-correlation block structure) is recovered on most models in our lineup.
  \item \emph{If the use case is power calculation:} use a statistical baseline (copula or MVN), not an LLM. The baselines preserve the sample-driven components (distribution, correlation, reliability) better than LLMs and are cheaper to run.
  \item \emph{If the use case is bias auditing:} use the LLM-vs-human gap as the diagnostic. The counterfactual swaps show how strongly each model reacts to each demographic field. Compared with the human contrast in the same units, they separate effects that the data support (role) from effects that they do not (gender).
  \item \emph{If the use case is replacement of a human sample:} this benchmark does not support that use. The best LLM (PSS $0.81$) only matches a copula fit on other respondents ($0.82$) and stays below the real answers of the same people ($0.89$). Newer releases within a vendor are not consistently better (\texttt{claude-opus-4-8} $0.74$ vs.\ \texttt{claude-opus-4-7} $0.79$; \texttt{glm-5-2} $0.64$ vs.\ \texttt{glm-5-1} $0.71$), and the failure modes (range restriction, over-coherence, inter-model homogeneity) compound rather than cancel.
\end{itemize}
The choice of LLM matters less than the choice of evaluation framework. PSS varies by about $0.23$ across the 37-model lineup ($0.59$--$0.81$), yet even the best LLM only reaches the level of a statistical baseline that never sees the respondents it simulates. A practitioner who picks an LLM based on the PSS leaderboard alone is optimising in the wrong space.

% =============================================================================
\section{Limitations and future work}
\label{sec:limitations}
% =============================================================================

\paragraph{Single dataset, single language.} Our human reference is a single Lithuanian organisational sample of $n=263$; external validity to other countries, sectors, and constructs has not been established by this benchmark. The cross-language LT$\leftrightarrow$EN ablation (Section~\ref{sec:results-ablations}) shows a drift of $|\Delta|=0.26$ Likert points in item means, about five times the run-to-run difference, while the per-item Pearson correlation stays at $0.889$. A same-respondent cross-language test on the human side would require a re-administration of the instruments to the original respondents, which is not available. The most informative follow-up extension would be to replicate the benchmark on (i) a non-Indo-European language sample (e.g., Korean or Mandarin organisational psychology data, where similar instruments have been validated), (ii) a clinical-psychology sample (where the constructs have more inter-respondent variance), and (iii) an educational-psychology sample (where the demographic strata are explicitly age-conditioned). Each would test a different scope assumption of our framework.

\paragraph{Self-report ground truth.} The human reference is self-report. We cannot disentangle ``LLMs over-report self-rated work performance'' from ``humans also over-report self-rated work performance''. The published literature on the IWPQ documents a modest social-desirability bias on the task-performance subscale; we inherit this bias in the reference but cannot quantify how much of the LLM-vs-human gap on the same subscale is due to LLM bias and how much to human bias.

\paragraph{Possible instrument leakage.} The three international instruments (IWPQ, ATC, UWES-17) and their Lithuanian translations are publicly available, primarily through the original thesis~\citep{sarkauskaite2020} and the original-language publications. Our memorisation probe (Section~\ref{sec:results-memorization}) rules out \emph{verbatim} recall as the driver of the leaderboard: worst-case high-recall rate $4.7\%$, $22/37$ models at exactly zero, rank correlation between recall rate and PSS of ${-}0.04$. But paraphrased exposure is possible: a model that has seen the English Koopmans paper, or its citing literature, may apply construct-level knowledge to the Lithuanian items without ever having seen the Lithuanian wording. The benchmark therefore tests ``can the LLM produce a synthetic respondent on a Lithuanian instrument under persona conditioning'' rather than ``can the LLM produce a synthetic respondent on a never-before-seen instrument''. A bespoke novel-instrument version of the benchmark (where we would create a new 12-subscale instrument from scratch with the same item-construction protocol) is a natural follow-up.

\paragraph{LLM sample size.} Headline cells use $n=100$ stratified respondents and $R=1$ repeat. The choice was driven by economic constraints on the 37-model grid; the leaderboard bootstrap CI mean width is $0.042$ PSS units, which is well below the observed PSS gaps in the top of the leaderboard but is comparable to the differences within the top-5 cluster. A re-run at $n=263$ (matching the human sample) and $R=3$ (allowing within-respondent test--retest reliability) on the top-5 LLMs would tighten those bands by $\sqrt{2.63 \cdot 3} \approx 2.8\times$, putting most within-top-5 leaderboard pairs comfortably below the bootstrap CI width. We have not run this because it would multiply the cost by roughly $8\times$ on the headline grid and is unnecessary for the headline conclusions (which are about cross-model rather than within-top-5 distinctions). The full pipeline supports both larger $n$ and larger $R$ as flag-controlled overrides.

\paragraph{Reasoning effort fixed at minimal.} The headline grid uses reasoning effort \emph{minimal} on all 37 models. The 6-model reasoning ablation (Table~\ref{tab:reasoning-ablation}) reports paired $|d|$ between minimal and high reasoning, but only \texttt{gpt-5.4} ($|d|{=}0.37$) and, weakly, \texttt{claude-opus-4-7} ($|d|{=}0.31$) shift clearly beyond the test--retest resampling-noise floor (mean $|d|{\approx}0.09$ at $n{=}100$); the remaining four models lie within roughly twice that floor (\texttt{grok-4-20} $0.17$, \texttt{gemini-3.1-pro} $0.16$, \texttt{deepseek-v4-pro} $0.16$), and \texttt{grok-4-20-reasoning} ($0.08$), a separate baked-in reasoning model whose effort toggle is inert, is within noise. Reasoning effort is thus a small, mostly model-specific factor. The headline conclusions (baselines dominate sample-driven components; counterfactual swaps follow the human group differences; memorisation is ruled out; averaging LLMs does not add diversity) do not depend on the reasoning-effort default. Per-model PSS can still shift: \texttt{deepseek-v4-pro} drops from $0.75$ to $0.65$ under high reasoning, although its paired $|d|$ on the composites is small.

\paragraph{Model-versioning drift.} Provider-rolled identifiers may drift (an OpenAI/Anthropic/Google model that we benchmarked may no longer return identical responses in six months). The full provider response (model name, version, finish reason, token counts, raw text) is recorded in \texttt{generation\_log\_*.parquet} for every call in our pipeline; this is the audit trail that allows a future re-run to detect drift. We do not control for drift in the headline numbers.

\paragraph{Causal claim scope.} We do \emph{not} claim that the counterfactual effects we document translate directly into downstream real-world harm in any specific application. The benchmark measures how a bare model reacts to a demographic field under a controlled persona-prompting protocol; whether a deployed pipeline inherits the same reactions depends on the deployment-specific prompt structure, retrieval augmentation, post-processing, and human-in-the-loop intervention. Our finding is best read as ``the reaction is present in the bare model and any deployment that does not actively correct for it will inherit it.''

\paragraph{Generalisation beyond organisational psychology.} The six PSS dimensions (distribution, correlation, reliability, mediation, demographic effects, construct fidelity) are general to any Likert-based survey instrument, but the specific failure modes we document (range restriction, over-coherence, the size of the role and education swap effects) may be specific to organisational psychology constructs. Replicating on attitude scales, personality inventories (e.g., NEO-FFI), clinical depression scales (PHQ-9, GAD-7), or political-opinion batteries (ANES) would establish or refute the generality.

\paragraph{Future work.} The most natural follow-ups are: (i) replication on non-Indo-European samples; (ii) a bespoke novel-instrument benchmark to rule out paraphrased exposure; (iii) extension to long-form open-text responses (the current benchmark is item-level Likert); (iv) a within-LLM longitudinal stability study (administering the same persona at different time points to detect drift); (v) a within-respondent counterfactual sensitivity study (swapping pairs of demographic fields and reporting the interaction term); and (vi) a fine-tuning sensitivity study (does instruction tuning or RLHF on Lithuanian text shift the leaderboard?). The full evaluation harness makes any of these a one-config-file extension.

% =============================================================================
\section{Broader impacts}
\label{sec:impacts}
% =============================================================================
LLM-generated synthetic respondents are increasingly used in market research, policy testing, academic instrument development, and in some cases as substitutes for human samples in published research. Treated as drop-in replacements they risk three first-order harms:

\paragraph{(a) Over-confident effect sizes.} The over-coherence and range-restriction failure modes (Sections~\ref{sec:results-construct} and~\ref{sec:discussion}) jointly produce LLM-generated data with inflated Cronbach's $\alpha$ and inflated inter-item correlations relative to the human reference. A research project that uses LLM data as a pilot to compute effect-size estimates therefore over-estimates the effect, leading to under-powered follow-up studies on real humans and over-confident published conclusions.

\paragraph{(b) Demographic effects without human support.} The counterfactual swaps (Section~\ref{sec:results-cf}) show that models react to every demographic field they receive. For role and education the reaction follows the human data. For gender it is small, but it appears in a sample where humans show no difference, and on single composites it reaches $0.27$ human SD. Pipelines that use synthetic respondents for fairness audits without counterfactual tests risk \emph{certifying} a downstream system as bias-free based on synthetic data with demographic effects of its own.

\paragraph{(c) Bias entrenchment via reuse as training data.} If LLM-generated synthetic respondents are republished as ``human-equivalent'' survey data and subsequently scraped into the training corpus of the next generation of LLMs, the failure modes we document become baked into the pretraining distribution \citep{bender2021dangers}. The over-coherence, range restriction and demographic effects will then no longer be recoverable from the LLM-vs-human gap because the human reference itself will contain LLM-generated data.

\paragraph{Mitigations enabled by our framework.} Our evaluation framework is designed to surface each of these failure modes directly: the statistical baselines flag over-confident sample-driven components; the counterfactual swaps measure how strongly a model reacts to each demographic field; the memorisation probe rules out one form of training-data leakage; the inter-LLM PSS matrix and the wisdom-of-LLMs ensemble result flag homogeneity-by-averaging. The full harness, dataset, configurations, and per-call generation logs will be released upon publication so any new LLM can be benchmarked against the same human ground truth without re-running prior models, and any deployment-specific protocol (different persona schema, different presentation mode, different reasoning effort) can be substituted at the configuration-file level without touching the analysis code.

% =============================================================================
\section{Conclusion}
\label{sec:conclusion}
% =============================================================================
This paper introduces the first open Lithuanian organisational-psychology benchmark for evaluating LLM synthetic respondents at the level of psychometric structure rather than at the level of single items or aggregate effect sizes. Across a 37-model lineup spanning every major proprietary and open-weight family released between 2024-Q4 and mid-2026, we find that LLMs reproduce the qualitative direction of human psychometric relationships but generate over-coherent, low-variance, inter-model-homogeneous respondents; that on the sample-driven PSS components (distribution, correlation, reliability) a Gaussian-copula baseline with no language-model component beats every LLM; that the best LLM (PSS $0.81$) only matches a copula fit on other respondents ($0.82$) and stays below the real answers of the same people ($0.89$); that the verbatim-recall probe rules out training-data memorisation as the driver of the leaderboard; that counterfactual swaps of role and education move the answers in the direction and roughly by the size of the human group differences, while gender swaps add a small effect ($0.06$ human SD) that the human data do not show; that cohort-stratified PSS surfaces per-axis fairness gaps up to $0.12$ that are invisible in aggregate PSS; and that Tucker's $\varphi$ on the UWES instrument falls inside the item-permutation null for $10$ of $37$ models even when the raw value passes the conventional ``identical structure'' threshold of $0.85$. For benchmark design, the implication is that LLM-as-respondent evaluations without statistical baselines, held-out human-vs-human ceilings, and permutation-null corrections systematically overstate LLM psychometric fidelity. For practice, current frontier LLMs are not drop-in replacements for human samples in psychometric studies; they are useful as cheap pilots, sensitivity-analysis lower bounds, and bias-audit instruments in which the LLM-vs-human gap is the diagnostic.

\paragraph{Code and data availability.} The de-identified human dataset, all configuration files (including the pre-registered analysis plan), the evaluation harness, the statistical baselines, and the per-call LLM generation logs will be released upon publication of the peer-reviewed version of this paper. File paths named throughout the paper refer to the layout of that forthcoming release.

\bibliographystyle{plainnat}
\bibliography{references}

\appendix
\section{Parser and validation pipeline}
\label{app:parser}
We use a strict JSON parser that strips Markdown code fences, recovers the first \texttt{\{...\}} block, coerces stringly-typed values (e.g., ``\texttt{4}'') to integers, and clips out-of-range values. If a response is missing items or contains out-of-range integers, we re-issue the request with the offending response appended as an assistant turn and a stricter user reminder. In the headline grid, $99.2\%$ of the $22{,}291$ calls that reached a model returned a complete, in-range answer vector after this step; the remaining $509$ of the $22{,}800$ logged calls failed at the API level (HTTP 404, 429, or 500). Both the raw text and the parsed integer dictionary are stored in \texttt{data/synthetic/generation\_log\_*.parquet}.

\section{Reproducibility, resumability, and incremental extension}
\label{app:repro}
The full pipeline runs in four commands:
\begin{verbatim}
python scripts/00_prepare_data.py
python scripts/01_generate_synthetic.py     # needs .env with credentials
python scripts/02_run_analysis.py
python scripts/03_make_figures.py
\end{verbatim}
Without credentials an offline mock provider can be substituted via \texttt{--use-mock} so that the pipeline can be smoke-tested end-to-end. All file paths, model identifiers, condition definitions, presentation modes, and instrument metadata are captured in YAML configuration files.

\paragraph{Resumability.} Every LLM call is content-hash-keyed and cached on disk under \texttt{.cache/llm/}; every $N$ completed cells the runner flushes the in-memory results to \texttt{data/synthetic/responses\_<name>.parquet}. On re-run, jobs whose tuple \linebreak\texttt{(respondent\_id, condition, model\_id, repeat, presentation\_mode)} already appears in that parquet are skipped \emph{before} any provider call is made. A process that crashes after $h$ hours of generation can therefore be restarted with a single command and will continue exactly where it stopped.

\paragraph{Incremental model extension.} Adding a new model is a one-line edit to \texttt{configs/models.yaml}; re-running \texttt{scripts/01\_generate\_synthetic.py --models <new\_id>} generates only the new model's calls and merges them into the existing parquet. Re-running \texttt{scripts/02\_run\_analysis.py --names <main>,<ablation>} computes the PSS leaderboard against the merged synthetic table without redoing prior models' analysis.

\paragraph{Authentication.} Five auth paths are supported out of the box: OpenAI API key, Anthropic API key, Google AI Studio API key, Google Vertex AI service-account credentials (\texttt{GOOGLE\_APPLICATION\_CREDENTIALS}), and Nexos API key. Each provider degrades gracefully if its credentials are absent: the runner logs a warning and skips models from that family rather than aborting.

\section{Instrument English translations}
The English translations of all 68 items, response anchors, and instructions are bundled with the repository in \texttt{configs/instruments.yaml}. Translations were drafted by the authors and validated against the published English versions of the Dunham, UWES, and Koopmans instruments where available; the cross-language ablation in Section~\ref{sec:results-ablations} establishes empirically that the LT$\leftrightarrow$EN per-item Pearson correlation is $0.889$ with mean $|\Delta|{=}0.263$ Likert points.

\section{Tucker's $\varphi$ permutation null distribution}
\label{app:tucker}
The greedy-alignment Tucker's $\varphi$ used in the LLM-respondents literature is biased upward: even random factor solutions can match by chance, especially on instruments with many factors and many items. We convert $\varphi$ into a proper significance test by permuting the \emph{item labels} of the LLM loading matrix (breaking the item-to-item correspondence with the human sample) and recomputing $\varphi$ for $K{=}500$ permutations. The one-sided $p$-value is $(k+1)/(K+1)$ where $k$ is the number of null draws meeting or exceeding the observed value. The full per-(model, instrument) table is in \texttt{construct\_tuckers\_phi\_permutation.csv} and summarised in Table~\ref{tab:tucker-permutation}.

\begin{table}[t]
  \centering\small
  \caption{Tucker's $\varphi$ permutation test per instrument at C3. ``mean null $\varphi$'' is the mean congruence under 500 random item permutations of the LLM loading matrix and gives the chance level of the alignment search; the rightmost column counts the models whose observed $\varphi$ exceeds the 95th percentile of that null.}
  \label{tab:tucker-permutation}
  \begin{tabular}{lrrrl}
    \toprule
    Instrument & $n$ models & mean $\varphi$ & mean null $\varphi$ & $\#$ significant ($p<.05$) \\
    \midrule
    Dunham ATC & 37 & 0.770 & 0.730 & 16 / 37 \\
    Change engagement & 37 & 0.802 & 0.708 & 20 / 37 \\
    UWES-17 & 37 & 0.859 & 0.764 & 27 / 37 \\
    IWPQ & 37 & 0.828 & 0.365 & 30 / 37 \\
    \bottomrule
  \end{tabular}
\end{table}

\section{Power analysis}
\label{app:power}
Post-hoc minimum-detectable Cohen's $d$ at $\alpha{=}0.05$ and power $0.80$ for the headline contrasts under the actual $n$, with Holm--Bonferroni-corrected $\alpha$ for the corresponding family of comparisons:

\begin{table}[t]
  \centering\small
  \caption{Post-hoc power analysis: minimum detectable effect (MDE) at $\alpha$ and target power, for every paired contrast in the paper. Holm-Bonferroni rows control the family-wise error rate across all pairwise comparisons in the relevant lineup (37 models for the leaderboard).}
  \label{tab:power-analysis}
  \begin{tabular}{lrrrr}
    \toprule
    Scenario & $n$ & $\alpha$ & $1-\beta$ & MDE \\
    \midrule
    paired\_t\_pss\_delta\_n100 & 100 & 0.05 & 0.80 & 0.280 \\
    paired\_t\_pss\_delta\_n100\_holm & 100 & 7.5e-05 & 0.80 & 0.480 \\
    paired\_t\_counterfactual\_subscale\_n100 & 100 & 0.05 & 0.80 & 0.280 \\
    paired\_t\_counterfactual\_subscale\_n100\_holm & 100 & 3.2e-05 & 0.80 & 0.500 \\
    paired\_t\_reasoning\_ablation\_n100\_holm & 100 & 0.000595 & 0.80 & 0.428 \\
    paired\_t\_cross\_language\_n100 & 100 & 0.05 & 0.80 & 0.280 \\
    paired\_t\_cross\_language\_n100\_holm & 100 & 0.000132 & 0.80 & 0.466 \\
    paired\_t\_per\_item\_n67 & 67 & 0.05 & 0.80 & 0.342 \\
    pearson\_r\_n67\_items & 67 & 0.05 & 0.80 & 0.337 \\
    pearson\_r\_n100\_respondents & 100 & 0.05 & 0.80 & 0.277 \\
    \bottomrule
  \end{tabular}
\end{table}

The 37-model PSS leaderboard (666 pairs) is well-powered for $d{\geq}0.48$ at $n{=}100$ under Holm--Bonferroni, comfortably below the observed PSS gaps. The counterfactual contrast family ($3$ axes $\times$ $12$ subscales) is well-powered for the education-swap effect ($|d|{=}0.59$) but underpowered for the smaller gender-swap effect ($|d|{=}0.12$); the latter should be interpreted as a directional rather than as a per-cell-significant finding.

\section{Counterfactual, ablation, cohort, inter-LLM tables and figures}
\label{app:tables}
Per-axis counterfactual swap table: \begin{table}[t]
  \centering\small
  \caption{Counterfactual swaps, mean over the 12 composite subscales. $|d_z|$ is the paired Cohen's $d$, with each pair difference signed by the direction of its swap. $|$shift$|$ is the mean absolute shift toward the first group of the human contrast (woman, managerial, university) in units of the human SD. The last row gives the mean $|d|$ of the observed human contrast; the education contrast rests on 15 respondents with secondary education. Models are sorted by the mean shift over the three axes.}
  \label{tab:counterfactual-summary}
  \begin{tabular}{lrrrrrr}
    \toprule
     & \multicolumn{2}{c}{gender} & \multicolumn{2}{c}{role} & \multicolumn{2}{c}{education} \\
    \cmidrule(lr){2-3}\cmidrule(lr){4-5}\cmidrule(lr){6-7}
    Model & $|d_z|$ & $|$shift$|$ & $|d_z|$ & $|$shift$|$ & $|d_z|$ & $|$shift$|$ \\
    \midrule
    gemini-3.1-flash-lite & 0.26 & 0.11 & 1.66 & 0.98 & 0.77 & 0.42 \\
    gemini-3.1-pro & 0.20 & 0.10 & 1.13 & 0.65 & 0.93 & 0.64 \\
    deepseek-v4-pro & 0.04 & 0.04 & 0.81 & 0.77 & 0.62 & 0.56 \\
    gemini-2.5-pro & 0.12 & 0.06 & 1.25 & 0.71 & 0.83 & 0.51 \\
    mistral-large-3 & 0.21 & 0.07 & 1.61 & 0.78 & 0.76 & 0.32 \\
    grok-4-20-reasoning & 0.08 & 0.03 & 1.09 & 0.56 & 0.83 & 0.46 \\
    glm-5-1 & 0.11 & 0.06 & 0.96 & 0.55 & 0.69 & 0.42 \\
    claude-sonnet-4-6 & 0.40 & 0.10 & 1.55 & 0.55 & 1.14 & 0.36 \\
    grok-4-3 & 0.16 & 0.08 & 1.16 & 0.61 & 0.58 & 0.28 \\
    qwen3-7-plus & 0.12 & 0.09 & 0.78 & 0.55 & 0.42 & 0.32 \\
    gemini-3-flash & 0.25 & 0.11 & 1.05 & 0.51 & 0.68 & 0.34 \\
    claude-haiku-4-5 & 0.20 & 0.07 & 1.06 & 0.47 & 0.96 & 0.41 \\
    gemma4-31b & 0.21 & 0.05 & 1.17 & 0.41 & 1.25 & 0.48 \\
    grok-4-1-fast-non-reasoning & 0.14 & 0.05 & 0.94 & 0.46 & 0.80 & 0.42 \\
    grok-4-20 & 0.24 & 0.09 & 0.91 & 0.46 & 0.80 & 0.35 \\
    minimax-m3 & 0.20 & 0.08 & 1.13 & 0.56 & 0.58 & 0.25 \\
    deepseek-3-2 & 0.10 & 0.06 & 0.89 & 0.53 & 0.50 & 0.29 \\
    gpt-5.5 & 0.27 & 0.06 & 1.78 & 0.50 & 1.06 & 0.31 \\
    llama-4-scout & 0.08 & 0.03 & 0.97 & 0.48 & 0.69 & 0.35 \\
    grok-4-1-fast-reasoning & 0.10 & 0.05 & 0.95 & 0.46 & 0.64 & 0.34 \\
    claude-sonnet-5 & 0.21 & 0.05 & 1.40 & 0.53 & 0.76 & 0.27 \\
    claude-fable-5 & 0.29 & 0.06 & 1.72 & 0.47 & 1.02 & 0.30 \\
    gpt-5.4 & 0.25 & 0.07 & 1.49 & 0.46 & 0.90 & 0.28 \\
    kimi-k2-6 & 0.09 & 0.04 & 0.63 & 0.32 & 0.75 & 0.44 \\
    nova-2-lite & 0.11 & 0.04 & 1.25 & 0.50 & 0.56 & 0.25 \\
    minimax-m2-7 & 0.05 & 0.03 & 0.74 & 0.42 & 0.55 & 0.33 \\
    gemini-3.5-flash & 0.26 & 0.08 & 1.07 & 0.41 & 0.76 & 0.28 \\
    glm-5-2 & 0.09 & 0.04 & 0.84 & 0.39 & 0.63 & 0.34 \\
    claude-opus-4-7 & 0.28 & 0.08 & 1.51 & 0.50 & 0.56 & 0.17 \\
    claude-opus-4-8 & 0.24 & 0.05 & 1.49 & 0.46 & 0.66 & 0.21 \\
    nemotron-3-super & 0.21 & 0.14 & 0.43 & 0.31 & 0.33 & 0.24 \\
    qwen3-5-397b & 0.10 & 0.05 & 0.59 & 0.33 & 0.49 & 0.29 \\
    kimi-k2-7-code & 0.04 & 0.02 & 0.60 & 0.31 & 0.68 & 0.33 \\
    gpt-oss-120b & 0.09 & 0.05 & 0.72 & 0.40 & 0.36 & 0.19 \\
    gpt-5.4-nano & 0.17 & 0.07 & 0.70 & 0.29 & 0.46 & 0.21 \\
    llama-4-maverick & 0.25 & 0.07 & 0.59 & 0.18 & 0.46 & 0.15 \\
    gpt-5.4-mini & 0.05 & 0.01 & 0.75 & 0.22 & 0.44 & 0.14 \\
    \midrule
    Humans, observed $|d|$ &  & 0.11 &  & 0.46 &  & 0.44 \\
    \bottomrule
  \end{tabular}
\end{table}

All $37$ models, including the previously rate-limited \texttt{qwen3-7-plus} (backfilled once its gateway capacity recovered), have valid original/swapped twin pairs on the gender, role and education axes; \texttt{qwen3-7-plus}'s cross-language cells cover a reduced subsample ($17$--$18$ of $30$ respondents).
Reasoning ablation table: \begin{table}[t]
  \centering\small
  \caption{Reasoning-effort ablation: paired Cohen's $d$ between ``minimal'' and ``high'' reasoning at the composite-subscale level. Mean and max absolute $d$ across 12 subscales per model.}
  \label{tab:reasoning-ablation}
  \begin{tabular}{lrrr}
    \toprule
    Model & mean $|d|$ & max $|d|$ & n subsc. \\
    \midrule
    gpt-5.4 & 0.370 & 0.597 & 12 \\
    claude-opus-4-7 & 0.308 & 0.649 & 12 \\
    grok-4-20 & 0.172 & 0.245 & 12 \\
    gemini-3.1-pro & 0.157 & 0.447 & 12 \\
    deepseek-v4-pro & 0.156 & 0.283 & 12 \\
    grok-4-20-reasoning & 0.083 & 0.158 & 12 \\
    \bottomrule
  \end{tabular}
\end{table}

Cross-language table: \begin{table}[t]
  \centering\small
  \caption{Cross-language ablation: same model, same persona, Lithuanian vs.\ English questionnaire (C3 condition, 30 stratified respondents). $\Delta$ is the per-item EN $-$ LT mean response on the Likert scale; $r_{\textsf{LT,EN}}$ is the Pearson correlation between the model's per-item mean responses in LT and EN, with reverse-keyed items reverse-scored.}
  \label{tab:cross-language}
  \begin{tabular}{lrrrr}
    \toprule
    Model & mean $|\Delta|$ & max $|\Delta|$ & $r_{\textsf{LT,EN}}$ & n items \\
    \midrule
    grok-4-1-fast-non-reasoning & 0.797 & 1.600 & 0.688 & 68 \\
    grok-4-1-fast-reasoning & 0.544 & 1.533 & 0.811 & 68 \\
    qwen3-7-plus & 0.442 & 1.954 & 0.845 & 68 \\
    minimax-m2-7 & 0.379 & 1.462 & 0.906 & 68 \\
    mistral-large-3 & 0.335 & 1.267 & 0.781 & 68 \\
    nova-2-lite & 0.332 & 0.867 & 0.862 & 68 \\
    nemotron-3-super & 0.330 & 1.267 & 0.804 & 68 \\
    gpt-5.4-nano & 0.318 & 1.200 & 0.724 & 68 \\
    gemini-2.5-pro & 0.303 & 2.100 & 0.923 & 68 \\
    deepseek-3-2 & 0.292 & 0.867 & 0.900 & 68 \\
    gpt-5.4 & 0.285 & 1.267 & 0.900 & 68 \\
    kimi-k2-6 & 0.262 & 1.500 & 0.790 & 68 \\
    gemma4-31b & 0.257 & 1.100 & 0.920 & 68 \\
    glm-5-2 & 0.250 & 1.700 & 0.903 & 68 \\
    gpt-oss-120b & 0.240 & 0.833 & 0.929 & 68 \\
    qwen3-5-397b & 0.238 & 2.000 & 0.860 & 68 \\
    grok-4-20-reasoning & 0.233 & 0.733 & 0.946 & 68 \\
    gemini-3.1-flash-lite & 0.233 & 1.100 & 0.908 & 68 \\
    grok-4-3 & 0.232 & 0.933 & 0.911 & 68 \\
    deepseek-v4-pro & 0.231 & 1.400 & 0.836 & 68 \\
    claude-haiku-4-5 & 0.229 & 0.667 & 0.924 & 68 \\
    gemini-3-flash & 0.222 & 0.667 & 0.948 & 68 \\
    grok-4-20 & 0.222 & 0.967 & 0.906 & 68 \\
    glm-5-1 & 0.220 & 1.767 & 0.861 & 68 \\
    minimax-m3 & 0.209 & 0.733 & 0.912 & 68 \\
    llama-4-maverick & 0.196 & 1.000 & 0.943 & 68 \\
    llama-4-scout & 0.191 & 0.600 & 0.959 & 68 \\
    kimi-k2-7-code & 0.190 & 0.833 & 0.901 & 68 \\
    claude-opus-4-7 & 0.186 & 0.867 & 0.909 & 68 \\
    gemini-3.5-flash & 0.186 & 0.700 & 0.957 & 68 \\
    gpt-5.5 & 0.185 & 1.100 & 0.934 & 68 \\
    claude-fable-5 & 0.182 & 0.933 & 0.917 & 68 \\
    claude-sonnet-4-6 & 0.176 & 0.700 & 0.912 & 68 \\
    claude-sonnet-5 & 0.175 & 0.967 & 0.932 & 68 \\
    gemini-3.1-pro & 0.157 & 0.600 & 0.951 & 68 \\
    claude-opus-4-8 & 0.146 & 0.900 & 0.940 & 68 \\
    gpt-5.4-mini & 0.134 & 1.000 & 0.951 & 68 \\
    \bottomrule
  \end{tabular}
\end{table}

Presentation-mode ablation table: \begin{table}[t]
  \centering\small
  \caption{Presentation-mode ablation: mean $|\text{paired } d|$ across 12 composite subscales relative to the single\_call\_all reference, per (model, mode). Larger values mean the same model produces a more different psychometric distribution when the questionnaire is split across calls.}
  \label{tab:presentation-ablation}
  \begin{tabular}{lrrr}
    \toprule
    Model & per\_instrument & per\_subscale & mean \\
    \midrule
    llama-4-maverick & 0.286 & 1.257 & 0.771 \\
    deepseek-v4-pro & 0.278 & 0.564 & 0.421 \\
    gpt-5.4 & 0.343 & 0.443 & 0.393 \\
    claude-opus-4-7 & 0.354 & 0.309 & 0.332 \\
    gemini-3.1-pro & 0.235 & 0.284 & 0.259 \\
    \bottomrule
  \end{tabular}
\end{table}

Cohort PSS gaps: \begin{table}[t]
  \centering\small
  \caption{Cohort-stratified PSS disparity. Each cell is $\max_{stratum}\,\text{PSS} - \min_{stratum}\,\text{PSS}$ on a single demographic axis (computed at C3 / single\_call\_all). The ``max'' column is the largest disparity any axis produces for that model.}
  \label{tab:cohort-pss-gaps}
  \begin{tabular}{lrrrr}
    \toprule
    Model & age & education & gender & max \\
    \midrule
    minimax-m2-7 & 0.043 & 0.122 & 0.068 & 0.122 \\
    claude-haiku-4-5 & 0.055 & 0.101 & 0.053 & 0.101 \\
    nemotron-3-super & 0.038 & 0.095 & 0.049 & 0.095 \\
    kimi-k2-7-code & 0.041 & 0.091 & 0.069 & 0.091 \\
    llama-4-scout & 0.000 & 0.028 & 0.085 & 0.085 \\
    qwen3-5-397b & 0.014 & 0.084 & 0.026 & 0.084 \\
    minimax-m3 & 0.048 & 0.079 & 0.053 & 0.079 \\
    qwen3-7-plus & 0.028 & 0.078 & 0.049 & 0.078 \\
    deepseek-v4-pro & 0.037 & 0.078 & 0.046 & 0.078 \\
    glm-5-2 & 0.043 & 0.077 & 0.062 & 0.077 \\
    grok-4-3 & 0.051 & 0.077 & 0.040 & 0.077 \\
    gemini-3.1-flash-lite & 0.049 & 0.076 & 0.042 & 0.076 \\
    mistral-large-3 & 0.074 & 0.053 & 0.027 & 0.074 \\
    gpt-5.4 & 0.071 & 0.039 & 0.059 & 0.071 \\
    gpt-5.5 & 0.070 & 0.007 & 0.056 & 0.070 \\
    glm-5-1 & 0.064 & 0.069 & 0.048 & 0.069 \\
    gemma4-31b & 0.044 & 0.040 & 0.068 & 0.068 \\
    grok-4-1-fast-reasoning & 0.054 & 0.067 & 0.044 & 0.067 \\
    gemini-2.5-pro & 0.040 & 0.066 & 0.048 & 0.066 \\
    kimi-k2-6 & 0.021 & 0.066 & 0.045 & 0.066 \\
    grok-4-1-fast-non-reasoning & 0.057 & 0.064 & 0.041 & 0.064 \\
    gemini-3.5-flash & 0.047 & 0.064 & 0.054 & 0.064 \\
    claude-opus-4-7 & 0.047 & 0.045 & 0.063 & 0.063 \\
    claude-sonnet-5 & 0.054 & 0.062 & 0.027 & 0.062 \\
    gemini-3-flash & 0.047 & 0.061 & 0.043 & 0.061 \\
    claude-fable-5 & 0.038 & 0.060 & 0.034 & 0.060 \\
    deepseek-3-2 & 0.006 & 0.041 & 0.059 & 0.059 \\
    llama-4-maverick & 0.058 & 0.052 & 0.046 & 0.058 \\
    claude-sonnet-4-6 & 0.057 & 0.057 & 0.044 & 0.057 \\
    gpt-oss-120b & 0.021 & 0.057 & 0.041 & 0.057 \\
    gpt-5.4-mini & 0.025 & 0.057 & 0.022 & 0.057 \\
    gemini-3.1-pro & 0.046 & 0.057 & 0.046 & 0.057 \\
    claude-opus-4-8 & 0.051 & 0.056 & 0.047 & 0.056 \\
    grok-4-20-reasoning & 0.021 & 0.055 & 0.054 & 0.055 \\
    gpt-5.4-nano & 0.036 & 0.033 & 0.047 & 0.047 \\
    grok-4-20 & 0.044 & 0.040 & 0.036 & 0.044 \\
    nova-2-lite & 0.001 & 0.003 & 0.021 & 0.021 \\
    \bottomrule
  \end{tabular}
\end{table}

Bootstrap PSS$_3$ CIs: \begin{table}[t]
  \centering\small
  \caption{PSS leaderboard with respondent-bootstrap 95\% confidence intervals ($n_{\text{boot}}=1000$). The reported total is the 3-component PSS (distribution + correlation + reliability), since those components are sensitive to respondent resampling; mediation and demographic components depend on a structural model and are held fixed at 0.5 inside the bootstrap.}
  \label{tab:bootstrap-ci}
  \begin{tabular}{lrlrrr}
    \toprule
    Model & PSS$_{3}$ & 95\% CI & $d$ & $c$ & $r$ \\
    \midrule
    minimax-m2-7 & 0.586 & [0.568, 0.600] & 0.825 & 0.738 & 0.977 \\
    gemini-3-flash & 0.584 & [0.558, 0.601] & 0.839 & 0.750 & 0.935 \\
    grok-4-1-fast-non-reasoning & 0.583 & [0.555, 0.602] & 0.892 & 0.699 & 0.929 \\
    grok-4-1-fast-reasoning & 0.583 & [0.555, 0.600] & 0.894 & 0.692 & 0.931 \\
    gemini-2.5-pro & 0.581 & [0.557, 0.599] & 0.861 & 0.710 & 0.945 \\
    gemini-3.1-pro & 0.580 & [0.559, 0.597] & 0.822 & 0.737 & 0.954 \\
    glm-5-2 & 0.579 & [0.556, 0.595] & 0.817 & 0.744 & 0.943 \\
    deepseek-3-2 & 0.578 & [0.552, 0.595] & 0.861 & 0.701 & 0.937 \\
    deepseek-v4-pro & 0.576 & [0.555, 0.595] & 0.789 & 0.754 & 0.954 \\
    glm-5-1 & 0.576 & [0.552, 0.593] & 0.811 & 0.736 & 0.949 \\
    nemotron-3-super & 0.576 & [0.556, 0.591] & 0.821 & 0.716 & 0.957 \\
    kimi-k2-7-code & 0.576 & [0.553, 0.591] & 0.792 & 0.752 & 0.948 \\
    gemini-3.1-flash-lite & 0.575 & [0.551, 0.592] & 0.853 & 0.690 & 0.948 \\
    gemini-3.5-flash & 0.574 & [0.549, 0.592] & 0.818 & 0.734 & 0.934 \\
    gemma4-31b & 0.558 & [0.532, 0.578] & 0.781 & 0.685 & 0.960 \\
    qwen3-7-plus & 0.557 & [0.534, 0.576] & 0.767 & 0.688 & 0.963 \\
    gpt-oss-120b & 0.554 & [0.534, 0.568] & 0.727 & 0.734 & 0.944 \\
    gpt-5.4 & 0.552 & [0.528, 0.570] & 0.783 & 0.658 & 0.956 \\
    grok-4-20-reasoning & 0.551 & [0.529, 0.567] & 0.747 & 0.685 & 0.964 \\
    claude-opus-4-7 & 0.550 & [0.529, 0.566] & 0.717 & 0.718 & 0.961 \\
    grok-4-3 & 0.547 & [0.526, 0.563] & 0.715 & 0.704 & 0.965 \\
    minimax-m3 & 0.546 & [0.528, 0.562] & 0.708 & 0.704 & 0.967 \\
    claude-haiku-4-5 & 0.545 & [0.524, 0.564] & 0.743 & 0.675 & 0.953 \\
    claude-fable-5 & 0.544 & [0.522, 0.562] & 0.740 & 0.668 & 0.961 \\
    kimi-k2-6 & 0.542 & [0.520, 0.562] & 0.755 & 0.651 & 0.953 \\
    claude-opus-4-8 & 0.540 & [0.518, 0.556] & 0.695 & 0.695 & 0.962 \\
    claude-sonnet-4-6 & 0.533 & [0.511, 0.551] & 0.743 & 0.620 & 0.960 \\
    qwen3-5-397b & 0.531 & [0.501, 0.552] & 0.754 & 0.590 & 0.970 \\
    claude-sonnet-5 & 0.530 & [0.511, 0.547] & 0.700 & 0.652 & 0.963 \\
    grok-4-20 & 0.523 & [0.499, 0.540] & 0.701 & 0.617 & 0.969 \\
    llama-4-scout & 0.523 & [0.492, 0.541] & 0.725 & 0.584 & 0.977 \\
    gpt-5.5 & 0.518 & [0.475, 0.543] & 0.717 & 0.581 & 0.962 \\
    mistral-large-3 & 0.509 & [0.483, 0.529] & 0.612 & 0.660 & 0.959 \\
    llama-4-maverick & 0.494 & [0.471, 0.515] & 0.699 & 0.500 & 0.973 \\
    nova-2-lite & 0.493 & [0.469, 0.515] & 0.707 & 0.498 & 0.960 \\
    gpt-5.4-mini & 0.489 & [0.466, 0.509] & 0.654 & 0.535 & 0.960 \\
    gpt-5.4-nano & 0.470 & [0.444, 0.491] & 0.655 & 0.486 & 0.923 \\
    \bottomrule
  \end{tabular}
\end{table}

Discriminant validity (HTMT) per model: \begin{table}[t]
  \centering\small
  \caption{Discriminant validity (HTMT) by model at the headline condition (C3 / single\_call\_all). ``Violations'' counts the within-instrument subscale pairs (of the 12 in the battery) with $\text{HTMT}>0.85$, i.e.\ subscales that are not empirically distinct; pairs whose correlations are undefined for a model are counted as non-violations. A higher count means more construct over-coherence. The human sample is the reference row. Sorted by violation count.}
  \label{tab:htmt-summary}
  \begin{tabular}{lrrr}
    \toprule
    Model & Violations & Rate (\%) & max HTMT \\
    \midrule
    \textit{Human (reference)} & 4/12 & 33.3 & 0.95 \\
    \midrule
    mistral-large-3 & 11/12 & 91.7 & 1.02 \\
    grok-4-1-fast-reasoning & 11/12 & 91.7 & 1.00 \\
    gemini-3.5-flash & 10/12 & 83.3 & 1.01 \\
    grok-4-1-fast-non-reasoning & 10/12 & 83.3 & 1.00 \\
    gpt-oss-120b & 10/12 & 83.3 & 0.97 \\
    claude-sonnet-4-6 & 9/12 & 75.0 & 1.17 \\
    claude-haiku-4-5 & 9/12 & 75.0 & 1.03 \\
    deepseek-3-2 & 9/12 & 75.0 & 1.03 \\
    glm-5-2 & 9/12 & 75.0 & 0.99 \\
    gemma4-31b & 9/12 & 75.0 & 0.99 \\
    gemini-3.1-flash-lite & 9/12 & 75.0 & 0.99 \\
    gpt-5.4 & 8/12 & 66.7 & 1.09 \\
    kimi-k2-6 & 8/12 & 66.7 & 1.09 \\
    claude-fable-5 & 8/10 & 80.0 & 1.07 \\
    llama-4-scout & 8/12 & 66.7 & 1.05 \\
    llama-4-maverick & 8/10 & 80.0 & 1.03 \\
    gemini-3-flash & 8/12 & 66.7 & 1.01 \\
    nemotron-3-super & 8/12 & 66.7 & 1.01 \\
    glm-5-1 & 8/12 & 66.7 & 1.00 \\
    kimi-k2-7-code & 8/12 & 66.7 & 0.97 \\
    nova-2-lite & 7/12 & 58.3 & 1.26 \\
    minimax-m3 & 7/12 & 58.3 & 1.01 \\
    gemini-3.1-pro & 7/12 & 58.3 & 1.00 \\
    minimax-m2-7 & 7/12 & 58.3 & 0.99 \\
    grok-4-20-reasoning & 7/12 & 58.3 & 0.99 \\
    grok-4-3 & 7/12 & 58.3 & 0.98 \\
    claude-opus-4-8 & 7/10 & 70.0 & 0.98 \\
    deepseek-v4-pro & 7/12 & 58.3 & 0.95 \\
    gpt-5.5 & 6/10 & 60.0 & 1.07 \\
    claude-sonnet-5 & 6/10 & 60.0 & 1.06 \\
    gpt-5.4-nano & 6/12 & 50.0 & 1.05 \\
    claude-opus-4-7 & 6/10 & 60.0 & 1.03 \\
    gemini-2.5-pro & 6/12 & 50.0 & 1.03 \\
    qwen3-7-plus & 6/12 & 50.0 & 1.01 \\
    grok-4-20 & 6/8 & 75.0 & 1.01 \\
    qwen3-5-397b & 4/12 & 33.3 & 0.94 \\
    gpt-5.4-mini & 3/5 & 60.0 & 1.01 \\
    \bottomrule
  \end{tabular}
\end{table}

Response-style bias vs humans: \begin{table}[t]
  \centering\small
  \caption{Response style relative to the human sample at C3 / single\_call\_all (model minus human). \emph{Acquiescence} is the balanced agreement index on the two instruments with both item keys (Dunham ATC and change engagement): the mean of the forward-keyed and reverse-keyed items on the as-answered, 0--1 rescaled responses; the difference is in units of the human between-respondent SD of the index. \emph{SD} is the within-respondent SD across all items (Likert points). \emph{Extreme} and \emph{midpoint} are the shares of endpoint and scale-centre answers. Sorted by acquiescence.}
  \label{tab:response-style}
  \begin{tabular}{lrrrr}
    \toprule
    Model & Acq.\ $\Delta$ (SD) & SD $\Delta$ & Extreme $\Delta$ & Midpoint $\Delta$ \\
    \midrule
    nova-2-lite & +1.26 & -0.50 & -0.26 & +0.29 \\
    gemini-3-flash & +0.91 & -0.29 & -0.16 & +0.07 \\
    gpt-5.4 & +0.89 & -0.37 & -0.23 & +0.10 \\
    minimax-m2-7 & +0.80 & -0.39 & -0.22 & +0.14 \\
    gpt-5.4-nano & +0.69 & -0.50 & -0.26 & +0.28 \\
    gpt-5.5 & +0.69 & -0.39 & -0.24 & +0.17 \\
    grok-4-20-reasoning & +0.54 & -0.31 & -0.25 & +0.18 \\
    gemma4-31b & +0.49 & -0.39 & -0.22 & +0.18 \\
    kimi-k2-7-code & +0.48 & -0.51 & -0.22 & +0.14 \\
    gemini-3.5-flash & +0.44 & -0.45 & -0.22 & +0.12 \\
    grok-4-1-fast-non-reasoning & +0.42 & -0.25 & -0.12 & +0.09 \\
    qwen3-5-397b & +0.36 & -0.37 & -0.23 & +0.19 \\
    grok-4-1-fast-reasoning & +0.34 & -0.27 & -0.15 & +0.09 \\
    kimi-k2-6 & +0.32 & -0.44 & -0.23 & +0.17 \\
    grok-4-20 & +0.31 & -0.35 & -0.26 & +0.19 \\
    qwen3-7-plus & +0.30 & -0.38 & -0.22 & +0.20 \\
    grok-4-3 & +0.27 & -0.42 & -0.26 & +0.27 \\
    glm-5-1 & +0.26 & -0.31 & -0.21 & +0.13 \\
    llama-4-maverick & +0.24 & -0.32 & -0.25 & +0.07 \\
    gemini-2.5-pro & +0.24 & -0.16 & -0.15 & +0.06 \\
    gpt-5.4-mini & +0.21 & -0.47 & -0.25 & +0.11 \\
    deepseek-3-2 & +0.14 & -0.31 & -0.14 & +0.05 \\
    deepseek-v4-pro & +0.11 & -0.41 & -0.23 & +0.15 \\
    mistral-large-3 & +0.10 & -0.47 & -0.26 & +0.26 \\
    gemini-3.1-pro & +0.09 & -0.35 & -0.23 & +0.13 \\
    glm-5-2 & +0.07 & -0.33 & -0.21 & +0.14 \\
    gpt-oss-120b & +0.07 & -0.35 & -0.25 & +0.13 \\
    nemotron-3-super & +0.07 & -0.43 & -0.21 & +0.10 \\
    llama-4-scout & +0.04 & -0.35 & -0.24 & +0.12 \\
    gemini-3.1-flash-lite & -0.03 & -0.36 & -0.16 & +0.11 \\
    claude-fable-5 & -0.13 & -0.47 & -0.25 & +0.23 \\
    claude-opus-4-7 & -0.19 & -0.52 & -0.26 & +0.23 \\
    minimax-m3 & -0.28 & -0.52 & -0.26 & +0.29 \\
    claude-haiku-4-5 & -0.28 & -0.47 & -0.25 & +0.23 \\
    claude-sonnet-4-6 & -0.35 & -0.48 & -0.23 & +0.25 \\
    claude-opus-4-8 & -0.36 & -0.53 & -0.27 & +0.27 \\
    claude-sonnet-5 & -0.54 & -0.44 & -0.26 & +0.23 \\
    \bottomrule
  \end{tabular}
\end{table}

Downstream predictive validity: \begin{table}[t]
  \centering\small
  \caption{Downstream predictive validity at C3 / single\_call\_all. An OLS regression of the performance total on the attitudes and engagement totals is fit on each model's synthetic respondents and tested on a fixed 30\% split of the humans; we report the out-of-sample $R^2$ and its gap to the same regression fit on the other 70\% of humans ($R^2_{\text{human}}=0.277$). $\Delta R^2<0$ means the synthetic data is a worse training set than real human data; $R^2_{\text{synth}}<0$ means it predicts worse than the mean. Sorted by synthetic $R^2$.}
  \label{tab:predictive-validity}
  \begin{tabular}{lrr}
    \toprule
    Model & Synthetic $R^2$ & $\Delta R^2$ vs human \\
    \midrule
    qwen3-7-plus & +0.303 & +0.026 \\
    gemini-2.5-pro & +0.296 & +0.019 \\
    gemini-3.5-flash & +0.290 & +0.013 \\
    minimax-m2-7 & +0.284 & +0.007 \\
    gemini-3-flash & +0.268 & -0.010 \\
    deepseek-v4-pro & +0.258 & -0.019 \\
    gemma4-31b & +0.240 & -0.037 \\
    grok-4-20-reasoning & +0.233 & -0.044 \\
    gemini-3.1-pro & +0.213 & -0.064 \\
    glm-5-1 & +0.207 & -0.070 \\
    kimi-k2-6 & +0.205 & -0.073 \\
    claude-fable-5 & +0.186 & -0.091 \\
    grok-4-3 & +0.142 & -0.135 \\
    gemini-3.1-flash-lite & +0.124 & -0.153 \\
    gpt-5.4 & +0.099 & -0.178 \\
    gpt-5.5 & +0.098 & -0.179 \\
    kimi-k2-7-code & +0.097 & -0.181 \\
    claude-opus-4-8 & -0.018 & -0.295 \\
    claude-sonnet-4-6 & -0.061 & -0.338 \\
    deepseek-3-2 & -0.063 & -0.341 \\
    claude-opus-4-7 & -0.094 & -0.371 \\
    glm-5-2 & -0.100 & -0.377 \\
    nova-2-lite & -0.115 & -0.392 \\
    claude-haiku-4-5 & -0.116 & -0.393 \\
    qwen3-5-397b & -0.149 & -0.426 \\
    grok-4-1-fast-reasoning & -0.152 & -0.429 \\
    nemotron-3-super & -0.157 & -0.434 \\
    gpt-5.4-nano & -0.169 & -0.446 \\
    llama-4-maverick & -0.181 & -0.459 \\
    claude-sonnet-5 & -0.187 & -0.465 \\
    minimax-m3 & -0.236 & -0.513 \\
    grok-4-1-fast-non-reasoning & -0.322 & -0.600 \\
    mistral-large-3 & -0.409 & -0.687 \\
    llama-4-scout & -0.977 & -1.255 \\
    grok-4-20 & -1.596 & -1.873 \\
    gpt-oss-120b & -1.760 & -2.037 \\
    gpt-5.4-mini & -1.791 & -2.068 \\
    \bottomrule
  \end{tabular}
\end{table}

Mediation negative control: \begin{table}[t]
  \centering\small
  \caption{Mediation negative control at C3. Each row is a placebo $x\to m\to y$ path between composites of attitudes toward change (ATC), work engagement (UWES) and work performance (IWPQ), whose human indirect effect has a bootstrap CI that includes zero and $|ab|<0.10$. The 3 drawn paths in which a total score contains another variable of the path are left out, because they do not test mediation. ``Pooled'' is the indirect effect on all LLM respondents stacked ($n=3673$); ``Fabricated'' flags paths where that CI excludes zero and $|ab|>0.10$. The last column counts the models that fabricate the path on their own sample ($n\approx100$ each). Per model, the median number of fabricated paths is 2 of 7.}
  \label{tab:mediation-negative-control}
  \begin{tabular}{lrrrr}
    \toprule
    Placebo path ($x\to m\to y$) & Human ind. & Pooled ind. & Fabricated? & Models \\
    \midrule
    UWES-ded$\to$ATC-beh$\to$IWPQ-task & +0.040 & +0.126 & \textbf{yes} & 10/37 \\
    ATC-cog$\to$ATC-aff$\to$UWES-abs & +0.092 & +0.218 & \textbf{yes} & 20/37 \\
    ATC-aff$\to$IWPQ-cprod$\to$ATC-beh & -0.025 & -0.032 & no & 6/37 \\
    IWPQ-task$\to$ATC-beh$\to$IWPQ-cprod & -0.017 & -0.111 & \textbf{yes} & 14/37 \\
    ATC-total$\to$IWPQ-task$\to$UWES-abs & +0.015 & +0.075 & no & 17/37 \\
    ATC-aff$\to$IWPQ-task$\to$ATC-beh & +0.011 & +0.061 & no & 5/37 \\
    ATC-cog$\to$ATC-aff$\to$IWPQ-task & +0.049 & +0.228 & \textbf{yes} & 4/37 \\
    \bottomrule
  \end{tabular}
\end{table}

Test--retest reliability (ICC) per model: \begin{table}[t]
  \centering\small
  \caption{Test--retest reliability of LLM synthetic respondents. Each model answered the same stability panel $k$ independent times (fresh per-call sampling); ICC(1) is the scale-averaged one-way intra-class correlation of the composite scores (reverse-keyed items reverse-scored), treating the repeat index as an interchangeable rater. ``Top-line $\rho$'' is the mean inter-repeat Spearman correlation across all 68 items. $\Delta$ vs human is the gap to the pooled published human test--retest reference ($\overline{\text{ICC}}_{\text{human}}=0.78$); $n/k$ is the number of respondents with complete repeats and the number of repeats. Sorted by ICC(1). Models with fewer than 10 complete-repeat respondents are omitted (\texttt{gpt-oss-120b}).}
  \label{tab:test-retest}
  \begin{tabular}{lrrrr}
    \toprule
    Model & ICC(1) & Top-line $\rho$ & $\Delta$ vs human & $n/k$ \\
    \midrule
    gpt-5.4 & 0.93 & 0.93 & +0.15 & 28/20 \\
    qwen3-7-plus & 0.93 & 0.90 & +0.15 & 21/20 \\
    claude-sonnet-4-6 & 0.93 & 0.93 & +0.15 & 28/20 \\
    gemini-3.5-flash & 0.93 & 0.88 & +0.14 & 30/20 \\
    mistral-large-3 & 0.92 & 0.91 & +0.14 & 27/20 \\
    gemma4-31b & 0.92 & 0.93 & +0.13 & 27/20 \\
    gemini-3.1-flash-lite & 0.91 & 0.85 & +0.13 & 30/20 \\
    claude-opus-4-7 & 0.91 & 0.90 & +0.12 & 28/20 \\
    claude-opus-4-8 & 0.90 & 0.90 & +0.12 & 28/20 \\
    gemini-3-flash & 0.89 & 0.88 & +0.11 & 30/20 \\
    glm-5-2 & 0.88 & 0.86 & +0.10 & 30/20 \\
    kimi-k2-7-code & 0.88 & 0.88 & +0.10 & 27/20 \\
    llama-4-maverick & 0.87 & 0.96 & +0.09 & 27/20 \\
    gpt-5.5 & 0.87 & 0.90 & +0.09 & 28/20 \\
    grok-4-20 & 0.87 & 0.91 & +0.09 & 27/20 \\
    glm-5-1 & 0.86 & 0.84 & +0.07 & 30/20 \\
    minimax-m3 & 0.85 & 0.91 & +0.07 & 30/20 \\
    nova-2-lite & 0.84 & 0.83 & +0.06 & 27/20 \\
    kimi-k2-6 & 0.81 & 0.83 & +0.03 & 29/20 \\
    grok-4-20-reasoning & 0.79 & 0.87 & +0.01 & 27/20 \\
    claude-haiku-4-5 & 0.79 & 0.84 & +0.01 & 28/20 \\
    gemini-2.5-pro & 0.79 & 0.80 & +0.01 & 30/20 \\
    minimax-m2-7 & 0.76 & 0.84 & -0.02 & 27/20 \\
    grok-4-3 & 0.75 & 0.81 & -0.03 & 27/20 \\
    gemini-3.1-pro & 0.73 & 0.73 & -0.05 & 30/20 \\
    nemotron-3-super & 0.72 & 0.77 & -0.06 & 30/20 \\
    gpt-5.4-mini & 0.67 & 0.88 & -0.11 & 28/20 \\
    qwen3-5-397b & 0.60 & 0.81 & -0.18 & 27/20 \\
    gpt-5.4-nano & 0.56 & 0.62 & -0.22 & 28/20 \\
    deepseek-3-2 & 0.52 & 0.74 & -0.27 & 29/20 \\
    \bottomrule
  \end{tabular}
\end{table}

Human-vs-synthetic distinguishability (classifier AUC): \begin{table}[t]
  \centering\small
  \caption{Human-vs-synthetic distinguishability. A logistic-regression and a random-forest discriminator are trained to separate real human respondents from each generator's synthetic respondents; we report the mean held-out AUC (chance $=0.5$). ``Indistinguishable'' flags generators that a discriminator cannot beat ($\text{AUC}\le0.55$). The LLM row gives the median and range across models.}
  \label{tab:distinguishability}
  \begin{tabular}{lrr}
    \toprule
    Generator & Mean AUC & Indistinguishable? \\
    \midrule
    Gaussian copula & 0.39 & \textbf{yes} \\
    Multivariate normal & 0.52 & \textbf{yes} \\
    Marginal & 0.68 & no \\
    Stratum mean & 0.69 & no \\
    $k$-NN & 0.89 & no \\
    Midpoint & 0.99 & no \\
    \midrule
    \emph{LLMs} (37 models, C3) & 0.992 [0.966, 1.000] & 0 of 37 \\
    \bottomrule
  \end{tabular}
\end{table}

PSS vs.\ synthetic sample size (scaling curve): \begin{table}[t]
  \centering\small
  \caption{PSS$_3$ (the three sample-driven components; mediation and demographic held at 0.5) as a function of the number of synthetic respondents $n$ (bootstrap means, 60 draws). LLM rows use the C3 cell, which has about 100 respondents per model, so the LLM curve stops at $n{=}100$. The best LLM is the top model of the leaderboard.}
  \label{tab:scaling-curve}
  \begin{tabular}{lrrrrr}
    \toprule
    Generator & $n{=}10$ & $n{=}25$ & $n{=}50$ & $n{=}100$ & $n{=}200$ \\
    \midrule
    Gaussian copula & 0.694 & 0.751 & 0.785 & 0.806 & 0.819 \\
    Multivariate normal & 0.684 & 0.744 & 0.778 & 0.804 & 0.817 \\
    Best LLM (minimax-m3) & 0.617 & 0.665 & 0.686 & 0.695 & 0.702 \\
    \midrule
    Gap (copula $-$ best LLM) & +0.077 & +0.086 & +0.100 & +0.111 & +0.117 \\
    \bottomrule
  \end{tabular}
\end{table}

Item-level fidelity by item feature (OLS): \begin{table}[t]
  \centering\small
  \caption{Which items are hard to reproduce? OLS of the per-item mean Jensen--Shannon divergence between the LLM and human response distributions (averaged over the LLM cells at C3) on item features, $n{=}68$ items ($R^2=0.61$). Higher divergence means worse reproduction, so a positive coefficient marks items that are reproduced worse. $^{*}p<.05$, $^{**}p<.01$, $^{***}p<.001$.}
  \label{tab:item-features}
  \begin{tabular}{lrrl}
    \toprule
    Predictor & Coef. & SE & Sig. \\
    \midrule
    Intercept & +0.281 & 0.030 & *** \\
    Instrument: Dunham ATC & -0.008 & 0.019 &  \\
    Instrument: IWPQ & +0.071 & 0.020 & *** \\
    Instrument: UWES-17 & +0.155 & 0.021 & *** \\
    Item length (chars) & -0.000 & 0.000 &  \\
    Reverse-keyed & +0.049 & 0.019 & * \\
    \bottomrule
  \end{tabular}
\end{table}

Bifactor structure (ECV / $\omega_h$) vs human: \begin{table}[t]
  \centering\small
  \caption{Bifactor structure, LLM mean at C3 (37 models) vs human. ECV is the share of common variance explained by the general factor; $\omega_h$ is the general-factor (hierarchical) omega. $\Delta$ is LLM minus human.}
  \label{tab:bifactor}
  \begin{tabular}{lrrrr}
    \toprule
    Instrument & Human ECV & LLM ECV & $\Delta$ECV & $\Delta\omega_h$ \\
    \midrule
    Dunham ATC & 0.67 & 0.76 & +0.08 & -0.00 \\
    Change engagement & 0.62 & 0.71 & +0.09 & -0.02 \\
    UWES-17 & 0.70 & 0.80 & +0.10 & -0.00 \\
    IWPQ & 0.45 & 0.58 & +0.12 & +0.09 \\
    \bottomrule
  \end{tabular}
\end{table}

Steerability (debias instruction / few-shot vs zero-shot): \begin{table}[t]
  \centering\small
  \caption{Steerability. Mean over the 6-model subset (\texttt{claude-sonnet-4-6}, \texttt{gemini-3.1-pro}, \texttt{gpt-5.5}, \texttt{glm-5-1}, \texttt{llama-4-maverick}, \texttt{qwen3-5-397b}) at $n{=}100$ matched respondents. C13 adds an explicit instruction to answer like a real, varied respondent; C14 adds three human exemplars. PSS$_\rho$ and PSS$_{\mathrm{dist}}$ are the correlation and distribution components. SD is the within-respondent SD; Acq.\ is the balanced acquiescence index relative to humans, in human SD units; ERS/MRS are the extreme and midpoint response rates.}
  \label{tab:steerability}
  \begin{tabular}{lccccccc}
    \toprule
    Condition & PSS & PSS$_\rho$ & PSS$_{\mathrm{dist}}$ & SD & Acq. & ERS & MRS \\
    \midrule
    \textit{Human reference} & n/a & n/a & n/a & 1.19 & 0.00 & 0.27 & 0.21 \\
    \midrule
    Zero-shot (C3) & 0.664 & 0.682 & 0.764 & 0.82 & +0.21 & 0.04 & 0.37 \\
    \;+ Debias instruction & 0.682 & 0.663 & 0.780 & 0.99 & +0.33 & 0.10 & 0.32 \\
    \;+ Few-shot ($k{=}3$ humans) & 0.658 & 0.664 & 0.770 & 0.81 & +0.18 & 0.04 & 0.38 \\
    \bottomrule
  \end{tabular}
\end{table}

Demographic-conditional copula (profile conditioning vs unconditional): \begin{table}[t]
  \centering\small
  \caption{Conditioning a statistical baseline on the LLM's profile. The unconditional Gaussian copula ignores the persona; the \emph{conditional} copula ridge-regresses each item on the same profile fields the LLM receives and keeps the residual correlation structure, in-sample and under 5-fold cross-validation. All rows are scored with the same PSS. ``PSS w/o $d$'' sets the distribution component, which the bounded-Likert copulas reproduce by construction, to zero and keeps the other weights unchanged; it is computed the same way for every row.}
  \label{tab:cond-copula}
  \begin{tabular}{lrrrrrrr}
    \toprule
    Generator & PSS & PSS w/o $d$ & $d$ & $c$ & $r$ & $m$ & $g$ \\
    \midrule
    copula (unconditional) & 0.936 & 0.688 & 0.994 & 0.954 & 0.986 & 1.000 & 0.522 \\
    \quad cond-copula (in-sample) & 0.968 & 0.719 & 0.996 & 0.950 & 0.996 & 0.978 & 0.871 \\
    \quad cond-copula (5-fold OOS) & 0.929 & 0.680 & 0.995 & 0.938 & 0.976 & 0.864 & 0.773 \\
    MVN baseline & 0.949 & 0.702 & 0.990 & 0.950 & 0.994 & 0.995 & 0.664 \\
    best LLM (minimax-m3) & 0.814 & 0.636 & 0.713 & 0.764 & 0.966 & 0.998 & 0.520 \\
    \bottomrule
  \end{tabular}
\end{table}

\clearpage

The full 37-model PSS leaderboard with all components is in Table~\ref{tab:pss-leaderboard}: \begin{table}[t]
  \centering\small
  \caption{Anchored PSS leaderboard at the C3 (full profile) condition under single\_call\_all presentation, sorted within each block. $d$ = distributional, $c$ = correlation, $r$ = reliability, $m$ = mediation, $g$ = demographic. Human references, statistical baselines, then LLMs. All rows are scored with the same five components against all 263 humans. LLM rows, the same-respondent reference, and the copula fit on the other 163 humans have $n{=}100$; the other baselines are generated for all 263 respondents.}
  \label{tab:pss-leaderboard}
  \begin{tabular}{lrrrrrr}
    \toprule
    Model / baseline & PSS & $d$ & $c$ & $r$ & $m$ & $g$ \\
    \midrule
    human, same 100 respondents & 0.886 & 0.996 & 0.923 & 0.990 & 0.707 & 0.672 \\
    human, held-out split & 0.825 & 0.981 & 0.762 & 0.986 & 0.720 & 0.480 \\
    baseline-mvn & 0.949 & 0.990 & 0.950 & 0.994 & 0.995 & 0.664 \\
    baseline-copula & 0.936 & 0.994 & 0.954 & 0.986 & 1.000 & 0.522 \\
    baseline-copula, fit on other 163 & 0.824 & 0.985 & 0.805 & 0.970 & 0.722 & 0.382 \\
    baseline-stratum & 0.589 & 0.989 & 0.169 & 0.336 & 0.805 & 0.714 \\
    baseline-knn & 0.485 & 0.907 & 0.070 & 0.323 & 0.714 & 0.334 \\
    baseline-marginal & 0.444 & 0.997 & 0.011 & 0.097 & 0.667 & 0.395 \\
    baseline-midpoint & 0.099 & 0.395 & 0.000 & n/a & n/a & 0.000 \\
    minimax-m3 & 0.814 & 0.713 & 0.764 & 0.966 & 0.998 & 0.520 \\
    minimax-m2-7 & 0.809 & 0.833 & 0.807 & 0.976 & 0.761 & 0.515 \\
    kimi-k2-7-code & 0.802 & 0.801 & 0.805 & 0.948 & 0.796 & 0.521 \\
    claude-opus-4-7 & 0.789 & 0.722 & 0.765 & 0.960 & 0.957 & 0.336 \\
    gemini-3.1-flash-lite & 0.779 & 0.861 & 0.725 & 0.947 & 0.753 & 0.421 \\
    qwen3-5-397b & 0.777 & 0.760 & 0.653 & 0.971 & 0.825 & 0.652 \\
    gemini-2.5-pro & 0.758 & 0.870 & 0.757 & 0.944 & 0.570 & 0.483 \\
    deepseek-v4-pro & 0.750 & 0.798 & 0.804 & 0.954 & 0.488 & 0.614 \\
    kimi-k2-6 & 0.747 & 0.763 & 0.701 & 0.953 & 0.714 & 0.477 \\
    gpt-5.4-mini & 0.743 & 0.659 & 0.602 & 0.962 & 0.929 & 0.489 \\
    claude-opus-4-8 & 0.740 & 0.700 & 0.742 & 0.962 & 0.792 & 0.288 \\
    claude-sonnet-5 & 0.739 & 0.703 & 0.704 & 0.962 & 0.770 & 0.412 \\
    gpt-oss-120b & 0.737 & 0.734 & 0.797 & 0.943 & 0.600 & 0.460 \\
    gemini-3.5-flash & 0.735 & 0.824 & 0.772 & 0.933 & 0.541 & 0.413 \\
    claude-haiku-4-5 & 0.720 & 0.750 & 0.726 & 0.953 & 0.589 & 0.428 \\
    qwen3-7-plus & 0.716 & 0.777 & 0.739 & 0.963 & 0.498 & 0.451 \\
    glm-5-1 & 0.712 & 0.819 & 0.777 & 0.949 & 0.388 & 0.461 \\
    grok-4-3 & 0.707 & 0.721 & 0.755 & 0.965 & 0.475 & 0.495 \\
    nemotron-3-super & 0.705 & 0.831 & 0.766 & 0.957 & 0.316 & 0.508 \\
    gemini-3-flash & 0.699 & 0.848 & 0.791 & 0.935 & 0.297 & 0.426 \\
    grok-4-20-reasoning & 0.683 & 0.753 & 0.737 & 0.963 & 0.405 & 0.363 \\
    claude-fable-5 & 0.682 & 0.746 & 0.716 & 0.960 & 0.456 & 0.329 \\
    grok-4-20 & 0.679 & 0.706 & 0.655 & 0.968 & 0.498 & 0.458 \\
    deepseek-3-2 & 0.655 & 0.870 & 0.750 & 0.936 & 0.121 & 0.391 \\
    gemini-3.1-pro & 0.654 & 0.830 & 0.784 & 0.953 & 0.099 & 0.395 \\
    gemma4-31b & 0.649 & 0.788 & 0.743 & 0.959 & 0.165 & 0.419 \\
    gpt-5.4-nano & 0.644 & 0.661 & 0.546 & 0.924 & 0.556 & 0.463 \\
    grok-4-1-fast-reasoning & 0.644 & 0.903 & 0.728 & 0.931 & 0.000 & 0.496 \\
    llama-4-maverick & 0.643 & 0.704 & 0.567 & 0.972 & 0.426 & 0.453 \\
    llama-4-scout & 0.640 & 0.731 & 0.629 & 0.976 & 0.483 & 0.083 \\
    grok-4-1-fast-non-reasoning & 0.639 & 0.901 & 0.732 & 0.929 & 0.000 & 0.448 \\
    glm-5-2 & 0.637 & 0.824 & 0.789 & 0.943 & 0.000 & 0.457 \\
    mistral-large-3 & 0.626 & 0.618 & 0.697 & 0.958 & 0.337 & 0.385 \\
    gpt-5.5 & 0.609 & 0.723 & 0.644 & 0.970 & 0.219 & 0.299 \\
    gpt-5.4 & 0.606 & 0.790 & 0.709 & 0.955 & 0.000 & 0.400 \\
    nova-2-lite & 0.605 & 0.713 & 0.541 & 0.960 & 0.275 & 0.448 \\
    claude-sonnet-4-6 & 0.586 & 0.749 & 0.666 & 0.959 & 0.000 & 0.408 \\
    ensemble-mean & 0.560 & 0.629 & 0.752 & 0.929 & 0.000 & 0.293 \\
    \bottomrule
  \end{tabular}
\end{table}

\begin{figure}[h]
  \centering
  \includegraphics[width=0.85\linewidth]{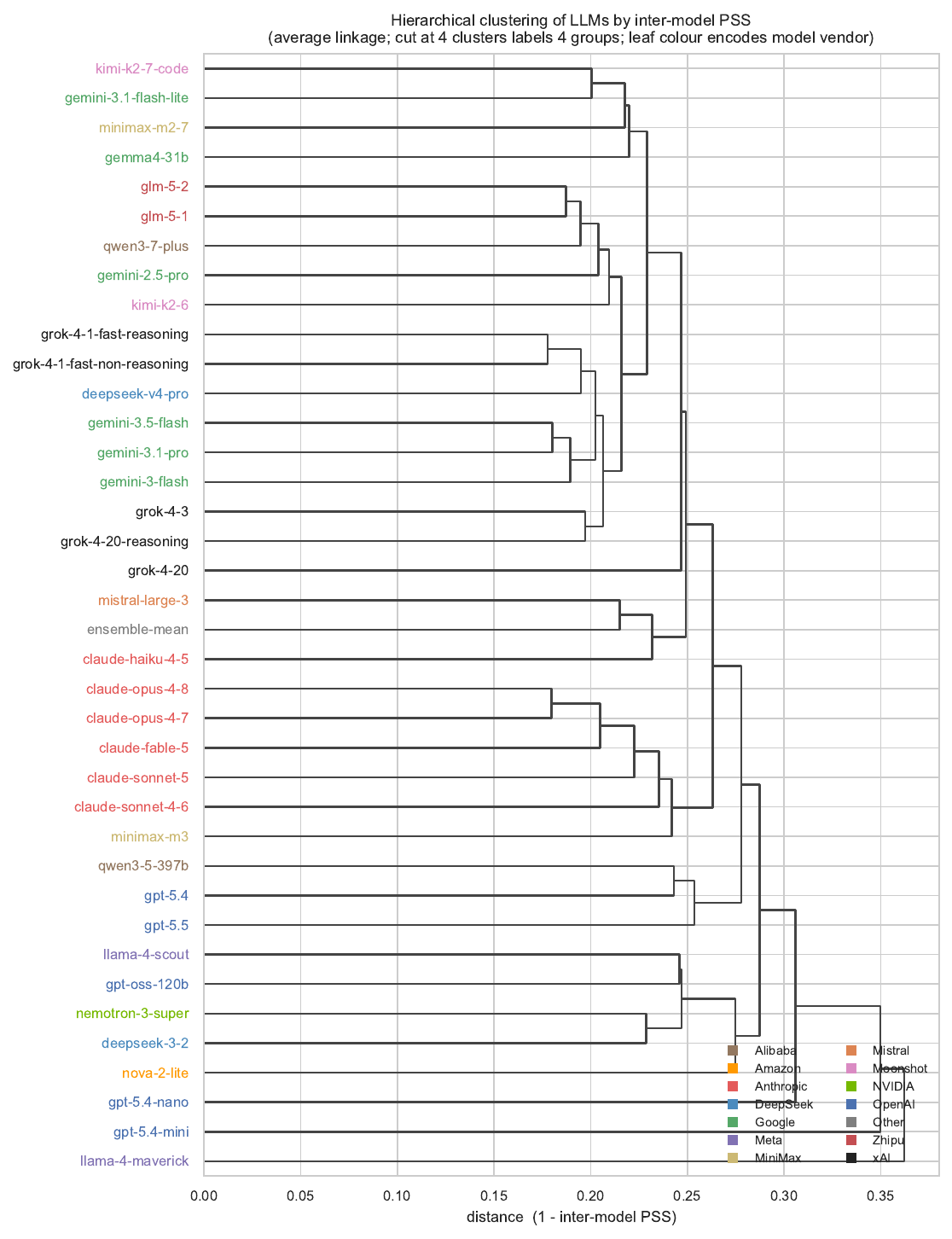}
  \caption{Hierarchical clustering of the 37 LLMs by inter-model PSS (distance $1{-}\text{PSS}$, leaf colour = model vendor). Most models form one large cluster that mixes vendors; \texttt{llama-4-maverick}, \texttt{gpt-5.4-mini}, \texttt{nova-2-lite}, \texttt{gpt-5.4-nano} and \texttt{mistral-large-3} lie outside it.}
  \label{fig:dendrogram}
\end{figure}

\begin{figure}[h]
  \centering
  \includegraphics[width=0.85\linewidth]{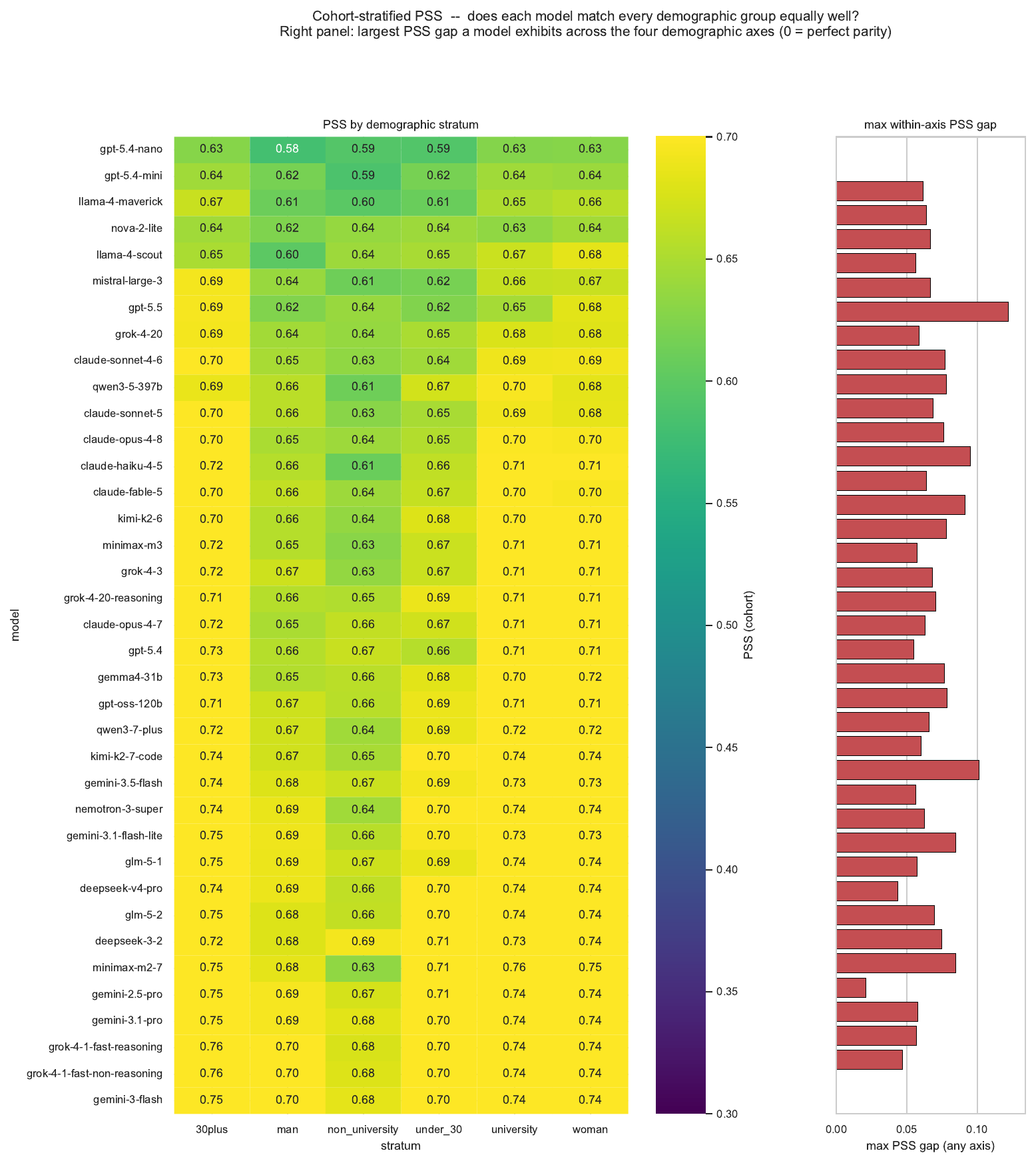}
  \caption{Cohort-stratified PSS: model $\times$ stratum heatmap (left) and worst-axis disparity per model (right).}
  \label{fig:cohort-app}
\end{figure}

\begin{figure}[h]
  \centering
  \includegraphics[width=0.85\linewidth]{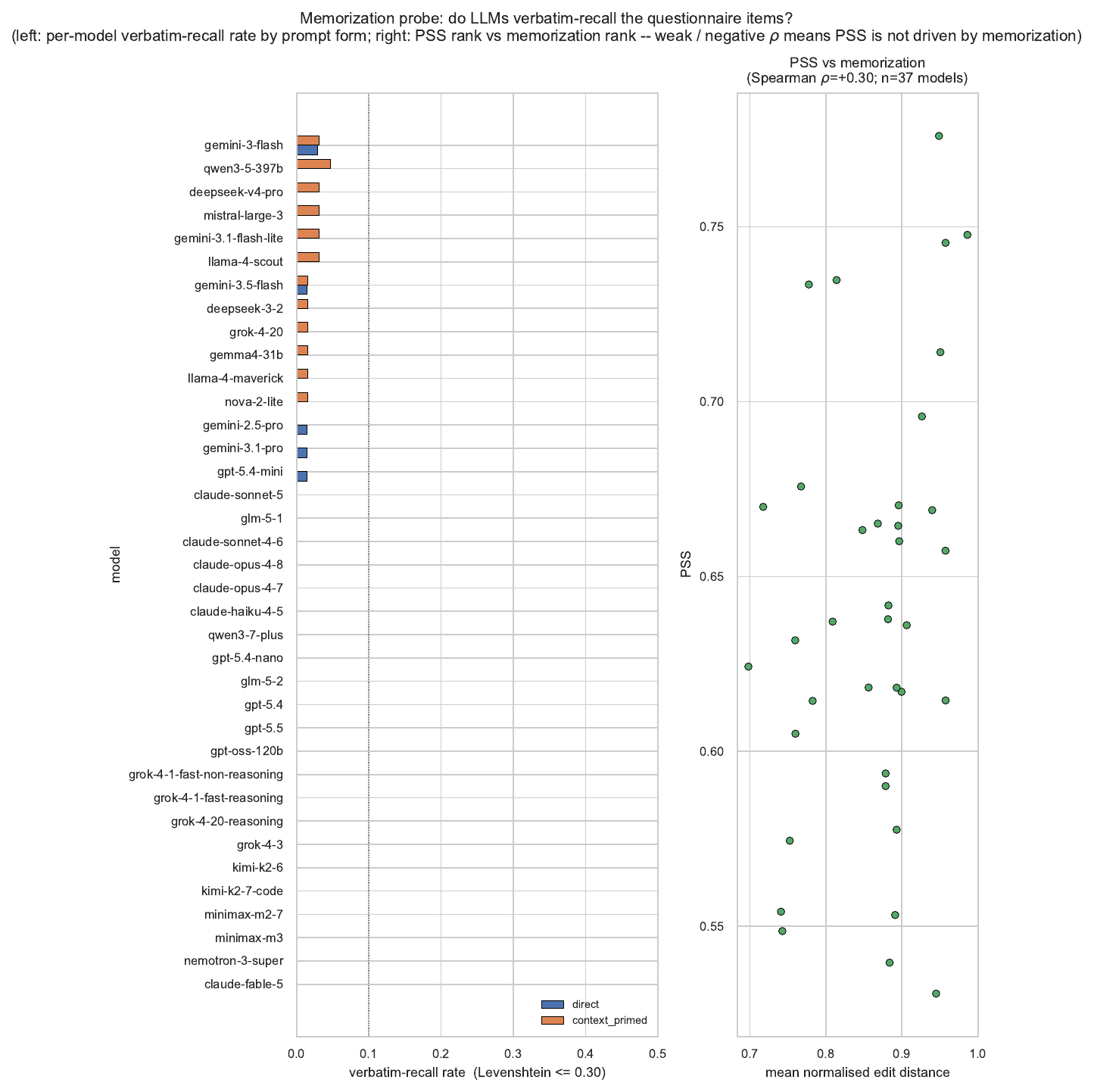}
  \caption{Memorization probe: per-model verbatim-recall rate (left) and PSS vs.\ recall scatter (right). Worst-case high-recall rate is $4.7\%$; $22/37$ models are exactly zero. Spearman rank correlation between recall rate and PSS is ${-}0.04$.}
  \label{fig:memorization-app}
\end{figure}

\section{Robustness ablations: numerical detail}
\label{app:robustness}
\paragraph{Persona-faithfulness recall by condition.} \texttt{persona\_recall.csv}: $0.32$ (C0) $\to$ $0.85$ (C2) on \texttt{claude-haiku-4-5}; recall peaks where the work context is most salient, with mean recall accuracy at C3 across the 37 models of $0.67$.
\paragraph{Reverse-keyed self-consistency.} On the raw answers in the test--retest panel, the mean forward-vs-reverse item correlation on the Dunham and change-engagement items is negative for all $31$ models (mean $r{=}{-}0.52$; the weakest is \texttt{gpt-5.4-mini} at ${-}0.08$), as expected from a faithful respondent. All other analyses reverse-score these items before comparison with humans.
\paragraph{Format-failure / refusal rate.} JSON-parse and refusal failures are negligible across the reachable lineup: the strict parser with structured retries recovers essentially every response, so no model's ranking is driven by parse failures. The only substantial per-model data loss is API-availability errors (HTTP 404/429) on the gateway deployments retired or rate-limited between waves (Section~\ref{sec:method}): \texttt{qwen3-7-plus}'s headline cells were backfilled once capacity recovered but retain a reduced effective $n$ ($72$--$94$ of $100$ respondents per condition), and four models (\texttt{gpt-oss-120b}, the \texttt{grok-4-1-fast} pair, \texttt{llama-4-scout}) lack the C9 narrative condition, which reduces effective $n$ and widens the affected bootstrap CIs. The lowest-ranked models (\texttt{claude-sonnet-4-6}, \texttt{nova-2-lite}, \texttt{gpt-5.4}) sit low because of an inflated mediation pathway or a weak correlation structure, not because of formatting failures.

% =============================================================================
\section{Psychometrics primer for machine-learning readers}
\label{app:psychometric-background}
% =============================================================================
This appendix provides a self-contained overview of the psychometric concepts used in the main paper. Readers with a background in measurement theory can skip it.

\paragraph{What a psychometric instrument is.} A psychometric instrument is a set of items (questions) designed to measure a latent construct (e.g., work engagement). Each item is answered on a fixed scale (typically a 5- or 7-point Likert: ``strongly disagree'' \dots\ ``strongly agree''). Items are grouped into \emph{subscales} that target distinguishable sub-constructs (e.g., the UWES-17 instrument has 3 subscales: vigour, dedication, absorption). A respondent's composite \emph{score} on a subscale is the mean of their item responses on that subscale; the full instrument score is sometimes a sum or mean across all items but is more commonly retained as the per-subscale vector.

\paragraph{Why psychometric structure matters more than item means.} A synthetic respondent that produces correct item means is not automatically a useful respondent for downstream analysis. The downstream uses of an instrument almost always involve quantities derived from the \emph{joint} distribution of items: subscale composites depend on per-subscale item correlations; reliability ($\alpha$, $\omega$) depends on within-subscale item covariances; mediation depends on the cross-subscale path coefficients; demographic-contrast Cohen's $d$ values depend on the within-stratum SD of composites. Each of these breaks if the LLM produces correct marginals but incorrect dependencies. This is why the PSS framework decomposes into six dimensions rather than reporting an aggregate.

\paragraph{Reliability: $\alpha$ and $\omega$.} Cronbach's $\alpha$ \citep{cronbach1951} is the classical internal-consistency reliability statistic:
\begin{equation}
\alpha = \frac{k}{k-1}\left(1 - \frac{\sum_{i=1}^k \sigma_i^2}{\sigma_T^2}\right),
\end{equation}
where $k$ is the number of items in the subscale, $\sigma_i^2$ is the variance of item $i$, and $\sigma_T^2$ is the variance of the sum of items. $\alpha$ has known biases (it assumes tau-equivalence of items and a single underlying factor) and is increasingly supplemented by McDonald's $\omega$ \citep{mcdonald1999}, which uses a factor model directly: $\omega = (\sum_i \lambda_i)^2 / [(\sum_i \lambda_i)^2 + \sum_i \theta_i^2]$, where $\lambda_i$ is item $i$'s loading on the latent factor and $\theta_i^2$ is its uniqueness. We report both, but note that the LLM-vs-human gap is essentially identical across the two statistics.

\paragraph{Factor congruence and Tucker's $\varphi$.} Two factor solutions can be compared by Tucker's congruence coefficient \citep{tucker1951method}: for two factor loading vectors $\mathbf{a}$ and $\mathbf{b}$, $\varphi(\mathbf{a},\mathbf{b}) = \mathbf{a}^\top \mathbf{b} / (\|\mathbf{a}\|\, \|\mathbf{b}\|)$, the cosine of the angle between them. With multiple factors, $\varphi$ is computed per factor pair and averaged after a greedy alignment that maximises the trace of the pairwise $\varphi$ matrix. Conventional thresholds in the human-respondent literature are $\varphi \in [0.85, 0.94]$ for ``fair'' factor match and $\varphi \geq 0.95$ for ``identical'' factor structure \citep{lorenzo2006tucker}. We show in the main paper that on LLM samples these thresholds are biased upward: random factor solutions can produce $\varphi > 0.85$ by chance on instruments with high-dimensional factor structures (17 items, 3 factors). The remedy is to replace the threshold with a permutation null distribution.

\paragraph{Mediation analysis.} The X$\rightarrow$M$\rightarrow$Y mediation model decomposes the total effect of X on Y into a direct path ($c'$) and an indirect path ($ab$) that runs through the mediator M. The classical Baron--Kenny three-equation derivation \citep{hayes2017introduction} yields point estimates of the standardised path coefficients; the indirect effect $ab$ is not normally distributed and the standard practice is to bootstrap its 95\% CI (percentile method, $n_{\text{boot}}=5000$). The X$\rightarrow$M$\rightarrow$Y interpretation requires temporal or theoretical ordering of the three variables; in our setting the published study's theoretical ordering is Attitudes$\rightarrow$Engagement$\rightarrow$Performance.

\paragraph{Discriminant validity and HTMT.} An instrument has good discriminant validity if its subscales measure distinguishable constructs. The HTMT ratio \citep{henseler2015new} compares the average between-subscale correlation to the average within-subscale correlation; values $<0.85$ are conventionally taken as adequate discriminant validity. The two subscales of an instrument that score $\text{HTMT}>0.85$ are not statistically distinguishable in the sample; on our LLM samples this happens to a mean of $67\%$ of the subscale pairs across the 37 models, against $4/12$ in humans.

\paragraph{Measurement invariance.} Measurement invariance is the property that an instrument measures the same construct across two samples (e.g., two languages, two demographic strata). It is tested in three steps: configural (same factor structure), metric (same factor loadings), and scalar (same item intercepts). Failure of metric invariance means the scale's items are weighted differently across groups; failure of scalar invariance means composite scores are not directly comparable. The $\Delta$CFI criterion of \citet{cheung2002evaluating} treats $|\Delta\text{CFI}| > 0.01$ between nested models as evidence against the more constrained model.

\paragraph{Bifactor decomposition.} A bifactor model represents item responses as the sum of a general factor (loading on every item) and group-specific factors (one per subscale). Two summary statistics from a bifactor fit are commonly reported: ECV (explained common variance, the fraction of common variance attributable to the general factor) and $\omega_h$ (the reliability of the general-factor score alone). High ECV ($>0.7$) is sometimes taken as evidence that the instrument is essentially unidimensional even when it has multiple subscales; low ECV indicates that the subscales contribute substantial unique variance.

% =============================================================================
\section{Compute and cost breakdown}
\label{app:compute-cost}
% =============================================================================
We report API costs and wall-clock time for the full set of experiments behind this paper.

\paragraph{Headline grid.} 37 models $\times$ 5 conditions (C0..C4) $\times$ 100 respondents $\times$ 1 repeat $=$ 18,500 calls. At an average of 4,500 prompt tokens and 1,200 completion tokens per call, the total token volume is $\sim 105$ million tokens. Per-provider costs (May 2026 list pricing):
\begin{itemize}[leftmargin=1.4em,itemsep=0pt]
  \item OpenAI (GPT-5.4/5.5 family, 4 models): $\sim\$30$.
  \item Anthropic (Claude family, 6 models): $\sim\$52$.
  \item Google (Gemini family, 5 models): $\sim\$14$ (mix of API-key and Vertex AI service-account billing).
  \item Open-weight via Nexos (22 models): $\sim\$40$.
\end{itemize}
Total headline grid: approximately \$136, completed in $\sim 6$ wall-clock hours with provider-specific concurrency limits.

\paragraph{Counterfactual swaps.} 3 swap axes $\times$ 37 models $\times$ 100 respondents $\times$ 1 repeat $=$ 11,100 additional calls, $\sim\$65$ additional cost, $\sim 4$ wall-clock hours.

\paragraph{Reasoning-effort ablation.} 6 models $\times$ 100 respondents $\times$ 2 reasoning levels (minimal, high) $=$ 1,200 additional calls; ``high'' reasoning costs roughly $5\times$ minimal on the OpenAI and Anthropic frontier models. Total $\sim\$18$, $\sim 2$ wall-clock hours.

\paragraph{Cross-language ablation.} 37 models $\times$ 30 respondents $\times$ 2 languages $=$ 2,220 calls, $\sim\$15$, $\sim 1$ wall-clock hour.

\paragraph{Presentation-mode ablation.} 5 models $\times$ 30 respondents $\times$ 3 modes (single-call, per-instrument, per-subscale). Per-subscale mode multiplies calls by $12\times$ so the total is $5 \times 30 \times (1 + 3 + 12) = 2,400$ calls, $\sim\$15$, $\sim 1$ wall-clock hour.

\paragraph{Memorisation probe.} 37 models $\times$ 65 instrument items $\times$ 2 prompt forms (direct, context-primed) $=$ 4,810 calls, $\sim\$8$, $\sim 4$ wall-clock hours (slowed by per-call thinking-trace timeouts on reasoning models).

\paragraph{Test--retest stability run.} 30 respondents $\times$ 20 repeats per model on the C3 condition ($\sim$17,500 completed cells across the lineup; $\sim$20,400 calls including retries), $\sim\$150$, $\sim 6$--$8$ wall-clock hours at per-provider concurrency 6--8. This is the second-largest run behind the headline grid and is what licenses the $R{=}1$ headline design (Section~\ref{sec:results-memorization}).

\paragraph{Total.} Approximately \$390 of API spend and $\sim 25$ wall-clock hours for the full set of experiments, dominated by the headline 37-model grid and the test--retest stability run. The per-call disk cache means that re-runs after the first execution cost nothing for already-completed cells, which is what makes incremental extension of the model lineup cheap.

\paragraph{Analysis-side compute.} All analysis runs on a single CPU node, and the full analysis script takes about one hour. The heaviest single operation is the per-cell mediation bootstrap (2,000 draws per model and condition, plus a subscale decomposition), which takes about 30 minutes. The Tucker-permutation null ($K=500$ relabelings $\times$ 37 models $\times$ 4 instruments) takes seconds, because the loading matrices are estimated once and only the item labels are permuted. The respondent-bootstrap CI ($n_{\text{boot}}=1000$ per model on the distribution, correlation and reliability components) takes a few minutes. Figure generation is $<5$ minutes for the full set.

% =============================================================================
\section{Worked qualitative example: one ATC item in the test--retest panel}
\label{app:case-studies}
% =============================================================================
This appendix follows a single item through the test--retest panel to show what the aggregate failure modes look like for individual respondents. The item is DUN\_07, ``A\v{s} nem\.{e}gstu poky\v{c}i\k{u}'' (``I don't like changes''), a reverse-keyed affective item of the ATC scale with a 1--5 response scale. We report raw answers, so a higher value means a stronger dislike of change. The data are the C3 test--retest run: the same 30 human respondents, 20 independent draws per respondent and model, and 31 models (Table~\ref{tab:test-retest}). The 30 humans answered 1 (8 respondents), 2 (13), 3 (8), and 4 (1), with mean $2.07$ and SD $0.83$.

\paragraph{Pattern across the lineup.} Three facts hold for almost every model. First, the 20 draws for one persona are nearly identical. The mean within-persona SD is $0.21$ Likert points (range $0.03$--$0.51$), so the spread between synthetic respondents comes almost entirely from the persona fields. Second, this spread is smaller than in humans. The SD of the per-persona means is below the human SD of $0.83$ for $29$ of the $31$ models (mean $0.55$). Third, the spread is attached to the wrong people. The per-persona mean correlates with the answer of the same human respondent at a mean $r{=}0.05$ (maximum $0.26$). It correlates much more strongly with the persona's stated personal relevance of change: mean $r{=}{-}0.72$, and $29$ of $31$ models are below ${-}0.30$. In the full human sample ($n{=}263$) the same correlation is only ${-}0.14$. The models therefore turn one context field into the answer: a respondent who reports that the changes are not relevant to them is assumed to dislike change.

\paragraph{Three respondents.} Three people from the panel make this concrete.
Respondent A (\texttt{H0030}) is a 46-year-old woman with a university degree, a manager in services with more than 10 years in the organisation, change intensity 7/10 and personal relevance 9/10. She answered 1.
Respondent B (\texttt{H0042}) is a 48-year-old woman with secondary education, a non-manager in manufacturing with 3--5 years in the organisation, change intensity 1/10 and personal relevance 2/10. She answered 3.
Respondent C (\texttt{H0124}) is a 29-year-old woman with a university degree, a non-manager in services with 1--3 years in the organisation, change intensity 7/10 and personal relevance 9/10. She answered 4, the strongest dislike of change in the panel.

\paragraph{\texttt{gpt-5.4-mini}.} The model answers 2 in $552$ of its $560$ draws on this item, and all three respondents receive 20 draws of 2. The model matches the human mean ($1.99$ against $2.07$) but removes the variation between people: the SD of its per-persona means is $0.07$. A correct item mean hides this kind of range restriction.

\paragraph{\texttt{gemini-3-flash}.} This model has the largest spread in the lineup (SD of per-persona means $0.95$) and uses all five scale points. Its spread follows the relevance field ($r{=}{-}0.94$) and not the respondents ($r{=}{-}0.10$). Respondent B, who reported low relevance, receives 12 draws of 5 and 8 draws of 4 against her answer of 3. Respondent C receives 19 draws of 2 and one draw of 1 against her answer of 4.

\paragraph{\texttt{grok-4-3}.} This model has the highest agreement with the human answers on this item ($r{=}0.26$), but it never uses the scale point 1, which 8 of the 30 humans chose. It is the only one of the four models shown here that comes close to Respondent C (11 draws of 4, 4 of 3, and 5 of 2). Respondent A receives 20 draws of 2 and Respondent B receives 16 draws of 4 and 4 of 3.

\paragraph{\texttt{claude-sonnet-4-6}.} This model is almost deterministic (within-persona SD $0.03$). Respondents A and C both receive 20 draws of 2. Respondent B receives ten draws of 3 and ten of 4.

The example matches the aggregate results. The models find a plausible mean, compress the spread between people, and attach the remaining spread to the most suggestive persona field. For this reason a correct item mean is weak evidence of a useful synthetic respondent. The distribution, correlation, and demographic-effect components of PSS are designed to detect these cases.

% =============================================================================
\section{Negative results}
\label{app:negative-results}
% =============================================================================
We report two experiments that did not produce the result we expected, but that are informative about the limits of the framework.

\paragraph{Negative result 1: temperature.} Decoding temperature has no effect on PSS detectable above sampling noise. We swept $T \in \{0.0, 0.3, 0.5, 0.7, 1.0, 1.5\}$ on five models at $n{=}30$ respondents per cell, which doubles as a natural control. Two of the five hold temperature fixed at the model default: \texttt{claude-opus-4-7}, whose API deprecates the parameter, and \texttt{gpt-5.4}, which our generation client pins to the default. Their six ``temperature'' cells are therefore six independent draws at one temperature, and their PSS spread ($0.12$ and $0.08$) measures the $n{=}30$ run-to-run noise floor directly. Two of the three models that do vary temperature stay within that floor (\texttt{gemini-3.1-pro} $0.03$, \texttt{llama-4-maverick} $0.10$). \texttt{qwen3-5-397b} exceeds it slightly ($0.15$), but its profile is non-monotonic, with the lowest PSS at $T{=}0.7$ and the highest at $T{=}0$. We conclude that temperature is negligible relative to model identity, persona disclosure, and presentation mode, and note that several frontier models no longer expose temperature as a tunable parameter at all. We request $T{=}0.7$ throughout; models that deprecate or pin the parameter run at their provider default.

\paragraph{Negative result 2: prompt-based debiasing and few-shot conditioning.} We next test whether the documented failure modes (range restriction, over-coherence, acquiescence) are merely a prompting artifact that disappears once the model is told to behave, or shown what humans actually do. We tested both fixes on a 6-model subset at $n{=}100$ matched respondents (Appendix Table~\ref{tab:steerability}): (i) an explicit \emph{debias instruction} (``use the full scale including the extremes; answer with the natural variability of a real person; do not be over-agreeable or consistent''), and (ii) \emph{few-shot conditioning} with $k{=}3$ real human answer vectors (disjoint from the targets) injected in-context. Neither closes the gap. The debias instruction partially repairs surface response style. It widens the response SD from $0.82$ toward the human value $1.19$ (to $0.99$), lowers midpoint overuse ($0.37\!\to\!0.32$) and raises endpoint use ($0.04\!\to\!0.10$), but it slightly increases acquiescence ($+0.21\!\to\!+0.33$ human SD). Overall PSS barely moves ($0.664\!\to\!0.682$), because the correlation component is the binding constraint and does not improve ($\mathrm{PSS}_\rho\ 0.682\!\to\!0.663$). Few-shot exemplars are almost entirely ignored: response style and PSS are indistinguishable from zero-shot ($\mathrm{PSS}\ 0.664\!\to\!0.658$; SD $0.81$). These results suggest the gap is structural rather than a prompt-conditioning artifact: the models do not absorb the human answer distribution even when it is shown explicitly in context.

% =============================================================================
\section{PSS weight sensitivity analysis}
\label{app:sensitivity}
% =============================================================================
The default PSS weights $(w_d, w_c, w_r, w_m, w_g) = (0.25, 0.25, 0.20, 0.20, 0.10)$ assign 50\% to the two sample-driven components (distribution + correlation) on which the LLMs underperform the copula baseline. A natural sensitivity check is to vary the weights. We re-rank the leaderboard under three alternative weightings:

\begin{itemize}[leftmargin=1.4em,itemsep=0pt]
  \item \emph{Equal weights} $(0.20, 0.20, 0.20, 0.20, 0.20)$. The Spearman rank correlation between the equal-weight ranking and the default ranking is $0.99$ across the 37 LLMs, and four of the top-5 models are shared. The in-sample copula drops from $0.936$ to $0.891$ and still leads every LLM (top LLM \texttt{minimax-m3}, $0.792$). The copula fit on other respondents scores $0.773$, now slightly below the top LLM.
  \item \emph{Correlation + reliability only} $(0, 0.50, 0.50, 0, 0)$. This restricts the leaderboard to two sample-driven components. The in-sample copula reaches $0.970$ (MVN $0.972$) and dominates every LLM by ${\geq}0.08$; the top LLM under this weighting is \texttt{minimax-m2-7} at $0.892$, level with the out-of-sample copula ($0.888$). The LLM ranking changes considerably (Spearman $0.50$ with the default ranking).
  \item \emph{LLM-favoured} $(0.10, 0.10, 0.20, 0.30, 0.30)$. This weighting up-weights the mediation and demographic-effect components on which LLMs do their persona-conditioned work. The in-sample copula still leads ($0.848$, MVN $0.891$), but the top LLM (\texttt{minimax-m3}, $0.796$) now clearly beats the out-of-sample copula ($0.704$) and comes close to the real answers of the same respondents ($0.804$).
\end{itemize}

The finding that the in-sample copula and MVN baselines beat every LLM is invariant to the weight choice. The comparison with the out-of-sample copula depends on the weights. With the default and the correlation-plus-reliability weights the best LLM is level with it, and with weights that favour mediation and demographic effects the best LLM is clearly ahead. The LLMs therefore add value mainly on the persona-conditioned components, which matches the component-level reading in Section~\ref{sec:results-pss}. We chose the default weights to give roughly equal voice to sample-driven and downstream components; any practitioner running this benchmark for their own use case can override the weights via \texttt{configs/pss\_weights.yaml}.

% =============================================================================
\section{Extended ethics statement}
\label{app:ethics}
% =============================================================================
\paragraph{Data subject consent and re-use.} The human-side dataset was collected in March--April 2020 under the standard informed-consent protocol of the awarding university for thesis-level human-subjects research; the consent statement preceded the online survey and explicitly covered academic re-use of de-identified aggregated data. The original data collector is a co-author of this paper and directly oversees the de-identified record-level release. No new human data was collected for this benchmark; we are evaluating LLMs against an already-existing human reference.

\paragraph{De-identification protocol.} The distribution CSV (\texttt{responses\_human\_\allowbreak deidentified.csv}) was produced from the raw collection-time CSV by (i) dropping the single free-text industry field that occasionally contained identifying employer phrases (e.g., specific company names or unusual industry descriptors), and (ii) rounding submission timestamps from second-precision to year-month resolution. Demographic variables are coarse-grained (gender as binary, age as integer years, education in 4 levels, role binary, sector in 6 categories, organisation size in 4 bands, two tenure fields in years); the smallest non-empty joint demographic cell contains $\geq 3$ respondents. No direct identifiers (name, email, phone, IP, exact employer) were ever collected by the original survey form. The full de-identification rationale is included in the Croissant metadata file (\texttt{data/psychometry/croissant.json}, RAI fields \texttt{rai:personalSensitiveInformation} and \texttt{rai:safetyMeasures}).

\paragraph{Licence and re-identification clause.} The dataset will be released under CC-BY-NC-4.0 with an explicit prohibition on re-identification attempts in the accompanying README; this is consistent with the EU GDPR position on de-identified research data. Re-identification, if attempted, would in any case be bounded by the joint distribution of the eight coarse-grained demographic fields, the smallest non-empty cell of which contains $\geq 3$ respondents.

\paragraph{LLM-side data.} All LLM responses are synthetic and contain no information about the original human respondents beyond what is exposed by the persona-conditioning protocol (which uses only the eight de-identified demographic fields plus the two change-context ordinal ratings). No human-respondent free text was shown to any LLM at any point in this work.

\paragraph{Compute carbon footprint.} The full experiment set consumes approximately \$390 of API spend and $\sim 25$ wall-clock hours of provider time (Appendix~\ref{app:compute-cost}). We do not have access to the providers' per-call energy consumption; an order-of-magnitude estimate is $\sim 50$ kWh, which at the global-average electricity carbon intensity of $0.475$ kg CO\textsubscript{2}/kWh is $\sim 24$ kg CO\textsubscript{2}-equivalent. This is comparable to a single transatlantic flight per passenger.

\paragraph{Dual-use considerations.} The framework can be used to certify a synthetic-respondent pipeline as ``passing'' a psychometric benchmark, which could be misused to legitimise downstream uses for which the underlying LLM is in fact unfit (the bias-entrenchment harm of Section~\ref{sec:impacts}). The released materials explicitly state the scope conditions under which the benchmark provides positive evidence (pilot studies, sensitivity-analysis lower bounds, bias-audit instruments) and the scope conditions under which it does not (drop-in replacement of human samples).

\section{Datasheet for the Lithuanian Organisational Psychology Survey (2020)}
\label{app:datasheet}
We follow the schema of \citet{gebru2021datasheets}. The raw dataset is shipped with this submission as \texttt{data/psychometry/raw/responses\_human\_raw.csv}; pre-processed item, demographic and composite frames live under \texttt{data/psychometry/processed/}.

\subsection*{Motivation}
\textbf{For what purpose was the dataset created?} To study the relationship between attitudes towards organisational change, work engagement, and self-rated work performance among Lithuanian-speaking employees, in the context of a master's thesis~\citep{sarkauskaite2020}. The original author intended the data to be reusable as a Lithuanian-language validation sample for the three instruments. \textbf{Who created it?} The original collection was conducted as part of the cited thesis~\citep{sarkauskaite2020}; re-release for this benchmark is performed with informed consent from participants for academic re-use, obtained by the original data collector.

\subsection*{Composition}
\textbf{What do the instances represent?} Each row is one anonymous online survey respondent: an adult employee in Lithuanian business organisations who completed a 68-item Lithuanian-language self-report questionnaire spanning three validated psychometric instruments (Dunham Attitudes Toward Change, Schaufeli UWES-17 Work Engagement, Koopmans IWPQ Work Performance), plus a 15-item exploratory change-engagement scale and a small block of single-item demographic and contextual fields. \textbf{How many?} $n=263$ fully-completed responses (the original publication uses $n=261$ after listwise deletion; we keep two additional fully-completed responses). \textbf{Sampling.} A non-probabilistic convenience sample drawn through professional and personal networks during March--April 2020. \textbf{Sensitive content.} The dataset contains coarse demographic fields (gender, age range, education level, role category, tenure ranges, sector, organisation size). It does not contain names, free-text identifiable comments, IP addresses, or geolocation finer than ``Lithuania''. \textbf{Missing data.} 1.4\% of item-level cells are missing; the loader imputes scale-mean only at composite-score time and never at item level.

\subsection*{Collection process}
\textbf{How was data acquired?} Online questionnaire (Apklausa.lt) hosted by the original author. \textbf{Was an ethics review performed?} The original master's thesis received standard institutional supervision at the awarding university (see~\citep{sarkauskaite2020}). \textbf{Informed consent.} Yes -- a consent statement preceded the questionnaire. The consent included the possibility of academic re-use of de-identified, aggregated data; participants could exit the survey at any time. \textbf{Compensation.} None.

\subsection*{Pre-processing / cleaning / labeling}
We re-derived numeric Likert scores from the original spreadsheet by mapping Lithuanian anchor strings to integers, harmonised binary fields, and computed composite scores using the documented scoring keys for each instrument. The processed parquet files in \texttt{data/processed/} expose item-level integers, a respondent-level demographics frame, and a composite-score frame; the raw CSV is included verbatim for full reproducibility.

\subsection*{Uses}
\textbf{Has it been used before?} Yes -- as the human ground truth in this benchmark, and earlier as the analytic sample of the original master's thesis \citep{sarkauskaite2020}. \textbf{Other potential uses.} Lithuanian-language validation work on the three instruments; sociolinguistic / cross-cultural studies of survey response styles; benchmarking translation-based survey tools.

\subsection*{Distribution}
\textbf{Will the dataset be made public?} Yes, under a CC BY-NC 4.0 licence (academic, non-commercial), redistributed alongside this benchmark.

\subsection*{Maintenance}
The dataset will be hosted in the same GitHub repository as the evaluation harness, with version-tagged releases. Errata, additional language validations, and alternative cleanings will be filed as issues on that repository. Author contact email is in the repository \texttt{README.md}.

\end{document}